\documentclass[a4paper,11pt]{article}
\pdfoutput=1 
\usepackage{graphicx}  
\usepackage{dcolumn}   
\usepackage{bm,relsize}        
\usepackage{amssymb, amsmath}
\usepackage{textcomp}
\usepackage{wasysym}
\usepackage{slashed}
\usepackage{caption, subcaption}
\usepackage{multirow}

\usepackage{lipsum, color}
\usepackage[usenames,dvipsnames,svgnames]{xcolor}
\usepackage{jheppub} 
\usepackage{booktabs}
\usepackage[utf8]{inputenc}

\def\beq{\begin{equation}}
\def\eeq{\end{equation}}
 \def\be{\begin{equation}} \def\ee{\end{equation}}
\def\bea{\begin{eqnarray}} \def\eea{\end{eqnarray}}

\usepackage{caption}
\usepackage{subcaption}
\usepackage{multirow}

\begin{document}

\title{Earth mover's distance as a measure of CP violation}

\author[a]{Adam Davis,}
\author[b]{Tony Menzo,}
\author[b]{Ahmed Youssef,}
\author[b]{Jure Zupan,}
\affiliation[a]{School of Physics and Astronomy, University of Manchester, M13 9PL, Manchester, UK}
\affiliation[b]{Department of Physics, University of Cincinnati, Cincinnati, Ohio 45221, USA}

\emailAdd{adam.davis@manchester.ac.uk}
\emailAdd{menzoad@mail.uc.edu}
\emailAdd{youssead@ucmail.uc.edu}
\emailAdd{zupanje@ucmail.uc.edu}

\date{\today}

\abstract{We introduce a new unbinned two sample test statistic sensitive to CP violation utilizing the optimal transport plan associated with the Wasserstein (earth mover's) distance. The efficacy of the test statistic is shown via two examples of CP asymmetric distributions with varying sample sizes: the Dalitz distributions of  $B^0 \rightarrow K^+\pi^-\pi^0$ and  of $D^0 \rightarrow \pi^+\pi^-\pi^0$ decays.
The windowed version of the Wasserstein distance test statistic is shown to have comparable sensitivity to CP violation as the commonly used 
energy test statistic, but also retains information about the localized distributions of CP asymmetry over the Dalitz plot. For large statistic datasets we introduce two modified Wasserstein distance based test statistics -- the binned and the sliced Wasserstein distance statistics, which show comparable sensitivity to CP violation, but  improved computing time and memory scalings. Finally, general extensions and applications of the introduced statistics are discussed.
}

\maketitle

\section{Introduction}
The Wasserstein distance or earth mover's distance (EMD) is a measure of similarity between two probability distributions, see, e.g., \cite{2019AnRSA...6..405P} as well as App.~\ref{app:optimal-transport}. The value of the EMD can be visualized as the work required to transport and reshape dirt (weighted samples) in the form of one distribution into the form of a second distribution. Similar distributions result in smaller values ($\approx$ zero) of the EMD while dissimilar distributions result in larger values. The EMD is thus sensitive to density asymmetries between samples and therefore well suited to be used as a test statistic that quantifies the amount of CP violation (CPV) in a physical system. 

Taking as an example the $B^0$ decays to a final state $f$, the direct CP asymmetry $A_f$ is defined as 
\beq
\label{eq:Cf:def:Br}
A_f=\frac{{\rm Br}(\bar B^0\to f)-{\rm Br}(B^0\to \bar f)}{{\rm Br}(\bar B^0\to f)+{\rm Br}(B^0 \to \bar f)}, 
\eeq
where $\bar f = {\rm CP}(f)$ is the final state CP conjugated to $f$. 
For two body $B^0$ decays such as $B^0\to K^+\pi^-$, the direct CPV is fully characterized by $A_f$. In the rest frame of the parent particle, the two final state particles are back to back and there is no dependence of the decay rate on their emission angle. Direct CPV is then simply given by the difference of observed $B^0\to f$ and $\bar B^0 \to \bar f$ decays as in Eq.~\eqref{eq:Cf:def:Br}. This is not the case, however, for three-body decays such as, for instance, $B^0\to K^+\pi^-\pi^0$ and its CP conjugate mode $\bar B^0\to K^-\pi^+\pi^0$~\cite{BABAR:2011ae}. In addition to the integrated CPV quantity, $A_f$, there is a continuous set of CP violating observables, namely the phase space dependent differential CP asymmetries
\beq
\label{eq:diff}
\begin{split}
\mathcal{A}_{\mathrm{CP}}(s_{12}, s_{13})=
\biggr(\frac{d \Gamma(\bar B^0\to \bar f)}{dp.s.}-\frac{d \Gamma ( B^0\to  f)}{d p.s.}\biggr)\biggr/ \biggr({\frac{d \Gamma(\bar B^0\to \bar f)}{dp.s.}+\frac{d \Gamma (B^0\to f)}{d p.s.}}\biggr),
\end{split}
\eeq
where $s_{12},s_{13}$ are the two Dalitz plot variables. 
To measure  $\mathcal{A}_{\mathrm{CP}}$ one can bin the Dalitz plot in large enough bins such that they contain reasonably large numbers of events, say $n_i, \bar n_i \sim {\mathcal O}(20)$,  and define $A_{f,i}$, Eq.~\eqref{eq:Cf:def:Br}, for each bin. In this way one could probe experimentally, if CP violation is present in the Dalitz plot distributions. 

Such an approach is not optimal, however, since the measurements depend on the choice of the binning.
If the primary goal is to test for the presence of phase space dependent CPV in the Dalitz plot distributions, not just in global $A_f$, two tests were put forward that improve on the binning method,  the $S_{CP}$ test (or the Miranda method) \cite{Bediaga:2009tr,BaBar:2008xzl} and the energy test \cite{Aslan:2004,Williams:2011cd,LHCb:2014nnj,Parkes:2016yie}, both of which have some drawbacks. The $S_{CP}$ test still relies on a binning procedure that, like $\mathcal{A}_{\text{CP}}$, leads to some loss of sensitivity to CPV and the energy test,  while being quite sensitive to the presence of CPV in the Dalitz plot distributions, is harder to interpret in terms of the underlying physics.

In this paper we propose an alternative approach -- the use of the Wasserstein distance, or EMD, as a measure of CPV in the Dalitz plot distributions. As we show below, the EMD based statistic combines the high sensitivity to CPV with easier interpretability, since it retains information about which part of the Dalitz plot the CPV originates from. The use of EMD in measuring CPV is reminiscent but distinct from the use of EMD to quantify the similarities between different LHC events, advocated in \cite{Komiske:2020qhg,CrispimRomao:2020ejk,Cai:2020vzx,Cai:2021hnn} (see also the related results in Refs. \cite{Komiske:2022vxg,Mitchell:2022dnf,Kitouni:2022qyr}). In particular, the optimal EMD based statistic for CPV involves reweighting (or filtering) of individual datapoint contributions to the EMD as we discuss in more detail in Sec. \ref{sec:win:EMD}.

The paper is organized as follows. In Sec. \ref{sec:EMD} we review the Wasserstein distance and introduce the relevant notation for its application to three body $B$ and $D$ decays. In Sec.~\ref{sec:three-bodyB} we analyze three body $B^0\to K^+\pi^-\pi^0$ decays and show that the Wasserstein distance is a sensitive probe of CP violation and introduce an optimized windowed Wasserstein distance statistic. In Sec.~\ref{sec:D:dec} we introduce two further Wasserstein distance based statistics, the binned Wasserstein distance and the sliced Wasserstein distance, which have improved computing complexity scalings and may be preferred when dealing with large datasets such as the three body $D$ decay data samples. We draw conclusions in Sec.~\ref{sec:concl}, while appendices contain details about the public code \texttt{EMD4CPV} (App.~\ref{sec:code}), on the computation complexity of the optimal transport problem (App.~\ref{app:optimal-transport}), further examples for EMD using Gaussian distributions (App. \ref{sec:Gauss:2D}), further results for probing $B\to K\pi\pi$ Dalitz plot CP asymmetries using Wasserstein distance based statistics (App.~\ref{sec:app:further:EMD}), and a review of the energy test (App.~\ref{app:energy_test}).

\section{Earth mover's distance as a measure of CPV}
\label{sec:EMD}
 
The Wasserstein distance, $W_q({\cal E}, \bar{\cal E})$, between the distributions of events, ${\cal E}$, in $B^0\to K^+\pi^-\pi^0$, and the distribution $\bar {\cal E}$ of $\bar B^0\to K^-\pi^+\pi^0$ decays
 is given by, see, e.g.,   \cite{Komiske:2020qhg,villani2008,Santa2015,ramdas2017wasserstein},
 \beq
 \label{eq:def:Wp}
 W_q({\cal E}, \bar{\cal E})=\biggr[\underset{\{f_{ij}\geq 0\}}{\min}\sum_{i=1}^N\sum_{j=1}^{\bar N}f_{ij}\big(\hat d_{ij}\big)^q\biggr]^{1/q},
 \eeq
 where $q\in (0,\infty)$, with $q=1$ defining the EMD.\footnote{In most works on  the optimal transport 
  $q$ is restricted to the convex cost functions, $q\in [1,\infty)$, such that its gradient is well defined everywhere, also at the $\hat d_{ij}=0$ point. An extension to the concave case, $q\in (0,1)$, requires an  introduction of an approximate gradient, however, a unique optimal transport still exists, see the discussion in chapter 3.3.2 of Ref.~\cite{Santa2015}. The network simplex algorithm as implemented in the Wasserstein {\tt Python} library~\cite{Komiske:2019fks,Komiske:2020qhg} can then be used without change to solve the optimal transport problem, in the same way as for $q\geq 1$.
 }
 The minimization is over the weights
   \beq
 \label{eq:constr:f:var}
 \sum_{i=1}^Nf_{ij}= \frac{1}{\bar N}, \qquad \sum_{j=1}^{\bar N}f_{ij}= \frac{1}{N}, \qquad \sum_{i, j=1}^{N,\bar N} f_{ij}=1,
 \eeq
where $N (\bar N)$ are the number of events in sample ${\cal E} (\bar {\cal E})$, and $\hat d_{ij}$ is the distance between the two events, $i$ in ${\cal E}$, and $j$ in $\bar {\cal E}$. The interpretation of $W_q({\cal E}, \bar{\cal E})$ is the cost incurred by moving in an optimal way the probability distribution corresponding to events in ${\cal E}$ into the probability distribution of event $\bar {\cal E}$, where the penalty is the distance $\hat d_{ij}$ between the events.

 Assuming that $N=\bar N$, so that that there is no integrated CP asymmetry, and that ${\cal E}$ and $\bar {\cal E}$ come from the same distribution (i.e. no CPV in distributions), then $W_q({\cal E}, \bar{\cal E})\to 0$ for large $N=\bar N$. In contrast, if ${\cal E}$ and $\bar {\cal E}$ differ (there is CPV), then  $W_q({\cal E}, \bar{\cal E})$ will tend to a nonzero value. For $d-$dimensional final phase space the parametric upper bound is $\langle W_p({\cal E}, \bar {\cal E})\rangle \lesssim C N^{-1/d}$  \cite{weed2017sharp}, with $C$ a constant that does not depend on $N$.\footnote{Note that for decays that are dominated by intermediate resonances the effective dimensionality is lower than the full dimensionality of the phase space. That is for a multibody decays where at most two resonances overlap we expect the same scaling as for the Dalitz plot $\langle W_p({\cal E}, \bar {\cal E})\rangle \lesssim C N^{-1/2}$.} For the Dalitz plot we have $d=2$ since it is fully described by two Dalitz variables, $s_{12}, s_{13}$, and thus $\langle W_p({\cal E}, \bar {\cal E}) \rangle \propto 1/\sqrt N$, i.e., it scales in the same way as the variance of the global direct CP asymmetry $\delta A_f\propto 1/\sqrt{N}$. Since we are mainly interested in CPV in distributions, we will assume for simplicity that $N= \bar N$ in the rest of the manuscript. However, the analyses we present below extend trivially to the  $N\ne \bar N$ case, with $W_q$ still probing the CPV in distributions and $A_f$ the integrated CPV.

For the 3D Dalitz plot we use the definition of the (dimensionless) distance $\hat d_{ij}$ that is symmetric in the Dalitz variables, $s_{12}, s_{13}, s_{23}$, 
\beq
\label{eq:hat:dij:Dalitz}
\hat d_{ij}\Big|_{\rm Dalitz}=\frac{1}{m^2}\Big\{\big|s_{12}(i)-\bar s_{12}(j)\big|^r+\big|s_{13}(i)-\bar s_{13}(j)\big|^r+\big|s_{23}(i)-\bar s_{23}(j)\big|^r\Big\}^{1/r},
\eeq
where, for example in the $B^0\to K^+\pi^-\pi^0$ system, $m=m_B$, and
\begin{align}
s_{12}&=(p_{K^+}+p_{\pi^-})^2, \qquad s_{13}=(p_{K^+}+p_{\pi^0})^2, \qquad s_{23}=(p_{\pi^-}+p_{\pi^0})^2,
\\
\bar s_{12}&=(p_{K^-}+p_{\pi^+})^2, \qquad \bar s_{13}=(p_{K^-}+p_{\pi^0})^2, \qquad \bar s_{23}=(p_{\pi^+}+p_{\pi^0})^2,
\end{align}
parametrize the $B^0\to K^+\pi^-\pi^0$ Dalitz plot and the CP conjugate variables in $\bar B^0\to K^-\pi^+\pi^0$ Dalitz plot, respectively. The normalization prefactor $1/m^2$ in Eq.~\eqref{eq:hat:dij:Dalitz} was chosen such that $\hat d_{ij}<1$. We use the Euclidean distance, i.e., $r = 2$, in the remainder of the paper. Other $r$--values were investigated but no significant changes to the sensitivity of CP violation were found.

\begin{figure}[t]
\begin{center}
\includegraphics[width=0.8\textwidth]{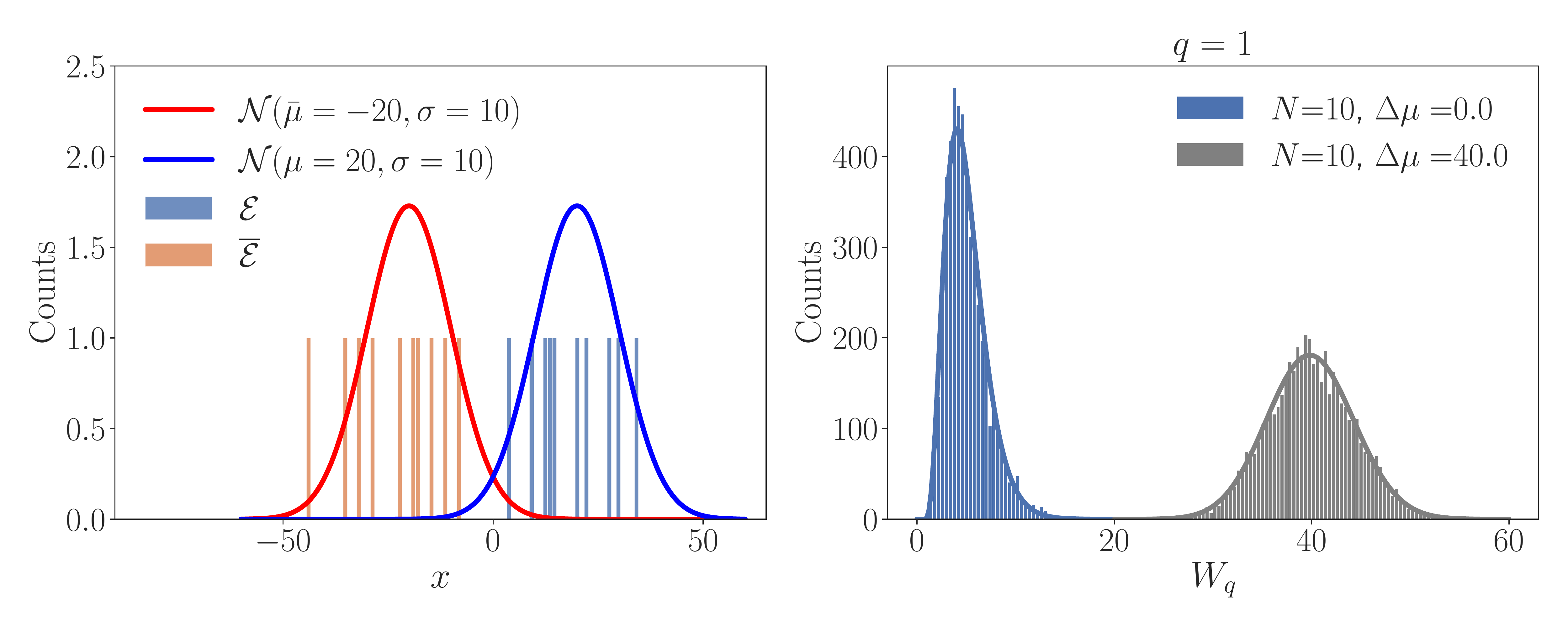}
\caption{
The nonzero displacement $\Delta\mu$  of the two Gaussian distributions (left) can be probed by using $W_q$, $q=1$, as the test statistic (right), see main text for details. }
\label{fig:wass_1d}
\end{center}
\end{figure}

Before discussing the more complicated case of $B$ and $D$ decays, let us first briefly consider a simpler toy example of two displaced Gaussian distributions, $G(x)={\cal N}(x|\Delta \mu/2,\sigma)$ and $\bar G(x)={\cal N}(x|-\Delta\mu/2,\sigma)$, i.e., two Gaussian distributions with equal widths, $\sigma$, but with their centers at $\mu=\Delta\mu/2$ and $\bar \mu=-\Delta \mu/2$ and thus displaced by $\Delta \mu$. In this toy example the question about CPV in multibody $B$ decays is replaced with a test whether or not $\Delta \mu\ne0$.
Drawing $N=10$ events ${\cal E}$ from $G$, as well as $\bar N=10$ events $\bar {\cal E}$ from $\bar G$,  and taking $\hat d_{ij}$ in Eq.~\eqref{eq:def:Wp} to be the Euclidean distance in 1D, gives a $W_1$ that is clustered around $\langle W_1\rangle \simeq \Delta \mu$, see the grey distribution in Fig.~\ref{fig:wass_1d} (right). This is appreciably larger than the distribution of $W_1$ values for $\Delta \mu=0$ (blue), even for relatively small event samples. In App.~\ref{sec:Gauss:2D} we show more illustrations of how the $W_1$ probes a difference between distributions, including an example of displaced 2D Gaussian distributions. In particular, we show numerically that $W_1$ can be used as a statistic, and that the CL intervals obtained from a known $\Delta \mu=0$  probability distribution for $W_1$ coincide with the expected exclusion intervals from negative log likelihood for $\Delta \mu$. 

\section{Application to three body $B$ decays}
\label{sec:three-bodyB}

\begin{figure}[t]
\begin{center}
\includegraphics[width=0.88\textwidth]{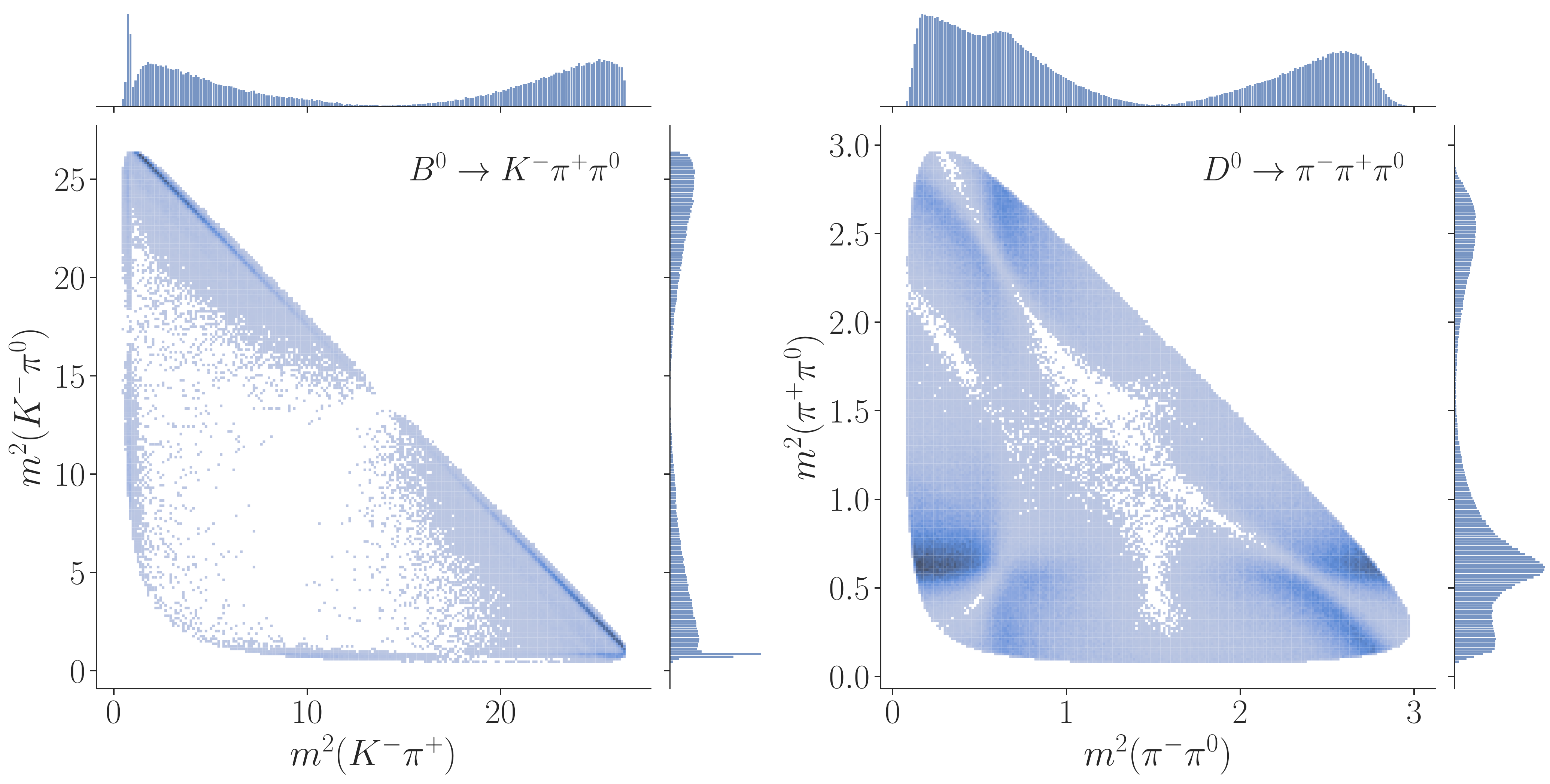}  
\caption{The 2--dimensional $B^0 \rightarrow K^+\pi^-\pi^0$ (left) and $D^0\to \pi^-\pi^+\pi^0$ (right) Dalitz plots and their respective 1--dimensional histograms along the borders for $10^6$ events.}
\label{fig:dalitz}
\end{center}
\end{figure}

As the first realistic example of using the Wasserstein distance to test for CP violation we use the $B^0 \rightarrow K^+\pi^-\pi^0$ and the CP conjugate $\bar{B}^0 \rightarrow K^-\pi^+\pi^0$ decays. The events are generated from the amplitude model by BaBar \cite{BABAR:2011ae} implemented in the \texttt{AmpGen} \cite{ampgen} framework. We create two data samples: the CP conserving (CPC) and the CP violating datasets. For the CPC datasets we use the central values of amplitudes and phases in the $B^0$ BaBar isobar model \cite{BABAR:2011ae} for both $B^0$ and $\bar B^0$ decays. For the CPV datasets, on the other hand, the amplitudes and phases for $B^0$ and $\bar B^0$ isobar models differ and are set to the central values of the measurements in Ref. \cite{BABAR:2011ae}. The $B-\bar B$ meson mixing is ignored in the generation of the samples.  The resulting $B^0 \rightarrow K^+\pi^-\pi^0$ Dalitz plot   with $10^{6}$ events is  shown  in Figure \ref{fig:dalitz} (left).

For three-body $B$ decays we highlight the use of $W_q(\hat{d}_{ij})$ on the low statistic datasets containing $N=10^3$ events in each of the samples, the $B$ and $\bar{B}$ decays ($N = \bar{N}$). This choice was made to roughly match the reported experimental sensitivity \cite{LHCb:2014mir}. 
The implementation and computation of the Wasserstein distance is done in two steps: first, the distances $\hat{d}_{ij}$, Eq.~\eqref{eq:hat:dij:Dalitz}, are computed using the \texttt{cdist} method within the \texttt{SciPy} framework \cite{2020SciPy-NMeth} which utilizes optimized \texttt{C} code to efficiently compute the distances. The computations of $W_q(\hat{d}_{ij})$ and the extraction of optimal transport data is then obtained using the \texttt{EMD} class within the \texttt{Wasserstein} library~\cite{Komiske:2019fks,Komiske:2020qhg}.
 There are two continuous parameters in the definition of $W_q(\hat{d}_{ij})$, $r$ and $q$, cf. Eqs.~\eqref{eq:def:Wp}, \eqref{eq:hat:dij:Dalitz}. These can be chosen such that the sensitivity to CPV is maximized. The optimal value of $q=0.1$ was chosen by finding, for $r=2$, the minimum average CL $p$--value for which the CPC hypothesis is excluded given the toy model CPV Dalitz plot distributions, as obtained from an ensemble of $N_e=500$ distinct datasets generated from the BaBar model \cite{BABAR:2011ae}, see further details in App.~\ref{sec:app:further:EMD}. In the analyzed examples, changing $r$ in the definition of the distance Eq.~\eqref{eq:hat:dij:Dalitz} did not lead to significant changes in the sensitivity. Thus, in the numerical results below we use the optimized values $\{r=2,q=0.1\}$, while in App.~\ref{sec:app:further:EMD} we also show the results for the non-optimal choices, $\{r=2,q=1\}$ and $\{r=2,q=10\}$.
 
 To determine the $p-$value with which the CPC  hypothesis is excluded for the particular CPV Dalitz plot sample, one needs the $W_q$ probability distribution functions (PDF) for the CPC Dalitz plot distributions. In the experiment one can determine the CPC PDF using the permutation method, which, as we show next, is estimated to lead to only a relatively small bias compared to the true CP conserving PDF. 
 
\subsection{Testing for bias in the permutation method}\label{subsec:gen_CPC_dist_methods}
\begin{figure}[t!]
\begin{center}
\includegraphics[width=0.6\textwidth]{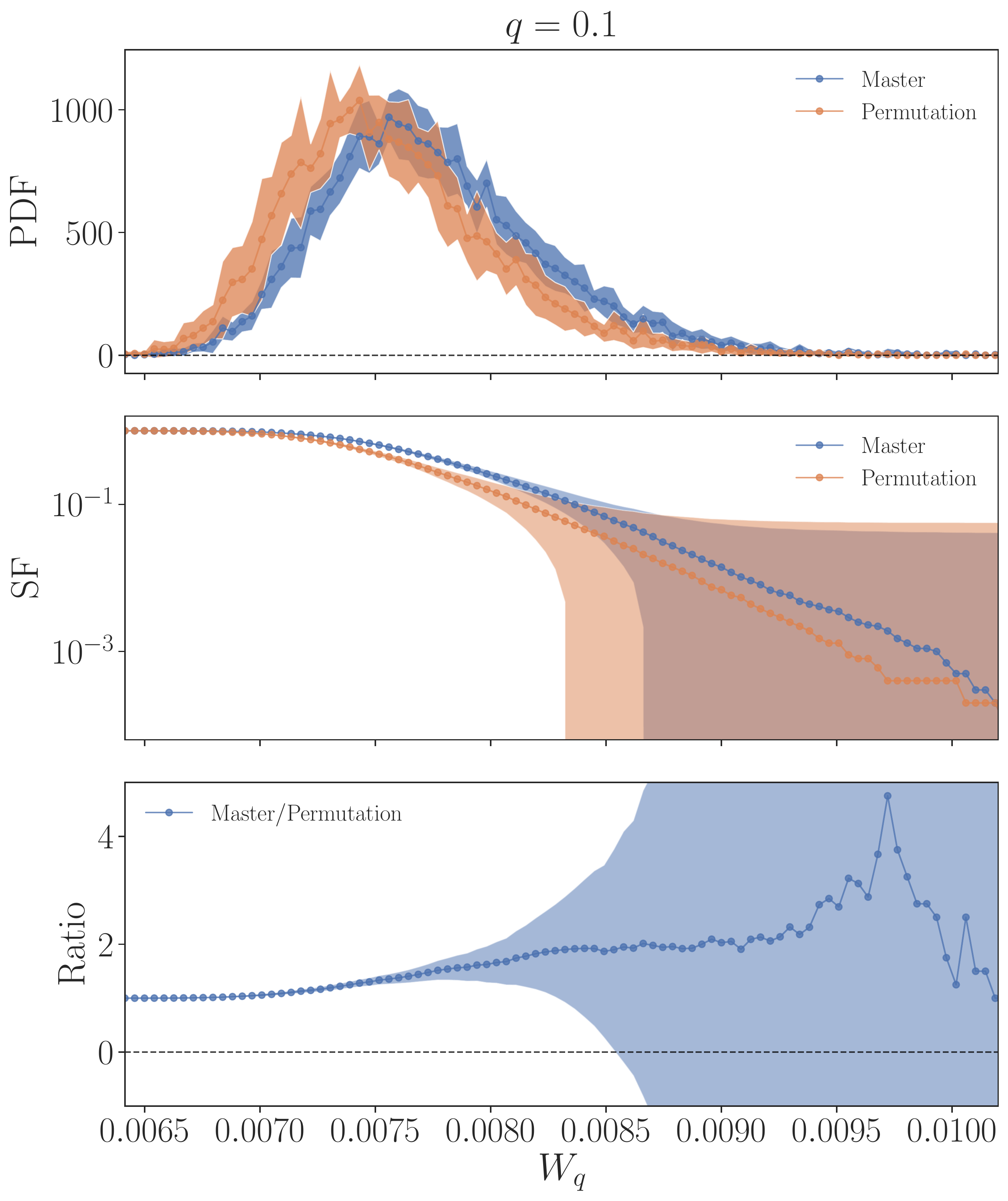}

\caption{The $W_q$ distribution function (PDF, top panel), and the survival factor (SF=1-CDF, where  CDF  is the cumulative distribution function, middle panel) obtained from the permutation method (orange) compared to the true CP conserving distribution (the master method, in blue), while the bottom panel shows the ratio of the SF obtained using the two methods. Each distribution consists of $10^3$ $W_q$ values with the solid curves and bands representing the average $\pm 1\sigma$ ranges for the bin counts obtained over 10 distinct distributions. 
}
\label{fig:scramble_vs_CPC_PDFs}
\end{center}
\end{figure}

In order to assign a $p-$value with which the CPC hypothesis is excluded, given  two samples of $B$ and $\bar{B}$ decays, one first calculates the Wasserstein distance between the two, $W_{q}^{\text{exp}}$. This encodes the dissimilarity between the two distributions of events. However, the value of $W_{q}^{\text{exp}}$ by itself is not particularly informative, except that smaller $W_q^{\rm exp}$ values  indicate more similar distributions. For a quantitative assessment of CPV we need the distribution of $W_q$ for the CP conserving case. We obtain this using two methods: 1. using the {\em permutation method}, i.e., by permuting the original $B$ and $\bar{B}$ samples (which have non-zero direct CPV) and then calculating $W_q$ for each such permutation and 2. using the {\em master method}, which is the true CP conserving PDF given our assumptions: we generate an ensemble of $B$ and $\bar B$ decay event samples, using the $B$ decay model for both,  and then calculate the corresponding $W_q$ probability density function (that is, we assume for simplicity that all the CP violating phases reside in the $\bar B^0$ decay amplitude). The permutation method can be implemented with experimental data, since it involves only the measured $B$ and $\bar B$ event samples. The master method, on the other hand, is only possible given a theoretical model of the decay amplitudes.  

The PDFs for the two methods, the permutation (orange) and master (blue), are shown in Fig.~\ref{fig:scramble_vs_CPC_PDFs}, as obtained from an ensemble of $N_e=10$ datasets containing $N=\bar N=10^3$ events in each sample. 
We see that the permutation method is a very good approximation of the true CP conserving PDF for $W_q$. Such a test of a possible bias in the permutation method 
can be performed for any multibody $B$ decay (or any multibody distribution in general) for which a reasonable description is available in terms of a resonance amplitude model. 

One can also test for a potential bias in the permutation method using only experimental data, but in this case only for $N$ that corresponds to a fraction, for instance half, of the measured sample size. That is, from data one can construct several distinct hypotheses for the CP conserving $W_q$ PDF.  The first CP conserving $W_q$ PDF hypothesis can be constructed by randomly splitting the measured $B$ decay sample into two halves and calculating the corresponding distribution of $W_q$. An alternative CP conserving $W_q$ PDF hypothesis is similarly obtained by randomly splitting the measured $\bar B$ decay sample. These can then be compared to the $W_q$ PDF that is obtained using the permutation method (but again using only half of the measured $B$ and $\bar B$ decay samples). The differences between the three PDFs should be a good proxy for the size of the possible bias in the permutation method when applied to the full dataset. 

\begin{figure}[t!]
\begin{center}
\includegraphics[width=0.65\textwidth]{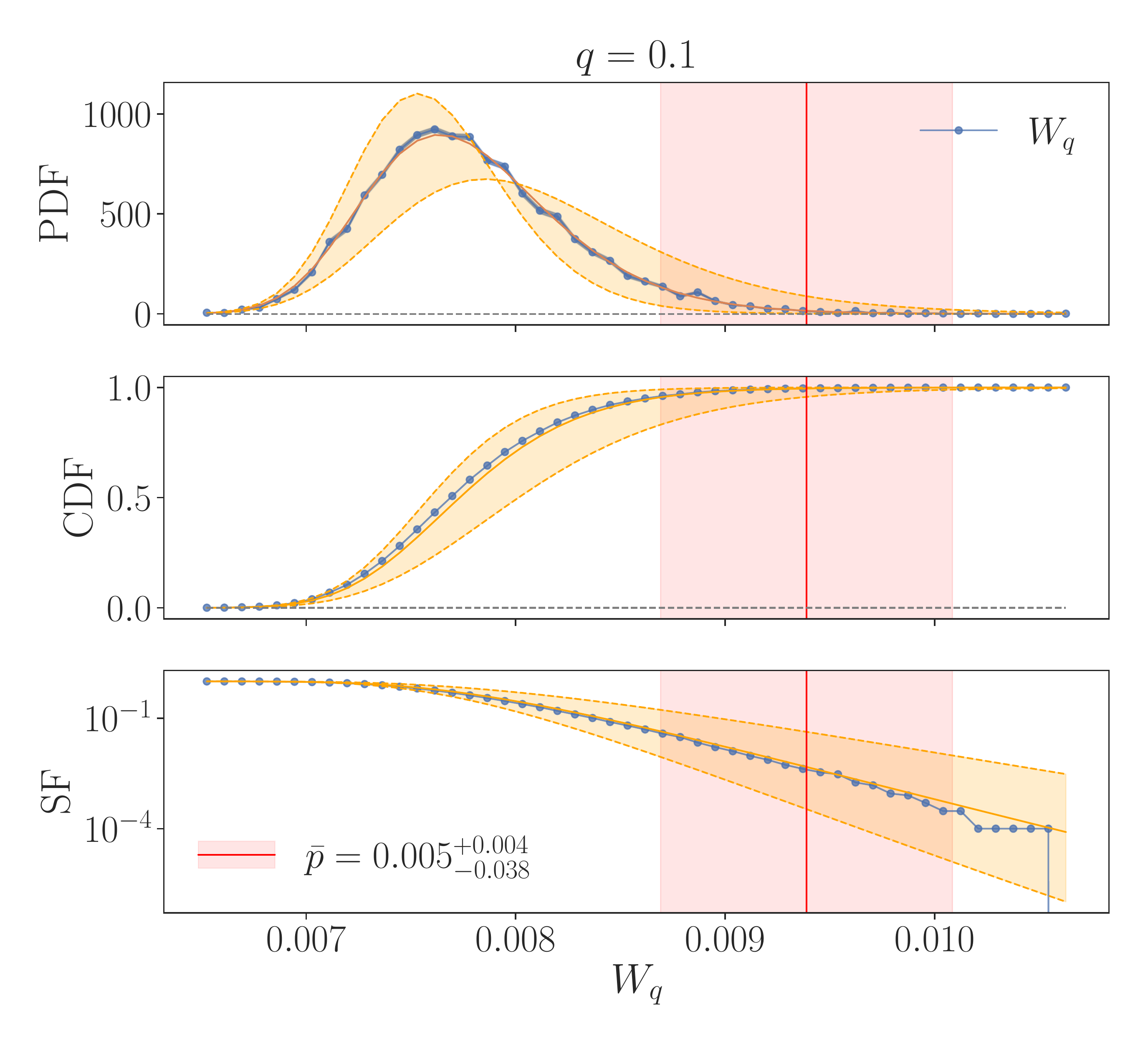}
\caption{The $W_q$  probability distribution function (PDF), the cumulative distribution function (CDF), and the survival factor (SF=1-CDF) for the CPC case and  $r=2,$ $q = 0.1$, obtained using the master method with a fit to the Johnson's $S_U$ distribution. The orange bands (blue band on top panel) denote the $\pm 1\sigma$ fit errors (statistical errors). 
The vertical red line (band) denotes the average $W_q$ value (the $\pm 1\sigma$ $W_q$ ranges) obtained from $10^3$ CPV datasets. We see that, on average, the CPC hypothesis is in this example excluded at the $\sim 3\sigma$ level, i.e., with a $p-$value of $\sim 0.005$. 
 }
\label{fig:PDF:CDF:SF}
\end{center}
\end{figure}

In the numerical results below we use the master method, i.e., the true CP conserving PDF for $W_q$ shown in Fig. \ref{fig:PDF:CDF:SF}, even though this is not accessible from experimental data. This choice was done for numerical expediency, and we expect it to introduce only small bias in the comparisons.

\subsection{Tracing CP violating phase space regions using EMD}
\begin{figure}[t]
\begin{center}
\includegraphics[width=\textwidth]{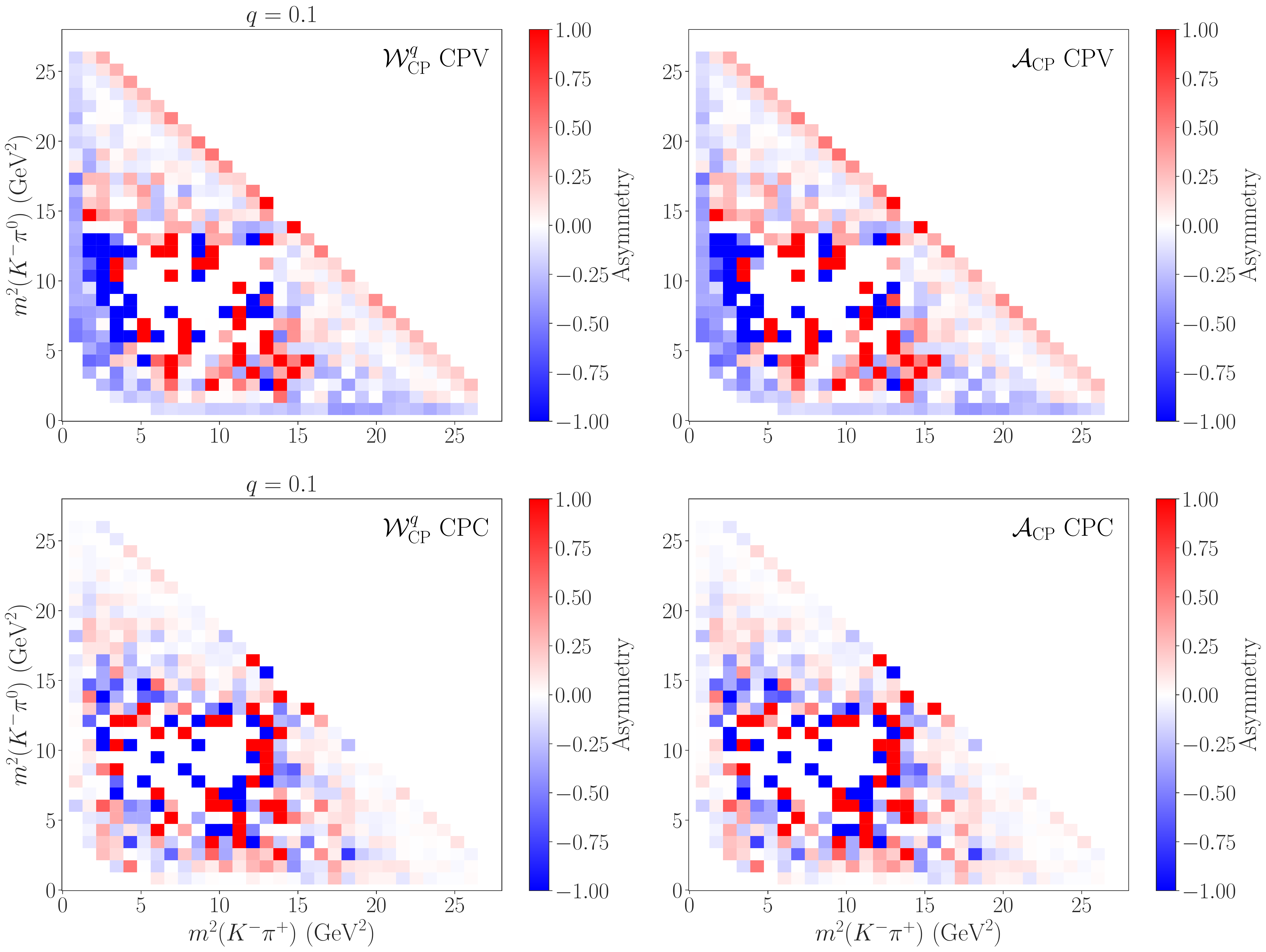}
\caption{
Binned Dalitz plot comparison between the Wasserstein asymmetry $\mathcal{W}_{\text{CP}}^q$ (left) and direct CP asymmetry $\mathcal{A}_{\text{CP}}$ (right), shown for CP violating $B^0\to K^+\pi^-\pi^0$ decays (top) and CP conserving decays (bottom), i.e., decays in which the asymmetries in the amplitude model were set to zero. The results shown are normalized and averaged over $100$ datasets, each containing $2N=2 \times 10^3$ ($B$ and $\bar{B}$) events. 
}
\label{fig:asymmetry_plots:q0.1}
\end{center}
\end{figure}

A benefit of the Wasserstein distance based statistic is that it traces in a straightforward fashion the variation of the CP asymmetry across the Dalitz plot. The standard definition of direct CP asymmetry, Eq.~\eqref{eq:Cf:def:Br}, also applies to the differential distributions, Eq. \eqref{eq:diff}, repeated here for convenience, 
\begin{equation}\label{eq:A_CP}
\mathcal{A}_{\mathrm{CP}}(s_{12}, s_{13})=\frac{d\bar{\Gamma}(\bar{s}_{12}, \bar{s}_{13})-d\Gamma (s_{12}, s_{13})}{d\bar{\Gamma}(\bar{s}_{12}, \bar{s}_{13})+d \Gamma ( s_{12}, s_{13})}, 
\end{equation} 
where $d\Gamma(s_{12}, s_{13})$ is the $B^0\to K^+\pi^-\pi^0$ partial decay width into the region of the Dalitz plot with $s_{12}=(p_{K^+}+p_{\pi^-})^2\equiv m^2(K^+\pi^-)$, $s_{13}=(p_{K^+}+p_{\pi^0})^2\equiv m^2(K^+\pi^0)$. Similarly,  $d\bar \Gamma(\bar s_{12}, \bar s_{13})$ is the CP conjugate partial decay width for $\bar B^0\to K^-\pi^+\pi^0 $, with $\bar s_{12}=(p_{K^-}+p_{\pi^+})^2\equiv m^2(K^-\pi^+)$, $\bar s_{13}=(p_{K^-}+p_{\pi^0})^2\equiv m^2(K^-\pi^0)$.  The binned version of the CP asymmetry ${\cal A}_{\rm CP}$ for the CP violating dataset, where we used the central values of the parameters for the BaBar amplitude model from \cite{BABAR:2011ae}, is shown in the upper-right panel in Fig.~\ref{fig:asymmetry_plots:q0.1}. The lower-right panel in Fig.~\ref{fig:asymmetry_plots:q0.1} shows the binned ${\cal A}_{\rm CP}$ for the CP conserving case, i.e., assuming that the $B^0\to K^+\pi^-\pi^0$ inputs in the amplitude model \cite{BABAR:2011ae} apply to both the $B^0$ and $\bar B^0$ decays. The panels in Fig.~\ref{fig:asymmetry_plots:q0.1} show expected CP asymmetries in each bin, obtained by averaging over an ensemble of $N_e=100$ datasets containing $N=\bar N=10^3$ $B$ and $\bar{B}$ pairwise samples. 

Next, we define the Wasserstein asymmetry utilizing the Wasserstein statistic $W_q$, Eq.~\eqref{eq:def:Wp}.  
We denote the contribution to $W_q$ from each datapoint $i$ in the $B^0$ Dalitz plot as $\delta W_q(i)$, and likewise $\delta \bar W_q (\bar i)$ denotes the contribution from datapoint $\bar i$ in the $\bar B^0$ Dalitz plot, such that
\begin{equation}
W_q^q=\sum_i \delta W_q(i)=\sum_{\bar i}\delta \bar W_q (\bar i).
\end{equation}
We define the binned Wasserstein asymmetry $\mathcal{W}^q_{\text{CP}}$ within each bin in Fig.~\ref{fig:asymmetry_plots:q0.1} as 
\begin{equation}
\label{eq:Wq:asymm}
    \mathcal{W}^q_{\text{CP}} (s_{12}, s_{13}) = \frac{\sum_{\bar i} \delta \bar W_q(\bar i) - \sum_{i} \delta {W}_q(i)}{\sum_{\bar i} \delta \bar W_q(\bar i) + \sum_i \delta {W}_q(i)},
\end{equation}
where the summation over $i$ $\left( \bar i \right)$ is only over the data-points contained in the bin centered at $(s_{12}, s_{13})$ (the CP conjugated $\bar B^0$ bin centered at $(\bar s_{12}, \bar s_{13})$). By construction, $ \mathcal{W}^q_{\text{CP}} $ vanishes when summed over the whole Dalitz plot, i.e., when there is only one bin encompassing the whole Dalitz plot. The Wasserstein asymmetry $ \mathcal{W}^q_{\text{CP}} $ is also statistically consistent with zero in the regions of the Dalitz plot that have vanishing CP asymmetry. Comparison of left and right panels in Fig.~\ref{fig:asymmetry_plots:q0.1} shows that $ \mathcal{W}^q_{\text{CP}} $ faithfully traces the variation of ${\cal A}_{\rm CP}$ over the Dalitz plot, including the statistical fluctuations, most readily visible in the CP conserving datasets shown in the lower panels in Fig.~\ref{fig:asymmetry_plots:q0.1}. This makes the  $\mathcal{W}^q_{\text{CP}}$ easily interpretable in terms of the underlying physics, i.e., which components of the resonant structure contribute most to the CP violation. 

\begin{figure}[t]
\begin{center}
\includegraphics[width=0.32\textwidth]{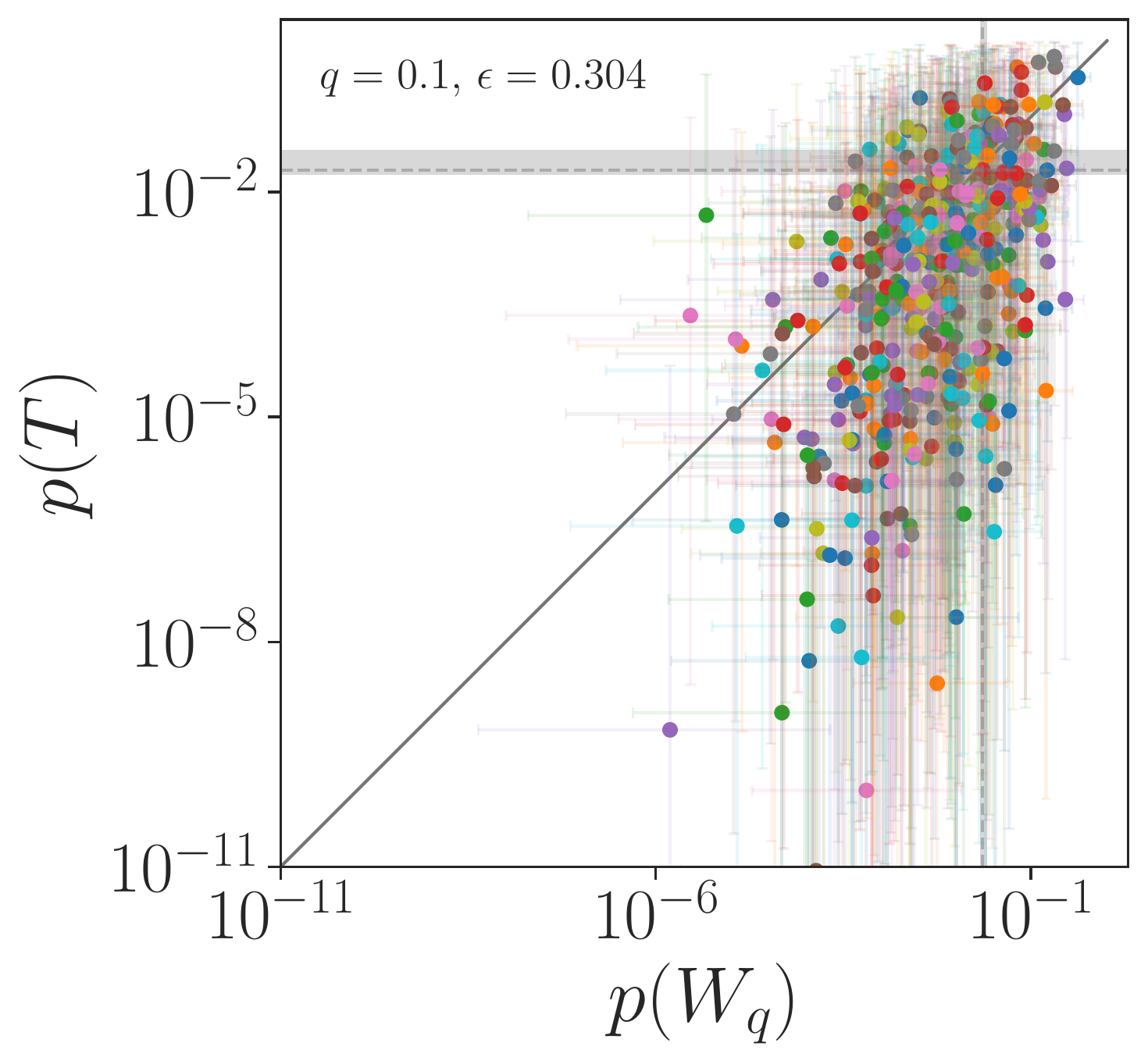} 
\includegraphics[width=0.32\textwidth]{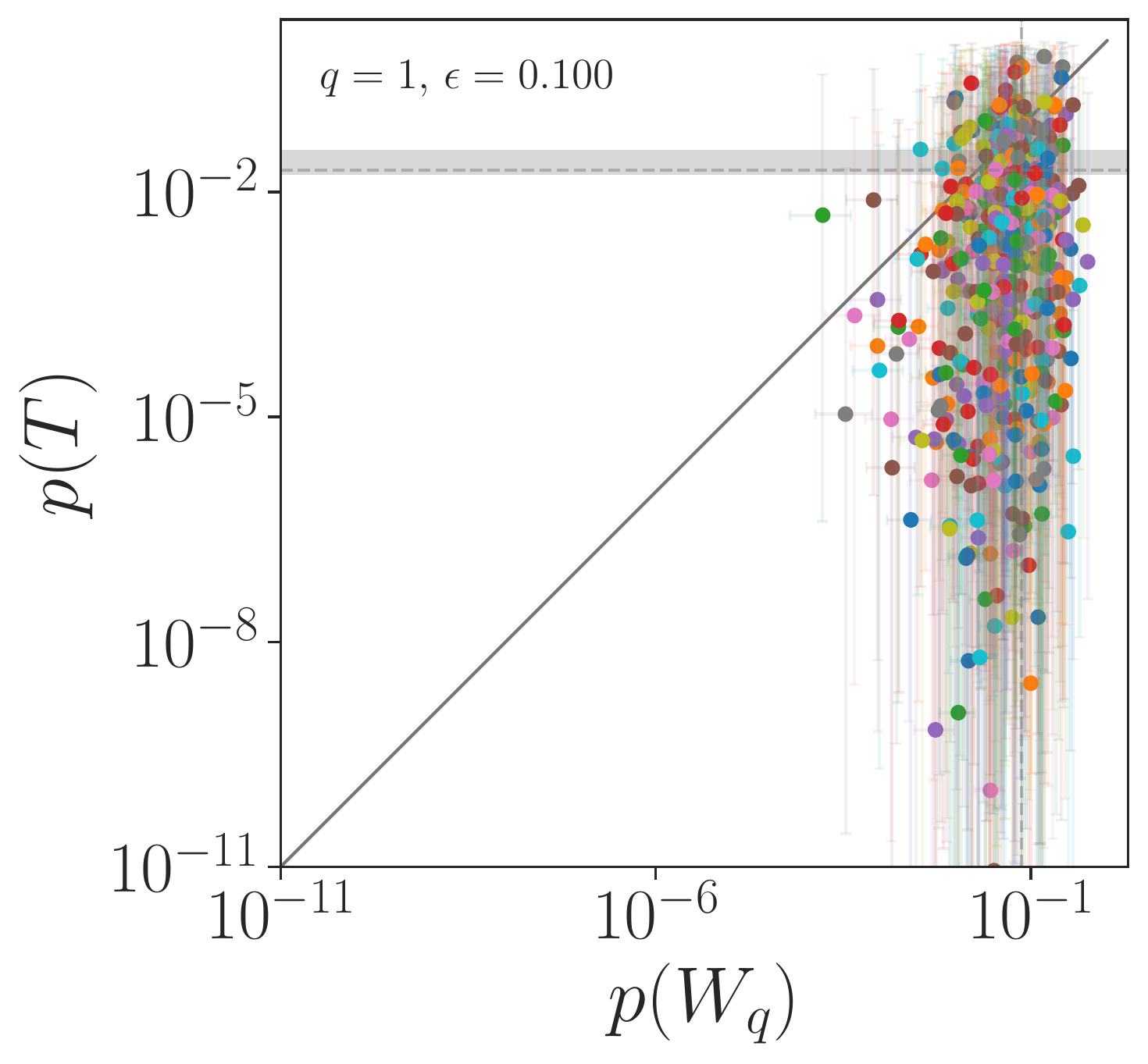}
\includegraphics[width=0.32\textwidth]{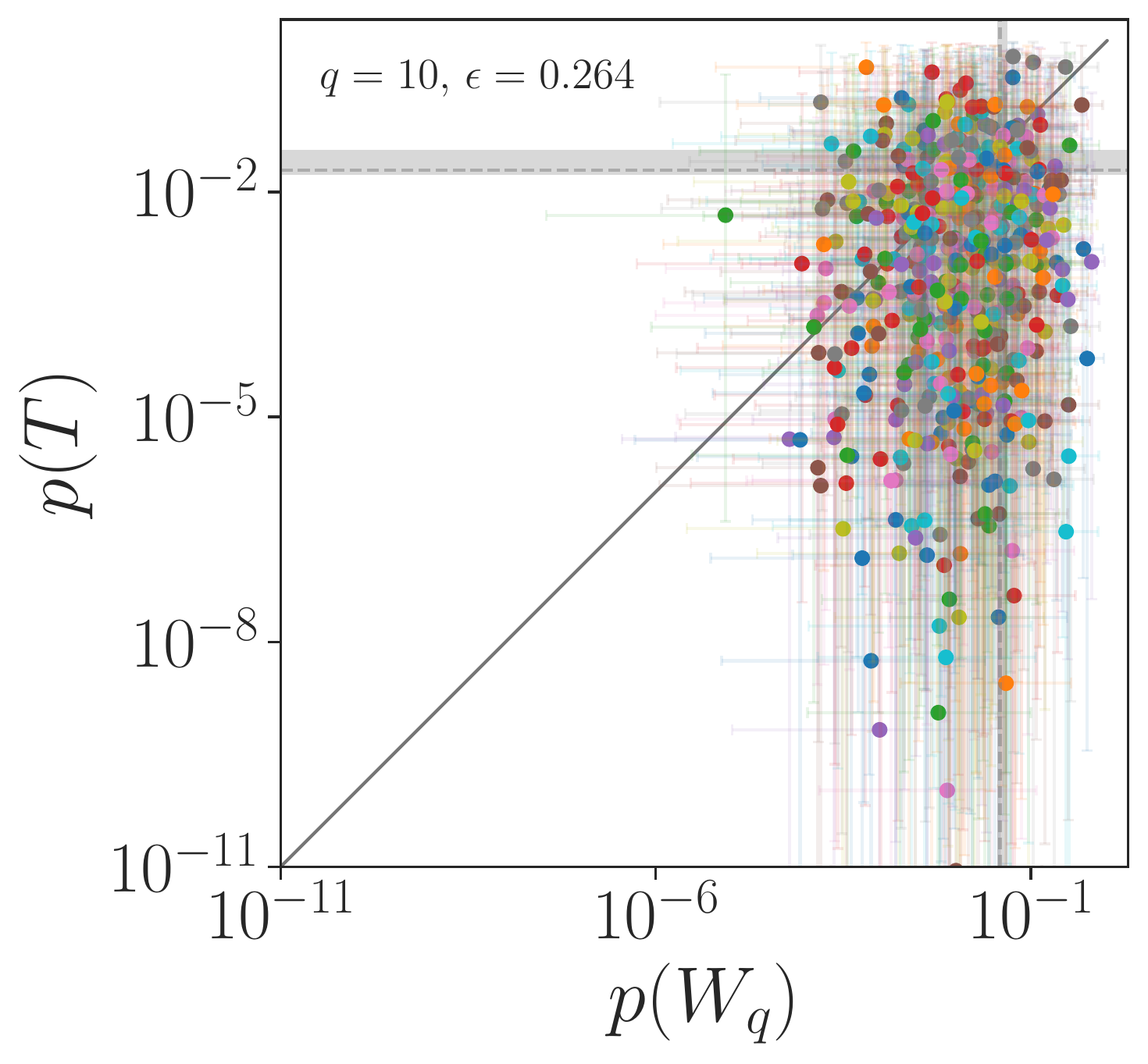} 
\caption{The scatter plot of $p$--values at which the CPC hypothesis is excluded, for the ensemble of $N_e=500$ samples with  $N=10^3$ $B^0\to K^+\pi^-\pi^0$ (and $\bar N=10^3$ CP conjugated $\bar B^0\to K^-\pi^+\pi^0$) decays, calculated using either the $W_{q}$ or $T$ statistics (dots), with $1\sigma$ fit error bars shown as lines, and setting $q=\{0.1, 1,10\}$ (from left to right).
The fraction $\epsilon$ of points above the $p(W_q)=p(T)$ diagonal line denotes the fraction of ensembles for which $W_q$ is more sensitive to CPV. 
The dotted gray lines  (solid bands) 
show the average ($1\sigma$ ranges of) $p-$values for the ensemble. 
}
\label{fig:Wq_vs_T_0.1}
\end{center}
\end{figure}

The advantage of Wasserstein distance over direct CP asymmetry, Eq.~\eqref{eq:A_CP}, as a measure of CP violation in the Dalitz plot distributions is that $W_q$ does not require binning. It is a global quantity that encodes the cumulative differences between the $B^0$ and $\bar B^0$ Dalitz plots. As such it can be used as a statistic sensitive to the CP violating Dalitz plot distributions. In Fig.~\ref{fig:Wq_vs_T_0.1} we compare the sensitivity of $W_q$ to CPV relative to another such unbinned statistic, the energy test statistic $T$ \cite{Aslan:2004,Williams:2011cd,Parkes:2016yie}, see App.~\ref{app:energy_test} for further details on the energy test. The energy test has already been successfully applied to search for CPV in multibody decays
\cite{LHCb:2014nnj}. 
On the other hand, we do not show comparisons with the $S_{\rm CP}$ test, a.k.a. the Miranda method \cite{Bediaga:2009tr,BaBar:2008xzl},  which uses optimized bins. In our numerical studies we found the $S_{\rm CP}$ test to always be less sensitive.

From Fig.~\ref{fig:Wq_vs_T_0.1} we see that the Wasserstein distance and the energy test have comparable sensitivity to CPV, but with $W_q$ somewhat less sensitive on average.   This can be quantified by introducing
\begin{equation}\label{eq:epsilon}
    \epsilon \equiv 
    \frac{1}{N_e}\sum_{i=1}^{N_e} 
    \begin{cases}
        +1 & p_i(W_q) < p_i(T),\\
        0 & \text{otherwise},
    \end{cases}
\end{equation}
where $N_e=500$ is the ensemble size for which the CPC exclusion CL  $p-$values were obtained either using the $W_q$ (giving $p(W_q)$) or the $T$ statistic (giving $p(T)$). 
That is, $\epsilon$ gives the fraction of randomly sampled datasets for which  $W_q$ statistic leads to stronger sensitivity to CPV than the energy test.  Since $\epsilon<0.5$ one may conclude that $W_q$ is on average less sensitive.  However, the average $p$--values for $W_q$ and $T$ test statistics (dashed lines) agree within $1\sigma$ ranges (gray bands). Similarly, many scatter points in Fig.~\ref{fig:Wq_vs_T_0.1} agree with the $p(W_q)=p(T)$ line within the error bars that are reflecting the uncertainties with which the $p-$values were determined from the fit. That is, for small $p-$values, $p\lesssim {\mathcal O}(10^{-4})$, we estimate the significance of the exclusion using an extrapolation of a fit to corresponding PDFs, where 
the fit distributions are chosen according to the minimization of a $\chi^2$. The energy test statistic is fit with a gamma distribution while for $q = 0.1$, we fit the $W_q$ master distribution with Johnson's $S_U$ distribution. Errors are assigned according to the 1$\sigma$ bands on the respective fit parameters, see App.~\ref{subsec:p_val_error_analysis} for further details.
The $\epsilon$ ratio does not take into account the error associated with our estimates of the $p-$value for each statistic. These errors can be large especially for small $p$-values, and as such $\epsilon$ should only be used as a cautious measure of performance.

The $T$ statistic has a continuous parameter, $\sigma$, which defines the scale of correlations probed by the energy test. For results shown in Fig.~\ref{fig:Wq_vs_T_0.1} the value of $\sigma$ was set to its (close to) optimal value  $\sigma =0.2 \text{ GeV}^2$, for which the energy test on average leads to the smallest expected $p-$values. Similarly, the parameter $q$ in $W_q$ was optimized, with the results in Fig.~\ref{fig:Wq_vs_T_0.1} shown for close to optimal value $q=0.1$. 
Note that in the actual experiment the above optimization should be performed on the mock data, using a model for $B\to K \pi\pi$ decay amplitudes, and not on actual experimental data, in order not to introduce bias. If the amplitude model does not describe well the data, this would lead to suboptimal choice for the continuous parameter and reduced sensitivity to CPV, but otherwise is not problematic.

 We expect that the somewhat reduced sensitivity of $W_q$ to CPV compared to the energy test is because $W_q$ also receives contributions from areas in the Dalitz plot that are CP conserving. This is in contrast to the energy test statistic $T$, which has a vanishing expectation value in those areas regardless of the number of events in the dataset. The contributions to $W_q$ from these regions, on the other hand,  only slowly tend to zero with increasing sample size $N$. That is, $W_q$ may be written as the sum of two contributions
\begin{equation}\label{eq:Wq_noise}
    W_q = \sum_i \delta W_q(i) = \sum_i \left[\delta W^{\text{signal}}_q(i) + \delta W^{\text{noise}}_q(i)\right] \text{ where } \lim_{N \rightarrow \infty} \sum_i \delta W^{\text{noise}}_q(i) = 0.
\end{equation} 
The term $\delta W_q^{\text{noise}}$ comes from CP conserving regions of the Dalitz plot, while $\delta W^{\text{signal}}_q$ is due to the presence of CPV and tends to a nonzero value for $N\to \infty$. If the signal and noise contributions preferentially occur at different length scales, one can construct a modified Wasserstein distance test with higher sensitivity to CPV, as shown in the next subsection.

\subsection{The windowed EMD}
\label{sec:win:EMD}
As discussed above, the disadvantage of the Wasserstein distance as a CPV test statistic is that, because all $\delta W_q(i)$ are positive, it includes an abundance of small nonzero contributions even in the absence of CPV,  generating a long--tailed CP conserving PDF for $W_q$. Within the Dalitz plot, CP violation manifests as local density differences between the $B$ and $\bar{B}$ datasets. If this CPV is either localized and/or relatively small, such as in $B^0\to K^+\pi^-\pi^0$ Dalitz plots, this translates into relatively small differences in the $\delta W_q(i)$ distributions between CPV and CP conserving $B^0$ decays. 

\begin{figure}[t]
\begin{center}
\includegraphics[width=0.75\textwidth]{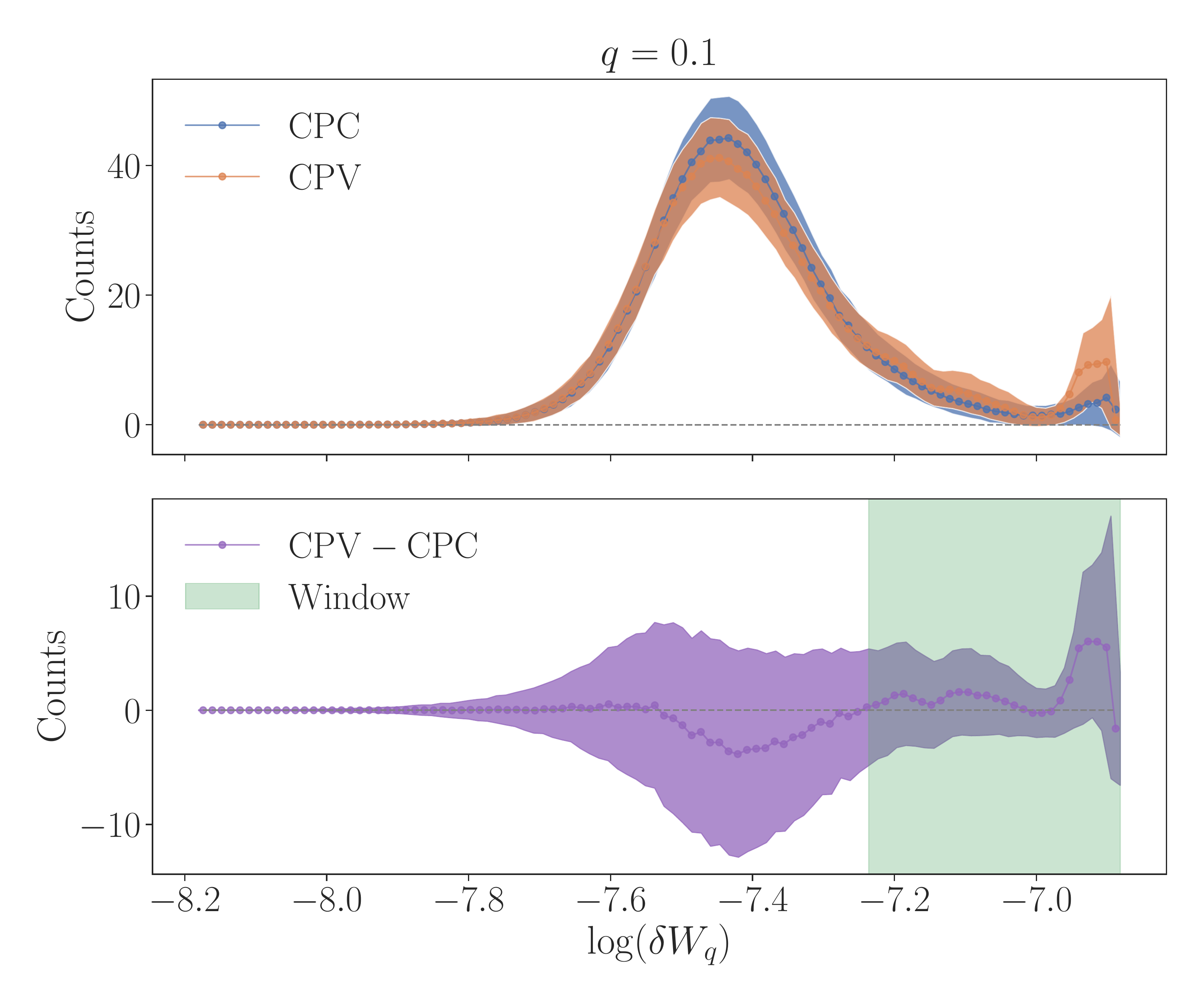}
\caption{\textbf{Top}: joined points (bands) represent the histogram of average ($1 \sigma$ range) $\log(\delta W_q)$ counts for $100$ bins, i.e., the binned counts of the log of pairwise optimal transport distances between $B$ and $\bar B$ sample events for $N=\bar N=10^3$ sample sizes, averaged over ensemble of $N_e=10^3$ samples, for CPC (blue) and CPV (orange) datasets. 
\textbf{Bottom}: joined points in purple (purple band) denote the average ($1\sigma$ ranges of) differences between CPC and CPV $\log(\delta W_q)$ counts, with the green band denoting the $[\delta W^{\text{win}}_{\text{min}}, \delta W^{\text{win}}_{\text{max}}]$ range for which sample events are used with positive weights in the construction of the windowed Wasserstein distance statistic, cf. Eqs.~\eqref{eq:windowed-Wq}--\eqref{eq:window} (in this example the range $[\delta W^{\overline{\text{win}}}_{\text{min}}, \delta W^{\overline{\text{win}}}_{\text{max}}]$ for negative weights   is taken to be zero). 
}
\label{fig:count_difference}
\end{center}
\end{figure}

This is illustrated in Fig.~\ref{fig:count_difference} (top), which
shows binned counts of $\log(\delta W_q)$, averaged over the  ensemble  of $N_e=10^3$ CPC (blue) and $N_e=10^3$ CPV (orange) samples, each containing $N=\bar N =10^3$ events, 
 with the bands denoting the $1\sigma$ ranges for bin counts.
 Fig.~\ref{fig:count_difference} (bottom)  shows the difference between the average CPC and CPV bin counts,  as well as the $1\sigma$ ranges. 
We see that the $\delta W_q$ distributions for CPC and CPV cases overlap significantly in many regions of pairwise $\delta W_q$ values. However, we also expect the CPC distributions to be more likely to lead to smaller $\delta W_q$, given that the $B^0$ and $\bar B^0$ Dalitz plot are more similar than in the CPV cases. Consequently, for the CPV case one would expect an excess of datapoints with larger $\delta W_q$ and a related excess of CPC bin counts at smaller $\delta W_q$ values, as shown in Fig.~\ref{fig:count_difference}. Depending on the details of the Dalitz plot the $\delta W_q$ distributions could exhibit other differences between the CPC and CPV cases not present in the example in Fig.~\ref{fig:count_difference}. For instance, if CPV is localized in a small region of the Dalitz plot containing $n$ events and of size $ \hat d$, cf. Eq.~\eqref{eq:hat:dij:Dalitz}, then we would expect an excess of CPV $\delta W_q$ bin counts over CPC in Fig.~\ref{fig:count_difference} at $\delta W_q\sim{\mathcal O}(\hat d/n)$. Once one sums over all $\delta W_q(i)$, and considers only the global Wasserstein distance $W_q=\sum_i W_q(i)$ as a measure of CPV, the information about such differences in the $\delta W_q$ distributions is lost. 

\begin{figure}[t]
\begin{center}
\includegraphics[width=0.7\textwidth]{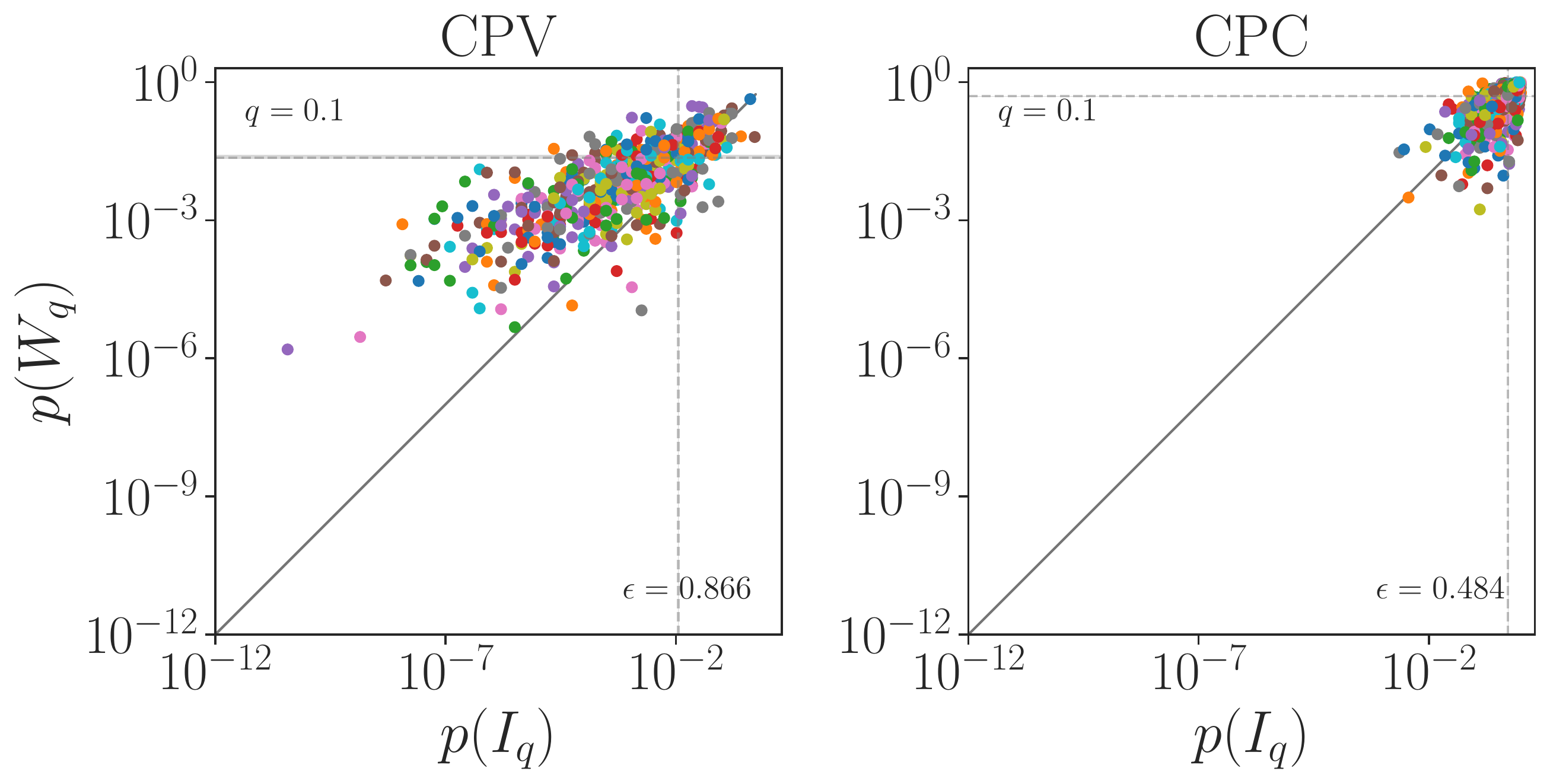}\\[3mm]
\includegraphics[width=0.7\textwidth]{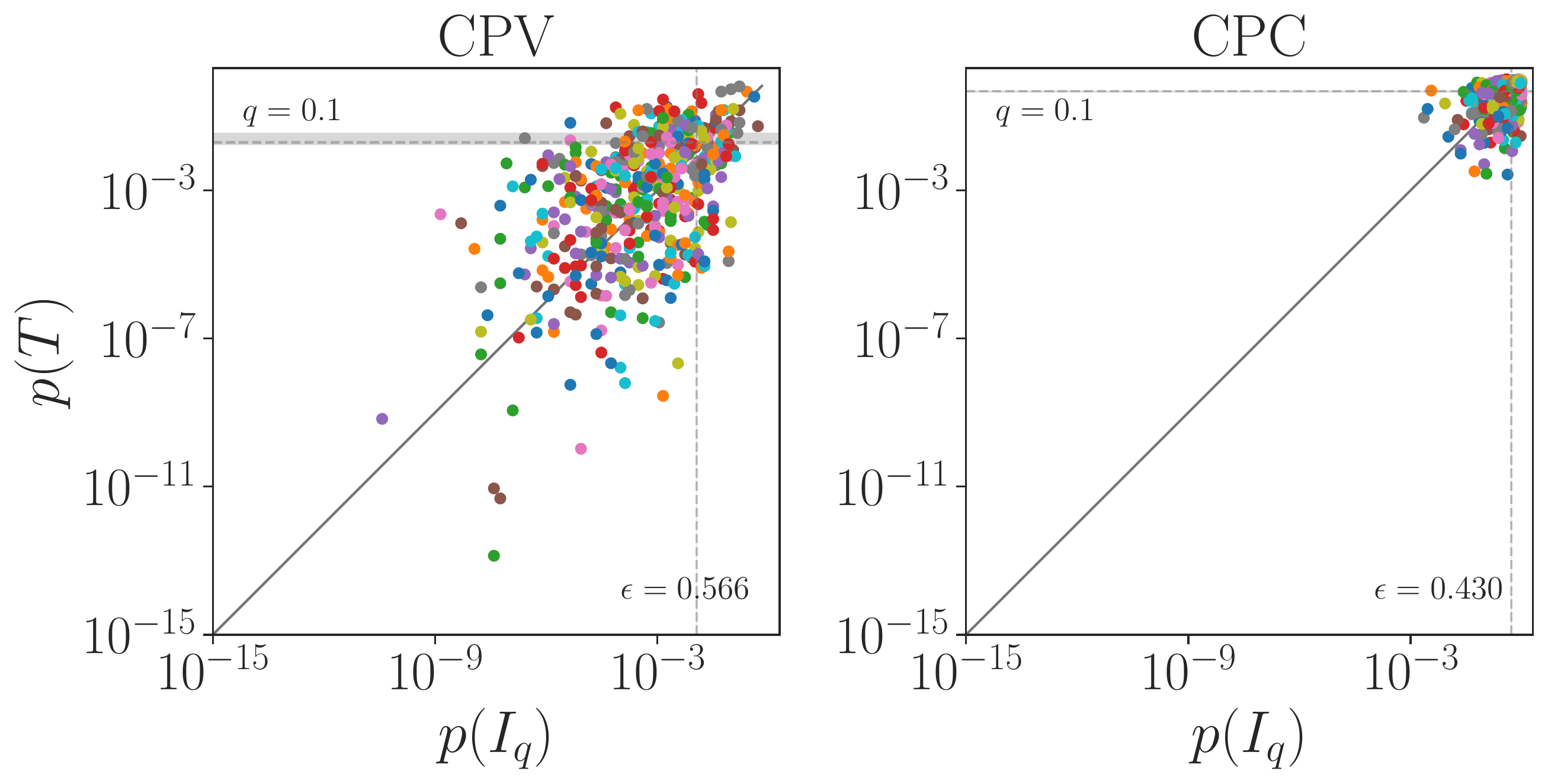}
\caption{The comparison of estimated $p$-value exclusions of CP conserving hypothesis for CPV (left) and CPC (right) $B^0\to K^+\pi^-\pi^0$ decays, comparing the windowed Wasserstein distance $I_q$ with either the global Wasserstein $W_q$ (top) or the energy test $T$ (bottom) statistic,  
for $q=0.1$, on $500$ distinct datasets.   
}
\label{fig:IWq_vs_Wq}
\end{center}
\end{figure}

Since there is more information in the $\delta W_q(i)$ distributions than in the global $W_q$ observable, we can define an 
  improved statistic ${I}_q$ 
\begin{equation}\label{eq:windowed-Wq}
    {I}_q \equiv \sum_i w\left(\delta W^{\text{win}}_{\text{min}}, \delta W^{\text{win}}_{\text{max}}, \delta W^{\overline{\text{win}}}_{\text{min}}, \delta W^{\overline{\text{win}}}_{\text{max}} ; \delta W_i \right),
\end{equation}
where for the example of $B\to K\pi\pi$ decays we define the window function as
\begin{equation}\label{eq:window}
    w(x) =  
    \begin{cases}
        +1 & x \in [\delta W^{\text{win}}_{\text{min}}, \delta W^{\text{win}}_{\text{max}}],\\
        -1 & x \in [\delta W^{\overline{\text{win}}}_{\text{min}}, \delta W^{\overline{\text{win}}}_{\text{max}}],\\
        0 & \text{otherwise}.
    \end{cases}
\end{equation}
The window function $w$ splits datapoints into three categories. 
The events in the high $\delta W_q$ values window $\delta W_q\in [\delta W^{\text{win}}_{\text{min}}, \delta W^{\text{win}}_{\text{max}}]$, and the events in the anti-window of mid-range $\delta W_q$ values, $\delta W_q(i) \in [\delta W^{\overline{\text{win}}}_{\text{min}}, \delta W^{\overline{\text{win}}}_{\text{max}}]$, are included in the windowed Wasserstein distance statistic ${I}_q$, but weighted with opposite signs, thus enhancing the difference between the CPC and CPV distributions. The remaining events, for which the CPC and CPV $\delta W_q$ distributions do not differ significantly, are instead not included in ${I}_q$. Keeping these events would only dilute the sensitivity to CPV. 

The optimization of window and anti-window ranges requires a model for $B^0$ and $\bar B^0$ amplitudes. Importantly, the $\delta W_q$ values depend on the sample size $N=\bar N$, and thus the optimization should be performed for the number of events actually measured in the experiment. One could attempt a data driven optimization of $w$ by splitting the measured dataset into subsets, correcting for the effect of smaller sample sizes, but we did not explore this further. For other decay channels, depending on the actual decay width distributions, other forms of window function could be better suited  than the one in Eq.~\eqref{eq:window}. For instance, one could define multiple disjoint window and anti-window regions, or use weights that are smooth functions of $\delta W_q$, not just the discrete values $\{-1,0,+1\}$. For $B^0\to K^+\pi^-\pi^0$ Dalitz plot and $q=0.1$, $N=\bar N=10^3$, there is on average an excess of CPC over CPV $\delta W_q$ distributions in the mid-value region $\log(\delta W_q) \in (-7.55, 7.3)$. However, it is accompanied with a large variability in bin counts, and thus for this case it proves advantageous to define  ${I}_q$ using only events in the window shown as the green band in Fig. \ref{fig:count_difference}, and drop all other events (that is, the anti-window range is shrunk to zero). For other values of $q$ both window and anti-window ranges are nonzero, see App.~\ref{sec:app:further:EMD}. 

\begin{figure}[t]
\begin{center}
\includegraphics[width=1.0\textwidth]{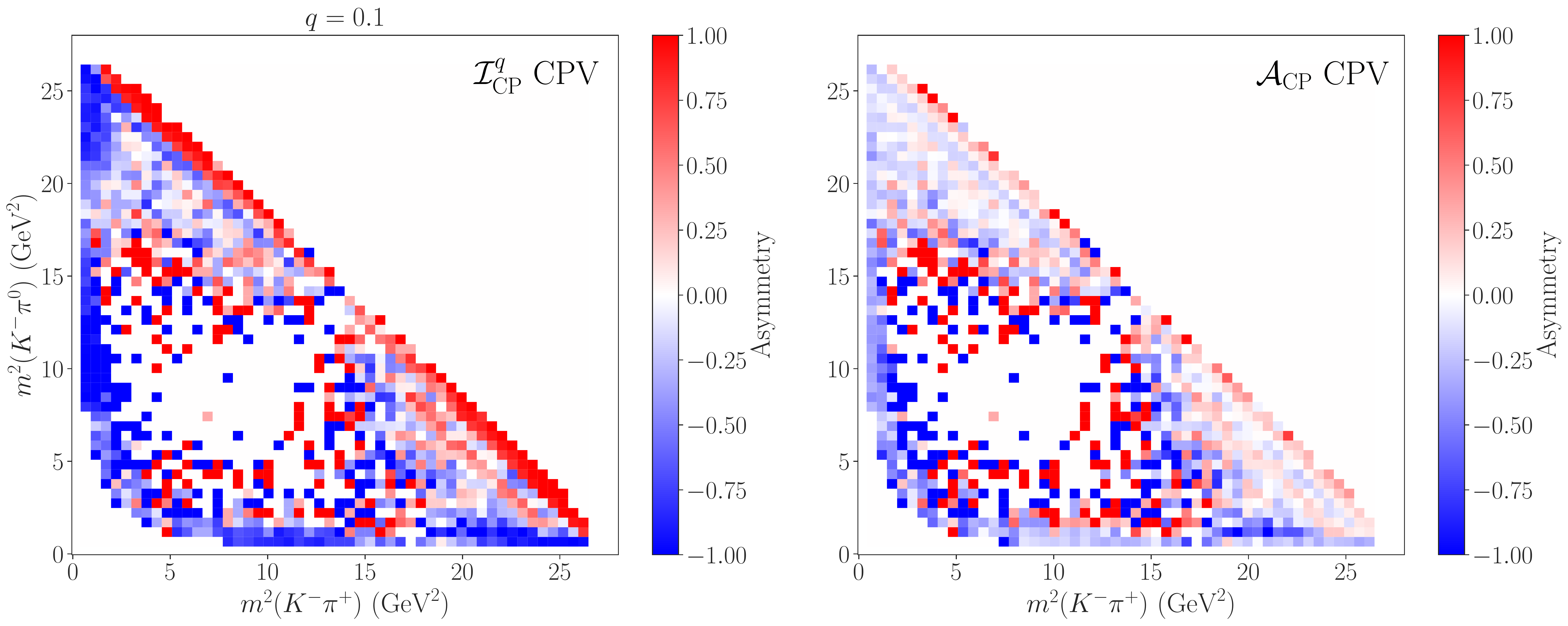}
\caption{The comparison of binned Dalitz plot asymmetry: the windowed Wasserstein asymmetry $\mathcal{I}^q_{\text{CP}}$ (left) and the fractional CP asymmetry $\mathcal{A}_{\text{CP}}$ (right), cf. also top panel in Fig.~\ref{fig:asymmetry_plots:q0.1}. When compared with the asymmetry significance's shown in Fig.~\ref{fig:asymm_sig_B} we see that the chosen window is correctly filtering CP conserving $\delta W_q$ values and retaining $\delta W_q$ values in the most significant regions of CPV.
}
\label{fig:IWq_dalitz}
\end{center}
\end{figure}

 Fig.~\ref{fig:IWq_vs_Wq} shows, for $q=0.1, N=\bar N=10^3$, the comparison of  $p$-values  at which the CPC hypothesis is excluded, when either the windowed Wasserstein statistic ${I}_q$ or the global Wasserstein statistic $W_q$ are used, Fig.~\ref{fig:IWq_vs_Wq} (top), or if the energy test statistic, $T$, is used instead,  Fig.~\ref{fig:IWq_vs_Wq} (bottom). 
 We see that the windowed Wasserstein distance statistics, ${ I}_q$, is as sensitive, or even slightly more sensitive, to the presence of CPV in the Dalitz plot distributions than the energy test, while both outperform the  global Wasserstein distance statistic. Fig.~\ref{fig:IWq_vs_Wq} also demonstrate that  $I_q$, like $W_q$ and the $T$ test statistic, does not introduce bias when CPC distributions are considered.  As an additional confirmation that the chosen windows are in fact selecting the relevant areas of the Dalitz plot associated with CPV and CPC we plot in Fig. \ref{fig:IWq_dalitz} the binned CP and Wasserstein asymmetries, but in the later only keeping the events that contribute to ${ I}_q$. That is, we define
 \begin{equation}
\label{eq:Iq:asymm}
    \mathcal{I}^q_{\text{CP}} (s_{12}, s_{13}) = \frac{\sum_{\bar i} w(\delta \bar W_q(\bar i) )- \sum_{i} w(\delta {W}_q(i))}{\sum_{\bar i} w(\delta \bar W_q(\bar i)) + \sum_i w(\delta {W}_q(i))},
\end{equation}
where each event is weighted according to the window function in Eq.~\eqref{eq:window}. The summation over $i$ $\left( \bar i \right)$ is only over the data-points contained in the bin centered at $(s_{12}, s_{13})$ (the CP conjugated $\bar B^0$ bin centered at $(\bar s_{12}, \bar s_{13})$). The comparison of left and right panels in Fig.~\ref{fig:IWq_dalitz} shows that the chosen window from Fig.~\ref{fig:count_difference} does indeed correctly select the regions of the Dalitz plot exhibiting CP violation and acts as a filter to better resolve CP asymmetries.

The shown results could be improved further. First of all, we did not perform a full optimization of the window function Eq.~\eqref{eq:window}, but rather only selected among several discrete, manually chosen, forms. It would also be interesting to explore if the features observed in the $\delta W_q$ distributions, Fig.~\ref{fig:count_difference}, can further inform amplitude models, in particular about the existence of CPV regions with resonances interfering.

\section{Application to three body $D$ decays}
\label{sec:D:dec}

Next, we apply the analysis to larger datasets with small but nonzero amount of CP violation. As a concrete example we consider the three body $D$ decay $D^0 \rightarrow \pi^+ \pi^- \pi^0$ and its CP conjugated channel,  $\bar{D}^0 \rightarrow \pi^- \pi^+ \pi^0$. The CP violation in $D$ decays is expected to be small, parametrically suppressed by ${\mathcal O}(V_{cb}V_{ub}/V_{cd}V_{ud})\sim 10^{-3}$ \cite{Grossman:2006jg,Brod:2011re,Li:2012cfa,Franco:2012ck,Feldmann:2012js,Cheng:2012wr,Bhattacharya:2012ah,Pirtskhalava:2011va,Dery:2022zkt} 
 and has only recently been measured to be nonzero \cite{LHCb:2019hro,LHCb:2022vcc}. Further searches for CP violation within the charm sector are highly motivated, since the discovery of enhanced CPV in specific modes, including multibody decays, could point to a discovery of new physics (for sum rules that the SM needs to satisfy see \cite{Grossman:2012eb,Dery:2021mll,Grossman:2012ry}).

 The $D^0\to\pi^+\pi^-\pi^0$ decays have been studied at the LHCb using the energy test, and found that the CPC hypothesis is excluded at the $p=(2.6 \pm 0.5)\%$~C.L.~\cite{LHCb:2014nnj}. Below, we show how the Wasserstein distance based statistics could be used as alternative analysis strategies to search for CPV in this and other multibody charm decays, taking $D^0\to\pi^+\pi^-\pi^0$ as a toy example.  
 
We generate the two datasets, for $D^0 \rightarrow \pi^+ \pi^- \pi^0$ and  $\bar{D}^0 \rightarrow \pi^- \pi^+ \pi^0$ decays, using the  BaBar amplitude model~\cite{BaBar:2007dro} implemented within the  {\tt Laura++} framework \cite{BACK2018198}, similarly to the case of $B^0\rightarrow K^+\pi^-\pi^0$ decays discussed in Sec. \ref{sec:three-bodyB}.  As a toy example of CP violation in the $D^0\to \pi^+\pi^-\pi^0$ Dalitz plot we follow Ref.~\cite{Parkes:2016yie} (where this was used to explore the sensitivity of the energy test), and  increase for the generation of CPV datasets the fit fraction of the $\rho(770)^-$ by 2\% and the phase of the corresponding decay amplitude by $2^\circ$. The $D-\bar D$ meson mixing is ignored in the generation of the samples.

The present experimental $D \rightarrow \pi \pi \pi$ decay samples are roughly $10^2-10^3$ times larger than the $B\to K\pi\pi$ decay samples. 
Because of the current implementation of the Wasserstein distance calculation that we use~\cite{Komiske:2019fks,Komiske:2020qhg}, large statistic datasets present a numerical problem. To solve the optimal transport problem utilizing the current publicly available linear programming libraries require the full cost matrix $\hat{d}_{ij}$ as the input. The cost matrix scales as $N\bar{N} \sim {\mathcal O}(N^2)$ and quickly demands more random access memory than available in an average personal computer. For example, the cost matrix for datasets containing $\sim 10^6$ events, i.e., comparable to the number of currently experimentally available $D^0 \rightarrow \pi^+ \pi^- \pi^0$ decays, requires roughly 7 TB of memory space. 

There are a number of solutions to the above memory problem. Below we develop two strategies, both of which use approximate calculations of (variants of) Wasserstein distance between the $D^0$ and $\bar D^0$ decay samples: a binned Wasserstein test in Sec. \ref{sec:binned:W}  and a sliced Wasserstein test in Sec. \ref{sec:SWD}. The two approximate approaches to the Wasserstein based statistic can be applied to large datasets, while continuing to use the publicly available and optimized software.  
Alternatively, one could attempt to create a new optimal transport algorithm geared toward large datasets, such as the $D$ decays, utilizing lazy evaluation and the sparseness of the transport matrix that does not require the full form of the cost matrix as an input. The latter, however, goes beyond the scope of the present manuscript. 

\begin{figure}[t]
\begin{center}
\includegraphics[width=1.0\textwidth]{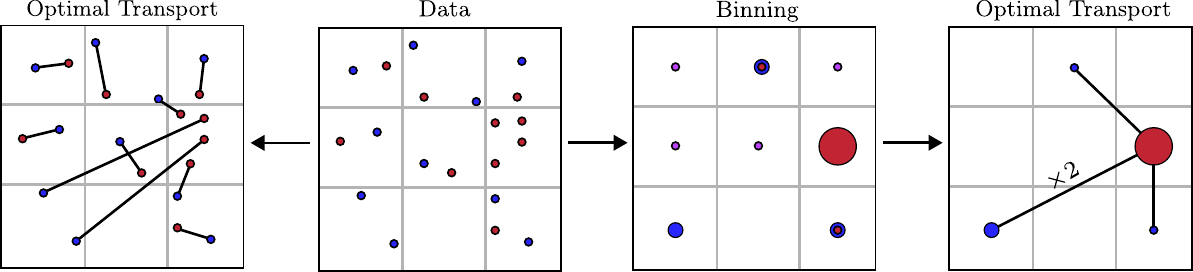}
\caption{Pictorial comparison between unbinned (left) and binned (right) Wasserstein statistic methods. Note how in the binned case, the optimal transport algorithm effectively sets to zero in the last step  the number counts in the bins that have the same counts between the two CP conjugate datasets (red and blue).}
\label{fig:unbinned_vs_binned}
\end{center}
\end{figure}

\subsection{Binned Wasserstein test}
\label{sec:binned:W}

Since the resonances in the $D^0\to\pi^+\pi^-\pi^0$ Dalitz plot have typical decay widths of ${\mathcal O}(100\,\text{MeV})$ or so, cf. Fig.~\ref{fig:dalitz}, we expect it is possible to capture well the change of the CP asymmetry across the Dalitz plot already with relatively modest numbers of bins. 
One can then apply the Wasserstein distance statistic to the binned Dalitz plot data in order to obtain a global measure of CPV in the distributions. While there is some loss of information due to binning compared to the Wasserstein distance statistic applied to full samples, we expect the loss to be small, if the binning is fine enough. In the limit of infinitely small bins one of course reverts to the case of unbinned statistic discussed in Sec. \ref{sec:three-bodyB}. 

 The binned Wasserstein distance is given by
\begin{equation}
    W_q^{\text{bin}}(\mathcal{E}, \bar{\mathcal{E}}) = \biggr[\underset{\{f_{ij}\geq 0\}}{\min}\sum_{i,j=1}^{N_b}f_{ij}\big(\hat d_{ij}\big)^q\biggr]^{1/q},
\end{equation}
where $N_b$ is the total number of bins in the $D$ (and $\bar D$) Dalitz plot, with bin counts $w_i$ ($\bar{w}_j$) in the $i-$th ($j-$th) bin. In the Dalitz plot we will use equal binning along each dimension, with $n_{\rm bins}$ in each direction, so that the number of bins with nonzero entries equals to $N_b\simeq n_{\rm bins}(n_{\rm bins}-1)/2$.\footnote{The equality sign applies in the $m_\pi\to 0$ limit or for large enough bins. In our numerical implementation we  use square $n_b\times n_b$ arrays that cover fully the Dalitz plot and take $N_b= n_{\rm bins}^2$ to be the total number of bins, including the ones containing zero events. The bins outside the kinematically allowed region are trivially zero, and do not add any complexity to the calculation of the binned Wasserstein distance, while this approach simplifies the encoding of the Dalitz plot in the binned array. }
The minimization of the weights $f_{ij}$ ($\bar{f}_{ij}$) gives the optimal transport from bins in $D$ to $\bar D$ Dalitz plot, subject to the constraints
\begin{equation}
    \sum_i^{n_{\text{bins}}} f_{ij} = \frac{\bar{w}_j}{\bar{N}}, \hspace{0.3in} \sum_j^{n_{\text{bins}}} {f}_{ij} = \frac{w_i}{N}, \hspace{0.3in} \sum_{i,j}^{n_{\text{bins}}} {f}_{ij} = 1
\end{equation}
with the distances $\hat{d}_{ij}$ taken to be between the centers of the $i-$th and $j-$th bins. The construction of the binned Wasserstein distance statistic $W_q^{\text{bin}}$ is illustrated in Fig.~\ref{fig:unbinned_vs_binned}. Since the binned versions of $\mathcal{E}$ and $\bar{\mathcal{E}}$ event samples  use the same binning, the optimal transport algorithm will always `zero' out the like counts in each bin between $\mathcal{E}$ and $\bar{\mathcal{E}}$, i.e., it takes no `work' to transport mass by zero distance. What is left is a representation of the local bin count density asymmetry between $\mathcal{E}$ and $\bar{\mathcal{E}}$. These count density asymmetries then get re-distributed by the optimal transport algorithm.
Thus, instead of encoding the CPV information via the distances between events in each dataset, as is done in $W_q$, the CPV is now encoded as the excess or overabundance of \textit{weight} between datasets (as well as how far these weight overabundances in $D$ Dalitz plot are from overabundances in the $\bar D$ Dalitz plot).

\begin{figure}[t]
\begin{center}
\includegraphics[width=0.9\textwidth]{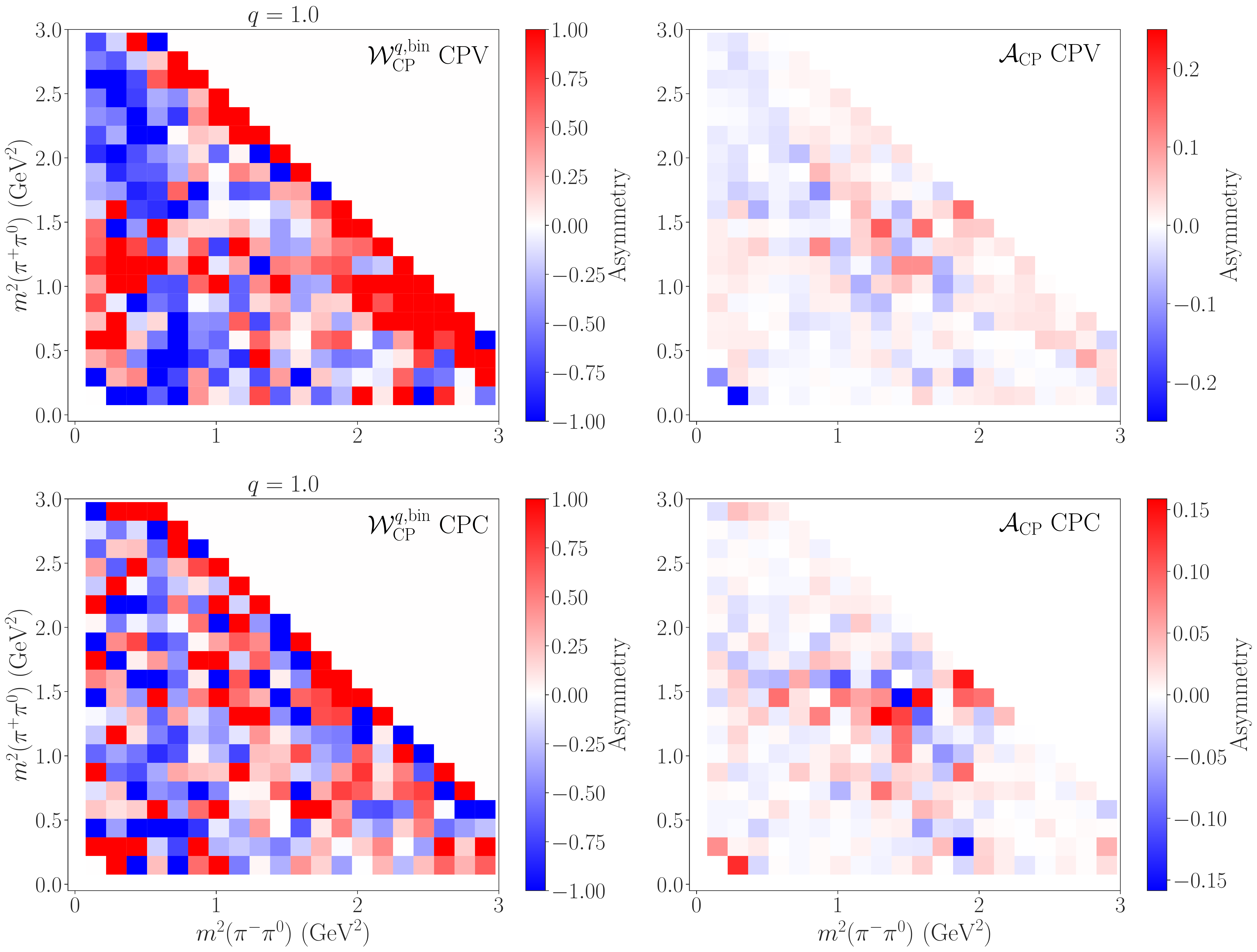}
\caption{Comparison between the binned Wasserstein asymmetry $\mathcal{W}_{\text{CP}}^{q,{\rm bin}}$ (left), cf. Eq. \eqref{eq:Wq_binned:asymm},  and the binned CP asymmetry $\mathcal{A}_{\text{CP}}$ (right) for the $D^0\to\pi^+\pi^-\pi^0$ Dalitz plot with 
$N=\bar N=10^6$ events in a sample, and $n_{\rm bins}=20$. 
}
\label{fig:D_binned_asymmetry}
\end{center}
\end{figure}

Denoting the contribution to $W^{\text{bin}}_q$ from the $i-$th bin in the $D^0$ Dalitz plot as $\delta W^{\text{bin}}_q(i)$, and likewise by $\delta \bar W^{\text{bin}}_q (\bar i)$ the contribution to $W^{\text{bin}}_q$  from $\bar i-$th bin in the $\bar D^0$ Dalitz plot, such that
\begin{equation}
(W^{\text{bin}}_q)^q = \sum_i \delta W^{\text{bin}}_q(i)=\sum_{\bar i}\delta \bar W^{\text{bin}}_q (\bar i),
\end{equation}
we define in analogy with Eq. \eqref{eq:Wq:asymm} the binned Wasserstein asymmetry $\mathcal{W}^{q, \text{bin}}_{\text{CP}}$ as 
\begin{equation}
\label{eq:Wq_binned:asymm}
    \mathcal{W}^{q, \text{bin}}_{\text{CP}} (i) = \frac{ \delta \bar W^{\text{bin}}_q(\bar i) - \delta {W}^{\text{bin}}_q(i)}{ \delta \bar W^{\text{bin}}_q(\bar i) + \delta {W}^{\text{bin}}_q(i)},
\end{equation}
where the $\bar i$-th bin in the $\bar D$ Dalitz plot is the CP-conjugate of the $i-$th bin in the $D$ Dalitz plot.  

 Fig.~\ref{fig:D_binned_asymmetry} shows a comparison between the binned Wasserstein distance asymmetry ${\mathcal W}_q^{\rm bin}$ (left panels) and the CP asymmetry  $\mathcal{A}_{\text{CP}}$ (right panels). We find that the binning results in enhanced asymmetries when data is represented using the ${\mathcal W}_q^{\rm bin}$ compared to $\mathcal{A}_{\text{CP}}$. This is true both for the CPV dataset, as well as for statistical fluctuations in the CPC example. Since direct CP violation in $D$ decays is small, it is hard to discern by eye whether or not there is CP violation in the Dalitz plot distributions, and one is forced to rely on a statistic sensitive to CPV in distributions such as $W_q^{\rm bin}$ or the energy test. 
 
 \begin{figure}[t]
\begin{center}
\includegraphics[width=0.6\textwidth]{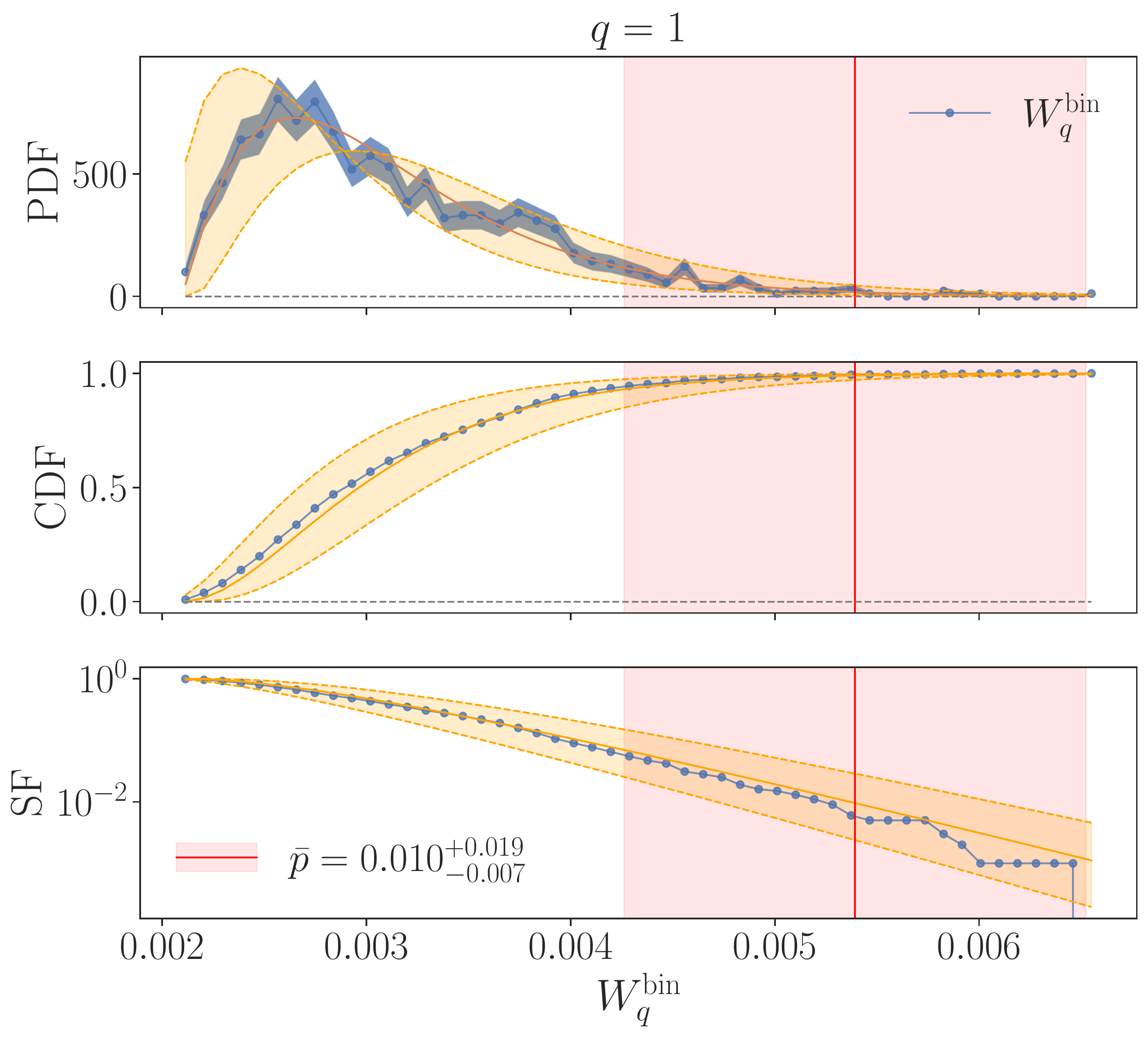}
\caption{The CPC probability distribution function (PDF), the cumulative distribution function (CDF), and the survival factor (SF=1-CDF) as functions of the binned Wasserstein statistic $W^{\text{bin}}_q$  for $r=2,$ $q = 1,$ and $n_{\text{bins}} = 50$, as obtained from the numerical master method result for the PDF, consequently fit to a gamma distribution, for  
$N=\bar N=10^5$ events in the sample, using an ensemble of $N_e=10^3$ samples. The orange bands (blue band in the top panel) denote the $\pm 1\sigma$ fit errors (statistical errors). The vertical red line (band) denotes the average $W_q^{\rm bin}$ value (the $\pm 1\sigma$ $W_q^{\rm bin}$ ranges) obtained from an ensemble of $N_e=10^2$ CPV datasets for our toy $D^0\to \pi^+\pi^-\pi^0$ amplitude model. 
}
\label{fig:PDF:CDF:SF_D}
\end{center}
\end{figure}
 
Fig.~\ref{fig:PDF:CDF:SF_D} shows that the Wasserstein test statistic is still sensitive to CP violation despite the binning procedure. The three panels show from top to bottom the probability distribution function (PDF), the cumulative distribution function (CDF), and the survival factor (SF=1-CDF) as functions of the binned Wasserstein statistic $W^{\text{bin}}_q$, for $r=2,$ $q = 1,$ and using  $n_{\text{bins}} = 50$, for CP conserving $D^0\to \pi^+\pi^-\pi^0$ Dalitz plot with $N=\bar N=10^5$ events in the sample. The average $W_q^{\rm bin}$ value (red vertical line) for our CPV toy $D$ decay model example is well above the bulk of the CP conserving $W_q^{\rm bin}$ PDF. We see that, on average, the CPC hypothesis is in this example expected to be excluded at the $\sim 2.5\sigma$ level, i.e., with a $p-$value of $\sim 0.01$. 

\begin{figure}[t]
\begin{center}
\includegraphics[width=0.4\textwidth]{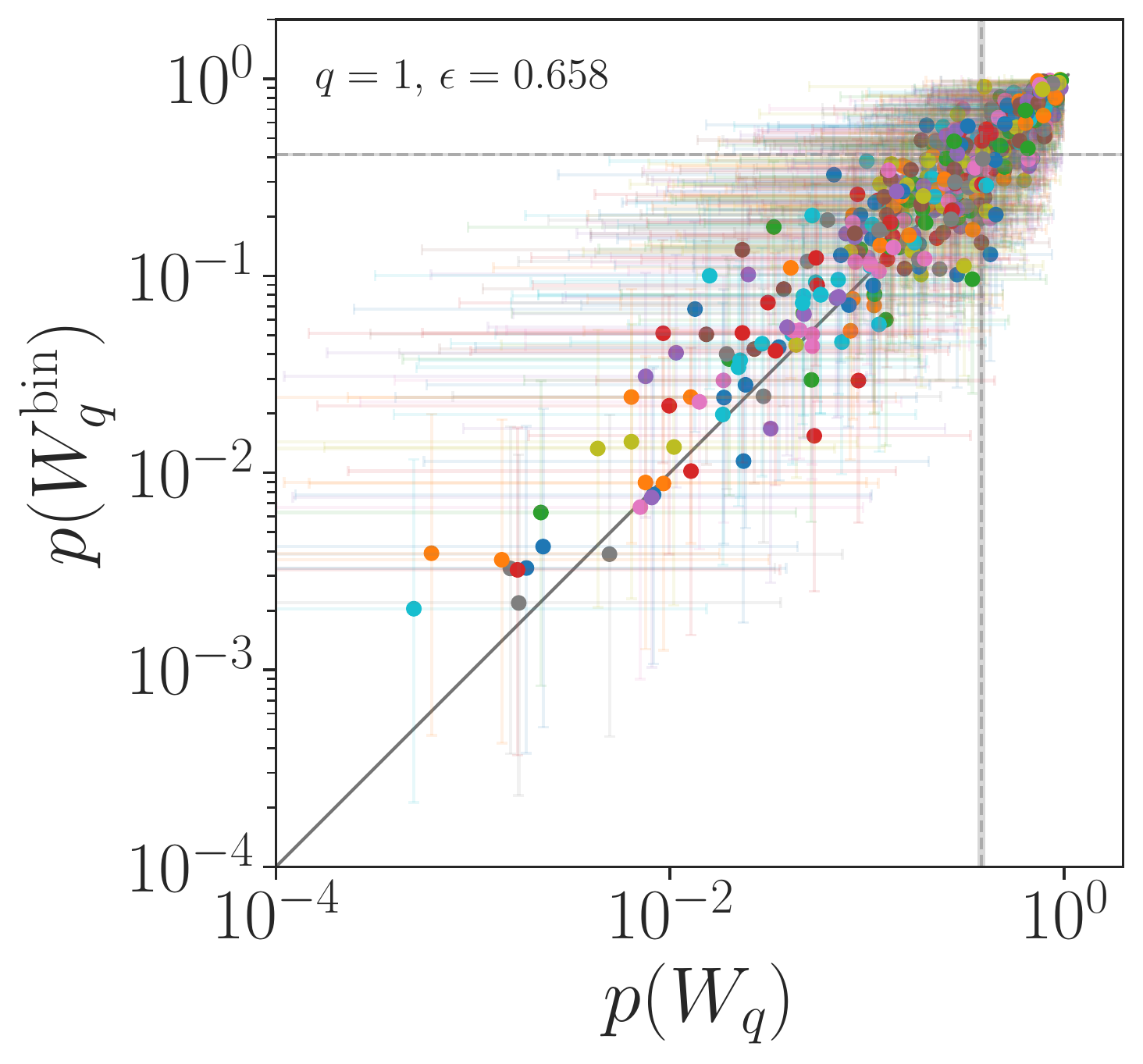}
\caption{The scatter plot of $p-$values at which the CP conserving hypothesis is excluded for our toy $D$ decay example. The plot shows an ensemble of  $N_e=500$ datasets with samples of  $N=10^4$ $D^0\to \pi^+\pi^-\pi^0$ (and $\bar N=10^4$ CP conjugated $\bar D^0\to \pi^-\pi^+\pi^0$) decays, with $p-$values calculated either using the unbinned $W_q$, giving $p(W_q)$, or using the binned $W_q^{\text{bin}}$ statistic with $n_{\text{bin}} = 50$, giving $p(W_q^{\rm bin})$  (dots), where in both cases we set $q=1$. The $1\sigma$ fit error bars on $p-$ values are shown as lines.
 The fraction $\epsilon$ of points above $p(W_q)=p(W_q^{\text{binned}})$ diagonal line denotes the fraction of ensemble samples for which $W_q$ is more sensitive to CPV than $W_q^{\text{binned}}$ is. The dotted gray horizontal and vertical lines (solid bands) show the average ($1\sigma$ ranges of) $p-$values for the ensemble. }
\label{fig:Wq_vs_binned_Wq}
\end{center}
\end{figure}

In fact, Fig.~\ref{fig:Wq_vs_binned_Wq} shows that the chosen binning size  $n_{\text{bins}} = 50$ (which was not optimized) is already fine enough for $N=\bar N=10^4$ that there is only little loss of sensitivity to CPV 
compared to the unbinned $W_q$.
In the scatter plot of $p-$values at which the CPC hypothesis is excluded, we see that the exclusion levels obtained by either using the $W_q$ or the $W_q^{\rm bin}$ statistic are comparable, and consistent within estimated errors (due to the systematic and statistical uncertainties  in the extrapolation of the fit to the CPC PDF).
 The binned Wasserstein statistic does have, however, the additional advantages of less memory consumption (space complexity) and computational efficiency (time complexity) due to the reduction of the dataset size from $N\bar{N}$ to $\sim n^2_{\text{bins}}$. Whether $n_{\text{bins}} = 50$ suffices also for sample sizes $10^6$, or whether fined binning will be required, should be tested when the method is applied to the actual $D$ decay data, however, we find the above results quite encouraging. 

\subsection{Sliced Wasserstein test}
\label{sec:SWD}

\begin{figure}[t]
\begin{center}
\includegraphics[width=0.8\textwidth]{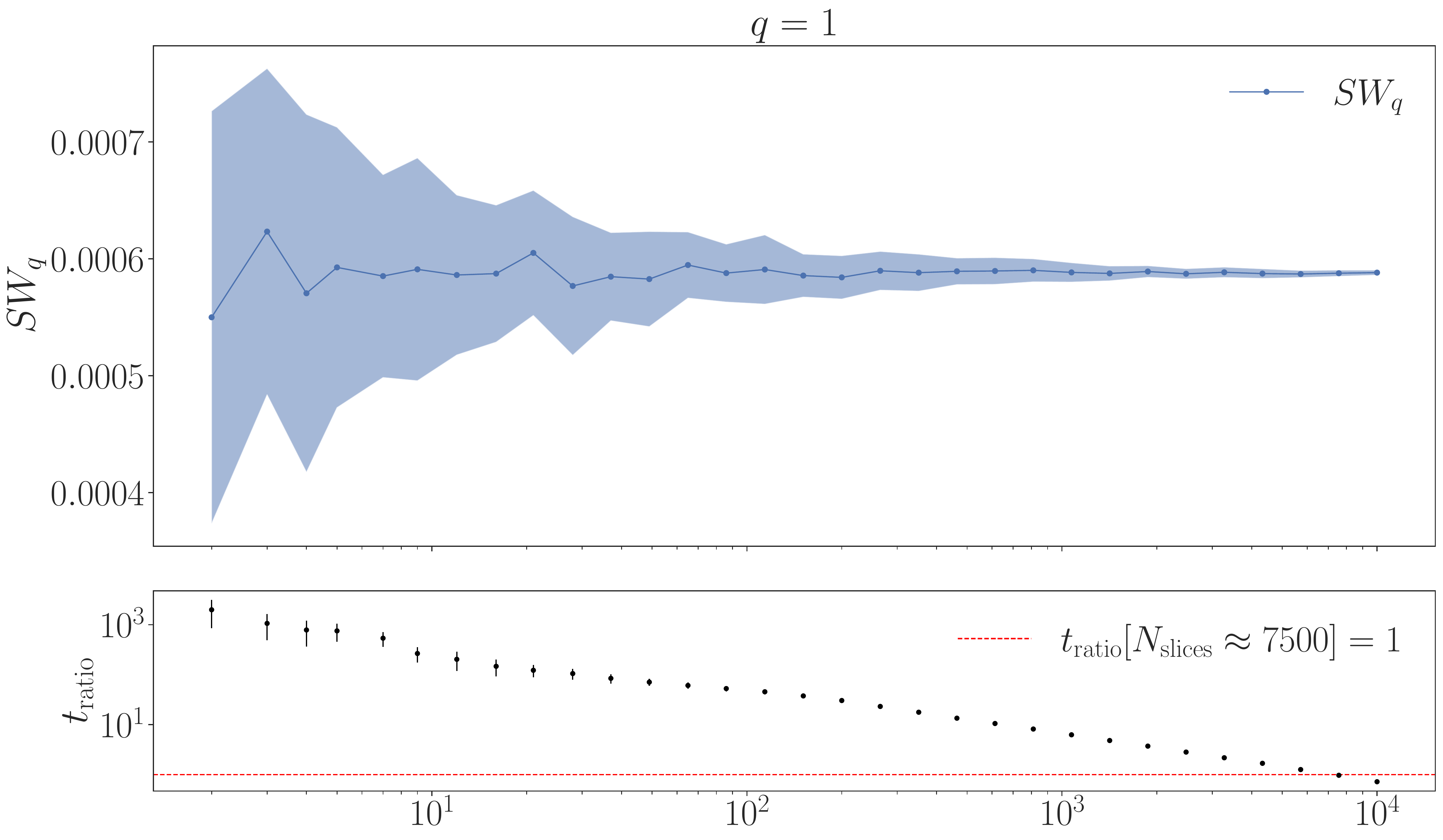}
\caption{
Top: The approximate evaluation of ${SW}_q$, Eq. \eqref{eq:SWq:approx}, as a function of the number of slices used, $N_{\rm slices}$, for a particular CPC $D^0\to \pi^+\pi^-\pi^0$ sample with $N=\bar N=10^4$ events, setting $q=1$. The solid blue line (blue band) shows the mean ($1\sigma$ range) of the ${SW}_q$ estimates obtained from an ensemble of $50$ different samplings of $N_{\rm slices}$ slices for each solid point. 
Bottom: the speed up of calculating $SW_q$ compared to $W_q$ defined as the ratio of computational times in the two cases,  $t_{\rm ratio} = t_{W_q}/ t_{SW_q}$. 
} 
\label{fig:SWDq_converging}
\end{center}
\end{figure}

The Sliced Wasserstein distance ($SW_q$) is a variant of the Wasserstein distance, in which the optimal transport in $d$-dimensions is replaced with a set of optimal transport problems on 1D slices, with the data points projected onto them. That is, the sliced Wasserstein distance $SW_q(g, f)$ between two distributions in $d-$dimension, $g(x)$ and $f(x)$, is given by \cite{Helgason2015}
\begin{equation}
\label{eq:SWq}
SW_q(g, f) = \bigg( \int_{\mathcal{S}^{d-1}} W_q(Rg(\cdot,\theta), Rf(\cdot, \theta) d\theta \bigg)^{\frac{1}{q}},
\end{equation}
where $Rg(\cdot,\theta)$ is the Radon transform of function $g(x)$, defined to be the projection of function $g(x)$ onto the line in the direction of the unit vector $\theta$, which then runs over the $d-1$ unit sphere $\mathcal{S}^{d-1}$. The $W_q$ in \eqref{eq:SWq} is therefore a 1D Wasserstein distance between functions $Rg(\cdot,\theta)$ and $Rf(\cdot, \theta)$. The 1D $W_q$ has a closed form solution, given by the integrated distance between the CDFs for the two functions, and can be efficiently calculated through a simple sorting algorithm. 

The sliced Wasserstein distance can thus be efficiently calculated, at least approximately, by performing a large enough number of slices, $N_{\rm slices}$,
\begin{equation}
\label{eq:SWq:approx}
SW_q(g,f) \approx \bigg( \frac{1}{N_{\rm slices}} \sum_{k=1}^{N_{\rm slices}} W_q(R q(\cdot, \theta_k), R f_{\nu}(\cdot, \theta_k) )   \bigg)^{\frac{1}{q}} ,
\end{equation}
where $\theta_k$ are random unit vectors uniformly distributed over the unit sphere $\mathcal{S}^{d-1}$. In the $N_{\rm slices}\to \infty$ limit  the l.h.s. approaches the r.h.s. in the above equation. 

Importantly for our purposes, both $W_q$ and $SW_q(g,f)$ are distances in the space of functions and both measure dissimilarity of $f$ and $g$ distributions. The $SW_q$ can therefore also be used as a test statistic, in the same way as we used the Wasserstein distance $W_q$ in the previous sections. Furthermore, $SW_q$ is closely related to the Wasserstein distance, $W_q$. For instance, for $q=2$ we have $SW_2(g,f)\leq W_2(g,f)/\sqrt{d}$, and in general $SW_q(g,f)\leq c_q W_q(g,f)$ with a known constant $c_q\leq 1$ (for $q\in [1,\infty)$).

\begin{figure}[t]
\begin{center}
\includegraphics[width=1.\textwidth]{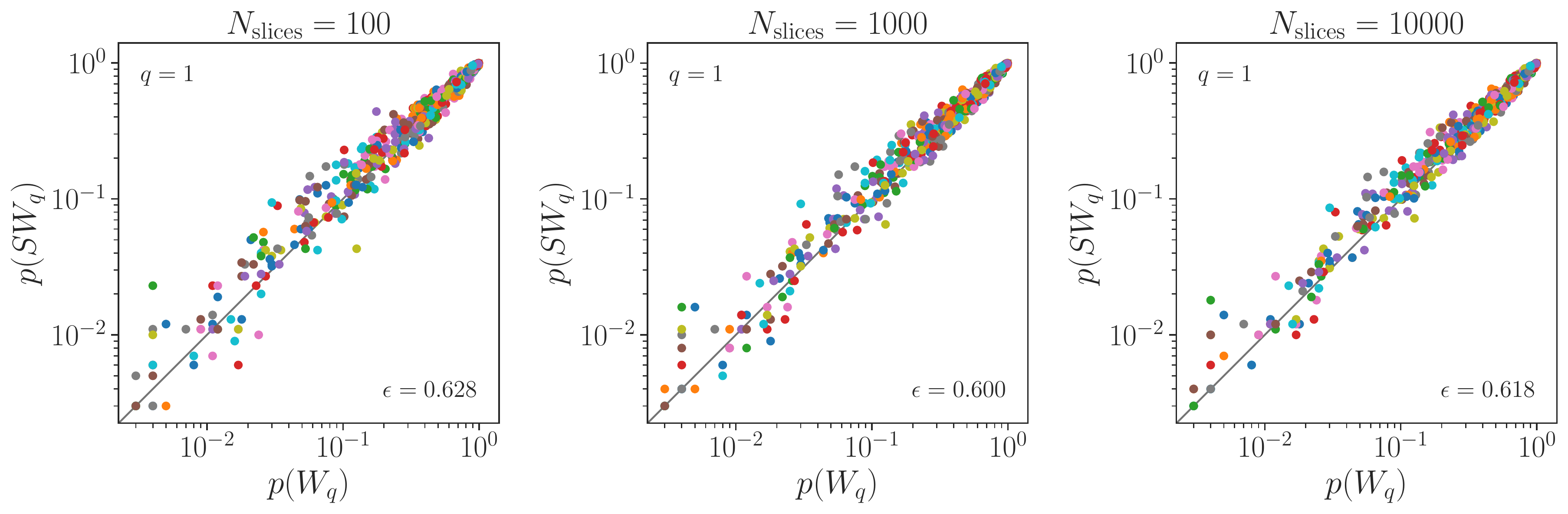}
\caption{The scatter plot of $p$--values at which no CPV hypothesis is excluded, calculated using  $SW_q$ 
and $W_{q}$  for $N_{\text{slices}}= 10^2, 10^3, 10^4$ (from left to right), for the ensemble of a $N_e=500$ datasets with samples of  $N=10^4$ $D^0\to \pi^+\pi^-\pi^0$ (and $\bar N=10^4$ CP conjugated $\bar D^0\to \pi^-\pi^+\pi^0$) decays.  The fraction $\epsilon$ of points above $p(W_q)=p(SW_q)$ diagonal line denotes the fraction of ensemble samples for which $W_q$ is more sensitive to CPV than $SW_q$. 
}
\label{fig:Wq_vs_SWDq}
\end{center}
\end{figure}

The improved computational efficiency for $SW_q$  relative to $W_q$ is shown in  Fig.~\ref{fig:SWDq_converging}. The ratio of the computing times, $t_{\rm ratio} = t_{W_q}/ t_{SW_q}$, where  $t_{W_q} (t_{SW_q})$ denotes the time required to calculate $W_q$ (approximate calculation of ${SW}_q$ using Eq. \eqref{eq:SWq:approx}) for a particular $D^0\to \pi^+\pi^-\pi^0$ sample with $N=\bar N=10^4$ events, where we take $q=1$. For small number of slices, $N_{\rm slices}\sim {\mathcal O}(10)$ the speed up is several orders of magnitude, however, at that point also the approximate evaluation of ${SW}_q$ still has a large uncertainty. The latter is denoted with the blue band, corresponding to $1\sigma$ range of ${SW}_q$ values obtained using Eq.~\eqref{eq:SWq:approx}, cycling through $50$ iterations. We observe that in this example the $SW_q$ evaluation is faster than the $W_q$ one for $N_{\text{slices}} \lesssim 7500$. We also observe that the approximate $SW_q$ evaluation converges to its limiting value for $N_{\text{slices}} \approx 1000$,  indicating a $\sim 7 \times$ speedup in the calculation of $SW_q$ compared to $W_q$. Beyond the speed-up, and maybe even more importantly for the scaling to large sample sizes, the evaluation of $SW_q$ does not require large memory resources. We have also checked that as the number of slices increases the $SW_q$ and $W_q$ distributions, obtained from an ensemble of $N=\bar N=10^4$ event samples, agree up to a scaling factor as expected.  Finally, since we are interested in the sensitivity to CPV and not in $SW_q$ itself, we show next that a high sensitivity to CPV can be achieved already with relatively approximate estimate of $SW_q$, relying on just a limited number of slices. 
 
Fig.~\ref{fig:Wq_vs_SWDq} shows the $p-$values at which the CPC hypothesis is excluded, either calculated using $W_q$ (giving $p(W_q)$) or via approximate evaluation of $SW_q$ using Eq. \eqref{eq:SWq:approx} (giving $p(SW_q)$) for three different values of slices, $N_{\text{slices}}= 10^2, 10^3, 10^4$ (from left to right). The fraction $\epsilon$ of points above $p(W_q)=p(SW_q)$ diagonal line denotes the fraction of $N_e=500$ datasets ensemble of $N=\bar N=10^4$ event samples for which $W_q$ is more sensitive to CPV than $SW_q$. We see that even for $N_{\text{slices}}= 10^2$ the obtained $p-$values are already comparable to the $p-$values obtained using full $W_q$, even though at that point the approximate evaluation of $SW_q$ still has a rather large spread, cf. Fig.~\ref{fig:SWDq_converging}. This is quite encouraging, and it would be interesting to explore in the future whether this feature remains for larger sample sizes.  Similarly, it would be interesting to explore where a windowed $\mathrm{SW}_q$, defined in analogy with the windowed Wasserstein distance statistic $I_q$, would lead to a similar increase in sensitivity to CPV that we saw in the case of full $W_q$.

\section{Conclusions}
\label{sec:concl}
The Wasserstein distance based test statistics are potentially powerful tools that can be used to search for the presence of CP violation in multibody decays. They combine the benefits of two alternative tests sensitive to CPV in distributions: (i) in a similar way as the binned CP asymmetry, the  Wasserstein distance based test statistics trace asymmetries to the regions of phase space the CPV resides in, while at the same time (ii) being a sensitive probe of CPV as an integrated measure, in a similar way as the energy test is. 

In this manuscript we introduced several such Wasserstein distance based test statistics, taking the multibody $B^0\to K^+\pi^-\pi^0$ and $D^0\to \pi^+\pi^-\pi^0$ decays as concrete examples for numerical studies. The simplest one is the Wasserstein distance, $W_q$, see Eq.~\eqref{eq:def:Wp} for the case of $B^0$ and $\bar B^0$ decays. The use of $W_q$ as a measure of CPV in principle requires no tuning, though there are optimizations that can be made regarding the exact definition of the distance in the Dalitz plot one uses, Eq.~\eqref{eq:hat:dij:Dalitz},  as well as the value of the continuous parameter $q$ in the definition of the Wasserstein distance, Eq.~\eqref{eq:def:Wp}. For instance, instead of the fully symmetric definition of the distance in Eq.~\eqref{eq:hat:dij:Dalitz} one could have used a simple Euclidean distance in the Dalitz plot, or the Euclidean distance in the square Dalitz plot. One can also tune the value of $q$ using an amplitude model to obtain the highest expected sensitivity to CPV, as we did in Sect.~\ref{sec:three-bodyB} (see also App.~\ref{sec:app:opt:q}). However, even without an amplitude model, origins of CPV across the Dalitz plot can be identified. Such tests allow for unbinned, model independent tests of CPV in the phase space of distributions, thereby informing future analyses. Its use with weighted datasets is also straightforward, as illustrated in Sect.~\ref{sec:binned:W}.

\begin{figure}[t]
\begin{center}
\includegraphics[width=0.5 \textwidth]{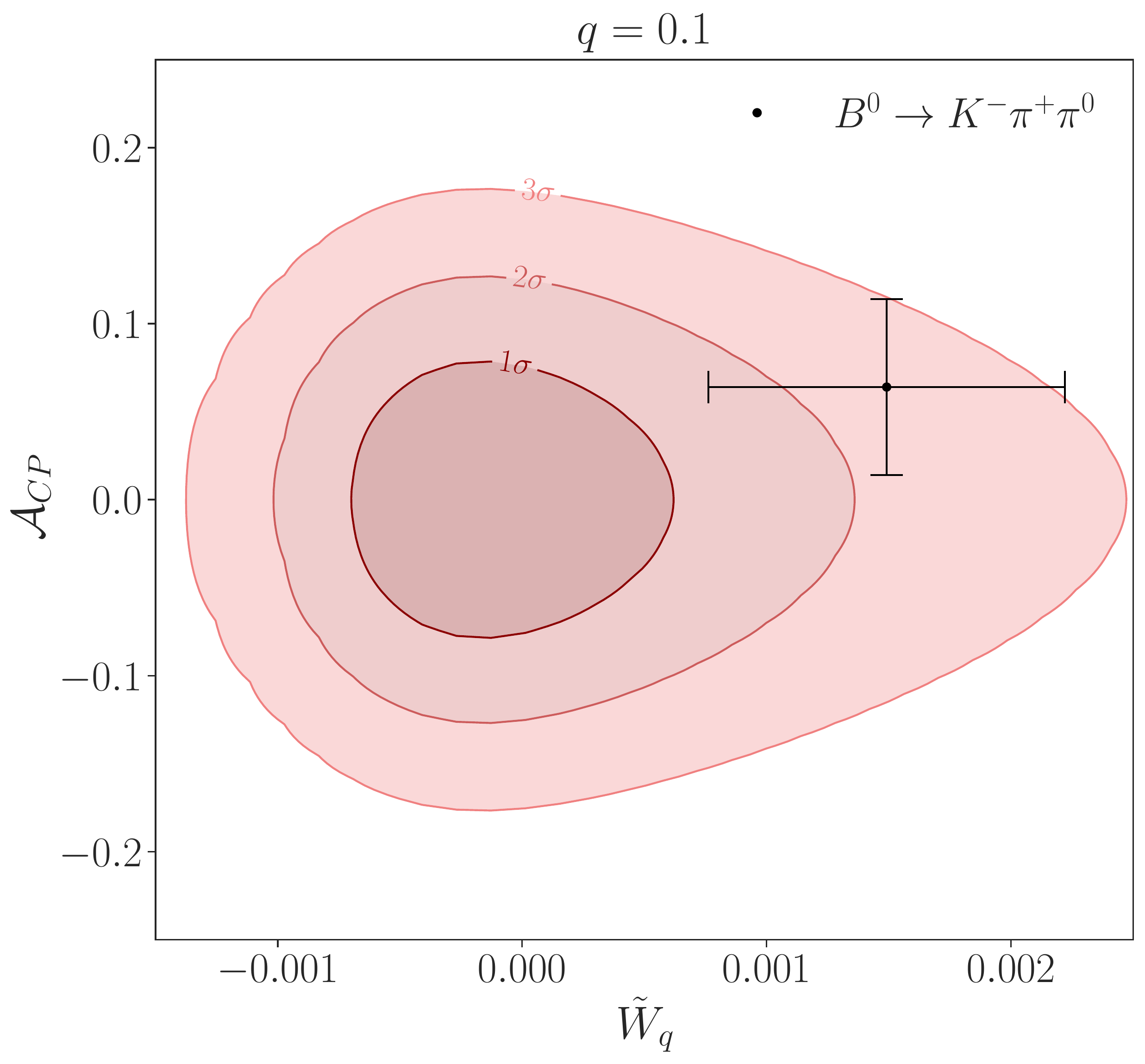}
\caption{
The expected $1\sigma, 2\sigma, 3\sigma$ C.L. contours (red shaded regions) in the $(\mathcal{A}_\text{CP},\tilde W_q=W_q-\overline{W}_q)$ plane, assuming CP conserving $B^0\to K^+\pi^-\pi^0$ decays. The black lines show the current global average for
$\mathcal{A}_{\text{CP}} = 0.064 \pm 0.050$ as well as the $1\sigma$ range for $\tilde{W}_q$ that follows from our toy model of CP violation in distributions,  introduced in Sect.~\ref{sec:three-bodyB}. The data was generated by first randomly sampling $A_{\text{CP}}$ via a Gaussian with mean $\mu = 0$ and standard deviation $\sigma = 0.05$ to mimic the uncertainty in the global average, selecting $B$ and $\bar{B}$ decays accordingly with the total number of events fixed to $2\times 10^3$, and computing $W_q$. This was repeated with $10^3$ distinct datasets and contours drawn according to a numerical integration of a 2--dimensional joint PDF where $\tilde{W}_q$ and $\mathcal{A}_{\text{CP}}$ were assumed to be independent. 
}
\label{fig:money_plot}
\end{center}
\end{figure}

Since $W_q$ measures the cummulative presence of CPV in the Dalitz plot one therefore needs only two observables to fully quantify the amount of direct CPV in a multibody $B$ decay: the total direct CP asymmetry, ${\cal A}_{\rm CP}$, and the Wasserstein distance $W_q$ (or a related Wasserstein distance test statistic such as the windowed Wasserstein distance $I_q$). This is illustrated in Fig.~\ref{fig:money_plot}, which shows a contour plot of $\mathcal{A}_\text{CP}$ vs $\tilde{W}_{q} = W_q - \overline{W}_q$,  where $\overline{W}_q$ is the median ${W}_q$  
expected for CP conserving $B$ decays (in this case obtained using the amplitude model, but could be obtained using permutation method). For CP conserving decays both $\mathcal{A}_\text{CP}$ and $\tilde{W}_{q}$ are consistent with zero within statistical uncertainties, and would be nonzero if there is significant CP violation. The two give complementary information: $\mathcal{A}_\text{CP}$ is nonzero if there is a difference in the partial decay widths between $B^0\to K^+\pi^-\pi^0$ and $\bar B^0\to K^-\pi^+\pi^0$ decay channels, while $\tilde{W}_{q}$ is nonzero, if there is a difference between the phase space distributions of the two CP conjugated decays.

For its simplicity, $W_q$ does have a drawback --- due to the CP conserving noise it usually results in a lower sensitivity to CPV compared to the energy test with an optimized regulator function. Applying filters on the optimal transports for each individual $B^0$ and $\bar B^0$ decay datapoints in the Dalitz plot, however, gives an optimized version of the Wasserstein test statistic, ${ I}_q$,  with sensitivity to CPV indistinguishable on statistical basis from the optimized energy test. In Sect.~\ref{sec:win:EMD} we focused on windowed filtering, with Heavyside step functions abruptly switching on and off (or assigning negative weights) to certain ranges of optimal transport distances, however, one could have also used other filtering variants with smooth versions of the window function Eq.~\eqref{eq:window}. 

The windowed Wasserstein distance statistic $I_q$ can match the extreme sensitivity of the energy test to the presence of CPV, a feature which can ultimately be attributed to the lack of long-tailed CP conserving probability distributions in both cases. That is, the energy test statistic successfully mitigates superfluous contributions from CP conserving variations among data samples. This comes at the price of additional $N(N-1) + \bar{N}(\bar{N} -1) \sim \mathcal{O}(N^2)$ computations (cf. the first two terms in Eq.~\eqref{eq:T:def}, encoding the CP conserving distance variations within each sample), as well as the need for a regulator function $\psi$, which restricts contributions to be only within a sphere of influence of radius $\sigma$, see App.~\ref{app:energy_test}. Such suppression of CP conserving variations is expected to be necessary for any metric based statistic with enhanced sensitivity to CPV. 
As we showed in Sect.~\ref{sec:win:EMD} the suppression of CP conserving variation can be implemented for the case of the Wasserstein distance based statistics by using windowed filtering. Again, this comes at the cost of additional computations required for the optimization of the filtering window function.

The computation requirements may become prohibitive when faced with large datasamples, such as the $D$ decays with  $N\gtrsim10^6$ events in a sample. In that case one can use approximate versions of Wasserstein distance to construct test statistics that scale better with $N$, at a rather small cost to sensitivity. In Sect.~\ref{sec:D:dec} we discussed two such possibilities, the binned Wasserstein test statistic, $W_q^{\rm bin}$,  and the sliced Wasserstein test statistic, $SW_q$. Both were shown to give similar sensitivities to CPV as $W_q$, when either the binning is fine enough (for $W_q^{\rm bin}$) or for large enough number of slices (for $SW_q$). 

The work presented in this manuscript could be extended in several directions. The extension to higher dimensional spaces, such as the $n$-body meson decays, $n\leq 4$, is straightforward with no changes to the formalism required. The main question in that case is the scaling with the number of particles in the final state, where the usual curse of dimensionality may be mitigated by the fact that the multi-body decays tend to have large quasi-two-body resonant decay structure. Less trivial extensions include time dependent weighting of decay rates in order to probe indirect CP violation. Finally, one could explore other deviations from the Wasserstein distance. For instance, an interesting direction would be to explore entropic smoothing of the Wasserstein distance, i.e., an entropic regularization of the optimal transportation problem. The resulting Sinkhorn divergence depends on a hyperparameter $\lambda$ which interpolates between the Wasserstein distance ($\lambda=\infty$) and the energy test ($\lambda=0$)~\cite{ramdas2017wasserstein}. 

Finally, we provide a public code \texttt{EMD4CPV} that allows a straightforward use of the introduced Wasserstein based statistics for two-sample tests, with further details about the code given in App.  \ref{sec:code},

\section*{Acknowledgements} 
We thank S. Bressler, J. Thaler for discussions on the two sample tests, and M. Gersabeck, D. White, G. Sarpis, S. Chen and Y. Brodzicka in particular for extended discussions on the energy test, as well as M. Szewc for comments on the manuscript. We thank T. Latham for help with the {\tt Laura++} framework, T. Evans for support using the {\tt AmpGen} framework, and J. Brod for providing access to the local computing resources.   
AD acknowledges support from STFC grants ST/S000925 and ST/W000601/1. AY, JZ and TM  acknowledge support in part by the DOE grant de-sc0011784 and NSF OAC-2103889.

\appendix
\section{Public code \texttt{EMD4CPV}}
\label{sec:code}
\begin{figure}[t]
\begin{center}
\includegraphics[width=0.4\textwidth]{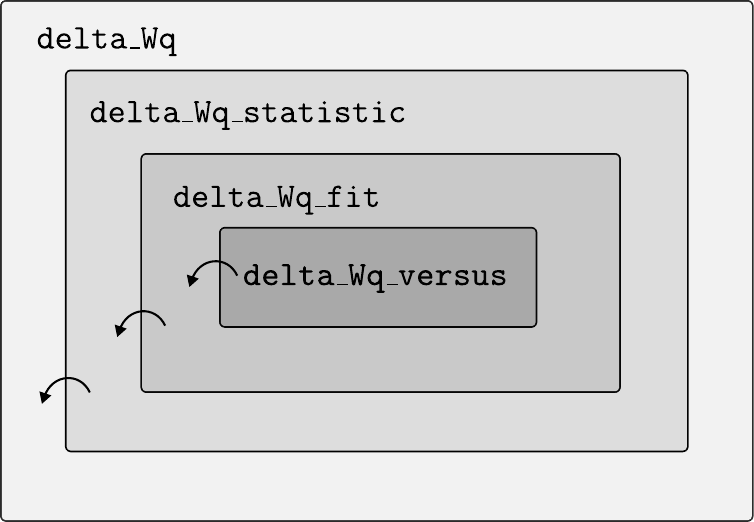}
\caption{Pictorial representation of the \texttt{EMD4CPV} program architecture. The largest box represents the highest level class, \texttt{delta\_Wq}, followed by lower level sub-classes contained within it. The arrows represent the inheritance of each sub-class.}
\label{fig:wass_dalitz}
\end{center}
\end{figure}

The public code and repository for this project may be found at
\begin{equation*}
    \text{\hyperlink{https://github.com/adamdddave/EMD4CPV}{\texttt{https://github.com/adamdddave/EMD4CPV}}}.
\end{equation*}

The program architecture is hierarchically structured, resembling a nested doll of classes and subclasses, as shown schematically in Fig.~\ref{fig:wass_dalitz},
\begin{equation}
    \texttt{delta\_Wq(delta\_Wq\_statistics(delta\_Wq\_fit(delta\_Wq\_versus)))}.
\end{equation}
The \texttt{delta\_Wq} is the highest level class and contains the sub-class \texttt{delta\_Wq\_statistics}, which in turn contains the sub-class \texttt{delta\_Wq\_fit},  which finally contains the sub-class \texttt{delta\_Wq\_versus}. This nested structure was implemented for three main reasons. Firstly,  the modularity improves readability and the ease of use, since the programs using the classes are structured as function calls from a software library. Secondly, this class--subclass structure follows the natural progression of the analysis pipeline used to compute and compare different $W_q$ based statistics.  For example, a typical usage of the library will follow a nested call of functions within each class, 
\begin{equation}
    \texttt{delta\_Wq} \rightarrow \texttt{delta\_Wq\_statistics} \rightarrow \texttt{delta\_Wq\_fit} \rightarrow \texttt{delta\_Wq\_versus}.
\end{equation}
Finally, since each sub-class inherits 
all functions 
from the previous class this allows the user to work at any level of the program architecture while only needing to initialize one class instance. While the use case of the program is oriented toward 3--body decays, the code is generic enough such that it can be used with any $n-$dimensional dataset. 

Below we summarize briefly the software pipeline (see the documentation as well as the example {\tt Python} notebook \texttt{example.ipynb} within the repository for more details):
\begin{itemize}
\item
The  \texttt{delta\_Wq} class contains functions which allow the user to input two $n$--dimensional distributions and obtain the associated binned or unbinned $\delta W_q$ values chosen by the optimization. Since in most cases the CP conserving distributions (functionals of $\delta W_q$)  need to be calculated, the class is set up such that the generation of the CP even distributions via the master or permutation methods can be done efficiently by randomly selecting a subset of unique datapoints from a larger datapool provided by the user in the form of a text file. In addition, this class may be used to compute the sliced Wasserstein distance $SW_q$.
\item
Once the $\delta W_q$ ensemble is obtained, the subclass \texttt{delta\_Wq\_statistics} can be used to compute the $W_q$, ${I}_q$, or any other user defined statistical distributions. 
\item
Oftentimes, when computing the $p-$values from the CPV datasets a fit is needed in order to extrapolate outside the ranges of explicitly calculated CP conserving distributions. These fits can be performed using the \texttt{delta\_Wq} subclass which allows the user to iteratively fit to any distribution within the \texttt{SciPy.stats} library and return the associated PDFs, CDFs, SFs, $\chi^2$--values, as well as the PDFs, CDFs, and SFs associated with the $\pm 1\sigma$ errors on the fit parameters. 
\item
Finally, the \texttt{delta\_Wq\_versus}\footnote{This subclass requires \texttt{Python 3.10+} while all other classes require \texttt{Python 3+}.} subclass may be used to iteratively compare the sensitivity of different statistics on ensembles of like datasets. 

\item 
Additionally, for convenience, the script \texttt{energy\_test.py} provides a \texttt{Python} implementation of the energy test statistic, i.e., the computation of the test statistic $T$ (see Eq.~\eqref{eq:T:def}) between two $n$--dimensional distributions for use when computing CPV statistic values in \texttt{delta\_Wq\_versus}. The program also includes an interface with \texttt{Manet} \cite{Parkes:2016yie} (which utilizes the \texttt{CUDA} API to parallelize the computation on NVIDIA GPUs) such that the user can efficiently generate large statistic CP conserving $T$ distributions if desired.

\end{itemize}

\section{The optimal transport problem}
\label{app:optimal-transport}
Consider two discrete samples $P$ and $\bar{P}$, the first consisting of $n$ points sampled from probability distribution $p$, each with weight $w_i$ such that  total weight is $W=\sum_i w_i$, and the second consisting of $\bar{n}$ points sampled from $\bar{p}$ with weights $\bar{w}_i$ and total weight $\bar{W}=\sum_i \bar w_i$. The problem optimal transport consists of transporting the weight $W$ of $P$ into the weight of $\bar{W}$ of $\bar{P}$, i.e., $P\rightarrow \bar{P}$, as efficiently as possible given some cost function related to the distances among the points. 

This requires minimizing an $n \times \bar{n}$ \textit{transport plan} ~matrix $T$, which contains information about the amount of work required to transporting $P\rightarrow \bar{P}$, such that the work required is minimal. The transport matrix thus requires knowledge of both the distances between $i$-th and $j$-th points as well as the amount of weight to be transported between the points of $P$ and $\bar{P}$. The distances between each point in $P$ and $\bar{P}$ can be encoded in an $n \times \bar{n}$ matrix $\mathbf{C}$. The information about the transportation of weights can be encoded via the $n \times \bar{n}$  `flow' matrix $\mathbf{F}$ subject to $\sum_i F_{ij} = \bar{w}_j, \sum_j F_{ij} = w_i$. 
That is,
 $\mathbf{F}$ specifies the fractional amount of weight to be transferred from $i$-th point in $P$ to the $j$-th point in $\bar P$, subject to the condition that the total weight from each point must be conserved.\footnote{Note that a particular transport configuration is \textit{not} required to be a bijective map, i.e., the weight of a particular point in $P$ can be partitioned to different points in $\bar{P}$.} 
 
 The total work or `cost' of a given configuration is then given by the inner product of the flow and distance matrices $T = \langle \mathbf{F}, \mathbf{C} \rangle$. Finding the most efficient plan amounts to finding the transport plan  $\mathbf{F}$ which minimizes the total cost. We denote the optimal flow matrix as $\mathbf{F}^*$ and define the Wasserstein distance as 
\begin{equation}
     W_q = \langle\mathbf{F}^*, \mathbf{C}^q \rangle^{1/q} \equiv \left( \sum_{i}^{n} \sum_j^{\bar{n}} F^*_{ij} C_{ij}^q \right)^{1/q}
 \end{equation}
Solving for $W_q$ optimally takes super-cubical time complexity with respect to the size of the input datasets $\mathcal{O}(N^3)$ \cite{pmlr-v97-atasu19a}.

\section{EMD analysis of Gaussian distributions}
\label{sec:Gauss:2D}

In this appendix we give further details on how the  Wasserstein distance $W_q$ can be used as a statistic sensitive to dissimilarities between two distributions. We use 
the toy example of two displaced Gaussian distributions, either in 1D or 2D, where the difference between distributions is taken to be controlled by a single ``CP violating'' parameter. We consider two limits: i) the two Gaussian peaks do not overlap, but are rather displaced by $\Delta\mu$, and ii) the peaks of the two Gaussian distributions overlap, while their widths differ, $\Delta \sigma\ne0$.

\begin{figure}[t]
\begin{center}
\includegraphics[width=1.0\textwidth]{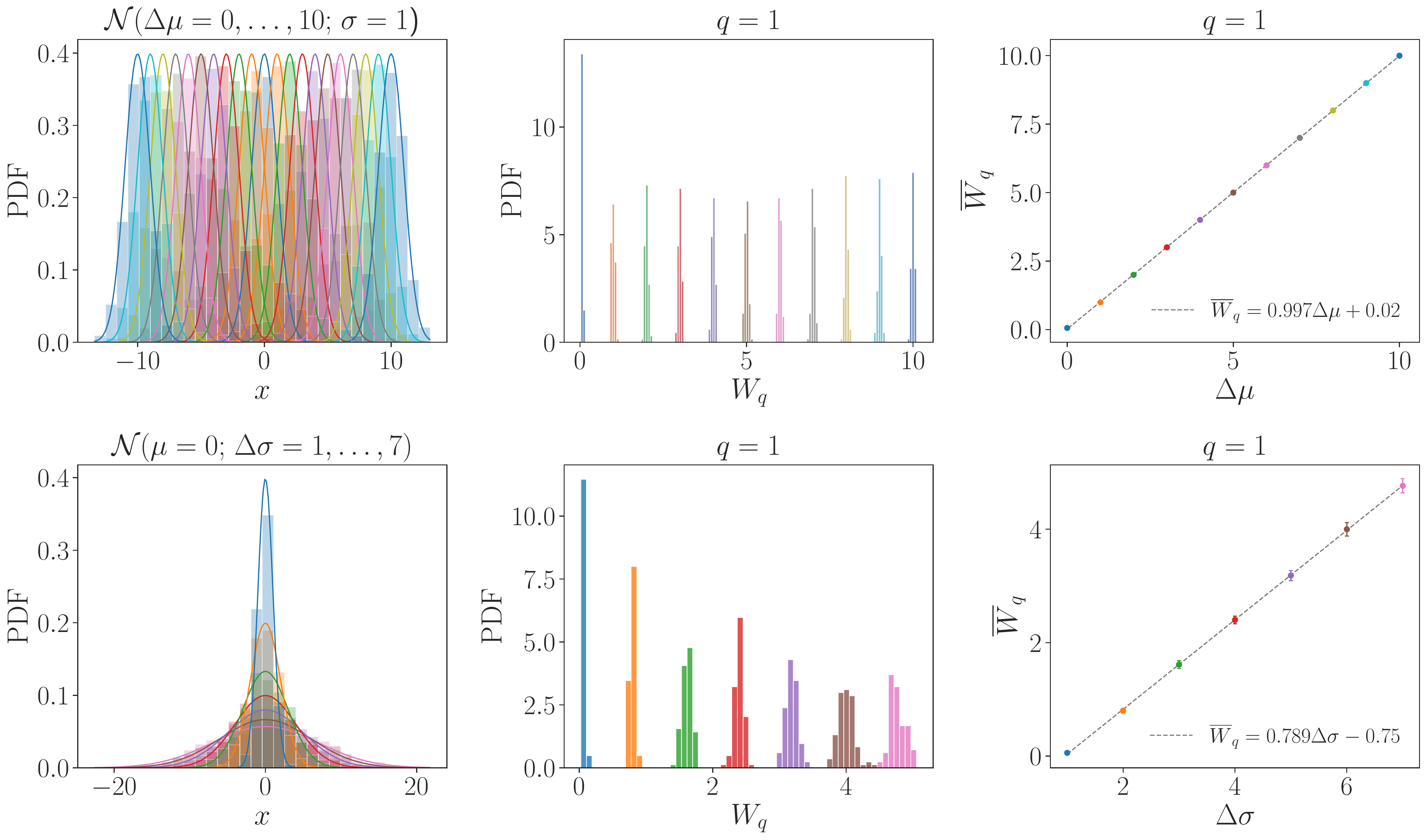}
\caption{The dependence of $W_1$ on the displacement $\Delta \mu$ (top row) or the width difference $\Delta \sigma$ (bottom)  of the two Gaussian distributions. Middle panels show the corresponding $W_1$ distributions, while the right panels show the linear dependence of the average value $\bar W_1$ on the displacement $\Delta \mu$ and width difference $\Delta \sigma$. The error bars shown in the right panels denote the $1\sigma$ bands of the $W_1$ distributions shown in the middle panels. 
}
\label{fig:wass_1d:more}
\end{center}
\end{figure}

In the main text we showed an example for the first choice where we considered two 1D Gaussians displaced by $\Delta \mu=40$ and with widths $\sigma=10$, cf. Fig.~\ref{fig:wass_1d}, where $\hat d_{ij}$ in  Eq.~\eqref{eq:def:Wp} here and below is taken to be the Euclidean distance. Since the optimal transport needs to move the points in the datasets sampled from the two distributions by a distance $\sim \Delta \mu$, the Wasserstein distance coincides with $\Delta\mu$, $W_1\to \Delta \mu$, for large $N$ (this is true to quite a good degree even for rather small values of $N$, cf. Fig.~\ref{fig:wass_1d}). This is also shown in the top panels in Fig.~\ref{fig:wass_1d:more}, where we consider 11 different  values $\Delta \mu=0,\ldots, 10$, while $\sigma=1$, cf. Fig.~\ref{fig:wass_1d:more} (top left). The distributions of $W_1$ straddle $\Delta\mu$ for $\Delta \mu$ sufficiently far away from zero (for $\Delta \mu=0$, $W_1\geq\Delta\mu$ since $W_1$ is nonnegative), shown for $N=\bar N=10^3$ in Fig.~\ref{fig:wass_1d:more} (top middle). Fig. \ref{fig:wass_1d:more} (top right) shows that the average value of the Wasserstein distance, $\bar W_1$, linearly increases with $\Delta \mu$, where for small values of $\Delta \mu/\sigma$ there is a deviation from this linear behavior, which however is almost imperceptible on the plot. 

The lower panels in Fig.~\ref{fig:wass_1d:more} show the dependence of Wasserstein distance on the width of the distributions. In this example one Gaussian distribution is held fixed, $G(x)={\cal N}(x|\mu=0,\sigma=1)$ while the other is taken to have different widths but a coinciding peak, $\bar G(x)={\cal N}(x|\bar\mu=0,\bar\sigma)$, $\bar\sigma=1,\ldots, 7$, see  Fig.~\ref{fig:wass_1d:more} (bottom left). With increasing $\bar \sigma$ the typical value of Wasserstein distance $\bar W_1$ increases, since the two distributions differ more and more, see  Fig.~\ref{fig:wass_1d:more} (bottom middle). The values of the $W_1$ also form a wider distribution for larger values of $\bar \sigma$, since the larger difference between the two Gaussians translates to a larger scatter of optimal transportation distances. The increase in the average value of the Wasserstein distance, $\bar W_1$, is linear in $\Delta\sigma=\bar\sigma-\sigma$, cf. Fig.~\ref{fig:wass_1d:more} (bottom right).

\begin{figure}[t]
\begin{center}
\includegraphics[width=0.8\textwidth]{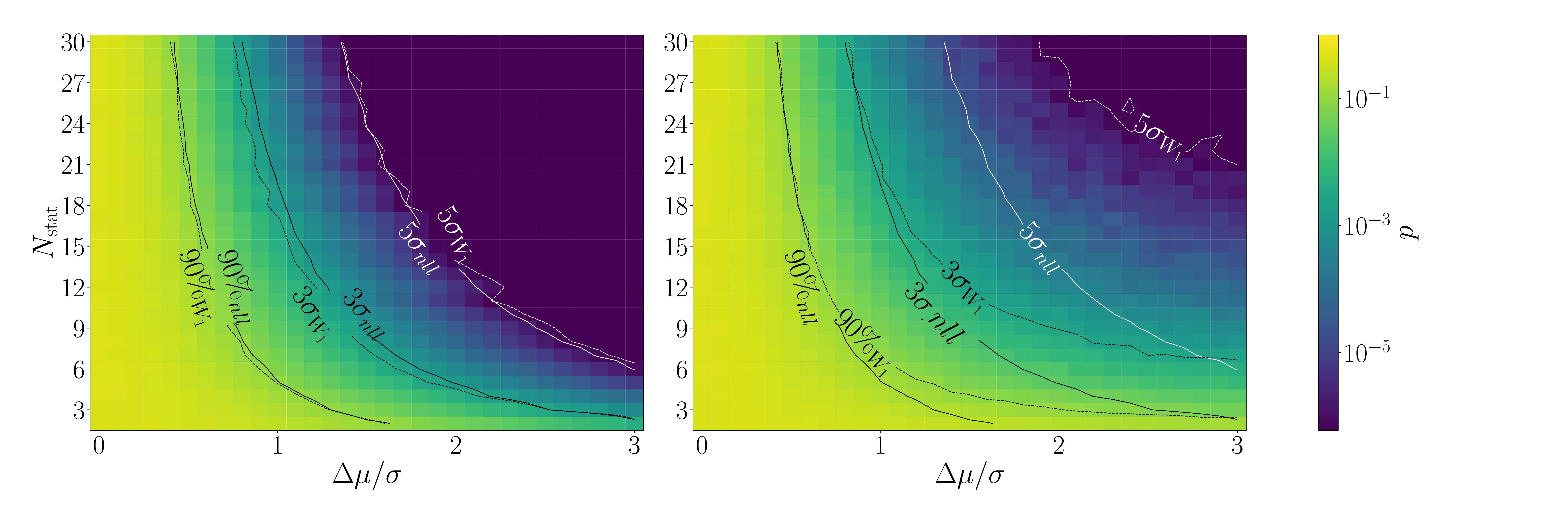}
\caption{ The expected 90\% C.L., $3\sigma$, and $5\sigma$ exclusion lines  for different Gaussian peak displacements,  $\Delta \mu/\sigma$, and sizes of statistical samples $N_{\rm stat}$ using EMD $W_1$ statistic (negative log-likelihood) are denoted with dashed (solid) lines, where one uses either true (left) or modeled (right) $\Delta \mu=0$ probability distributions.
}
\label{fig:contour_model:W1}
\end{center}
\end{figure}

Next, we check that the $W_1$ statistic leads to the same CL intervals as the negative log likelihood. Fig.~\ref{fig:contour_model:W1} (left) shows the expected $90\%$, $3\sigma$ and $5\sigma$ CL  for $\Delta \mu$ as a function of $N$ (solid contours) that follow from a known $\Delta \mu=0$  probability distribution for $W_1$.
These coincide with the expect CL
 exclusion intervals obtained from the negative log likelihood for $\Delta \mu$ (dotted contours). We see that in this case the $W_1$ statistic gives the correct coverage for all considered values of $N$ and $\Delta \mu$. 
 
 In Fig.~\ref{fig:contour_model:W1} (right) we also show the estimates of the exclusion contours that follow from a permutation (or re-randomization) test, i.e., where the symmetric ``CP-even'' $W_1$ probability distribution is modeled by randomly sampling events from ${\cal E}$ and $\bar {\cal E}$.  We see that the re-randomization estimate of the true $\Delta \mu=0$  probability distribution for $W_1$ results in a bias and thus in underestimated exclusion $p-$values. The benefit of the re-randomization is that such modeling of ``CP-even'' $W_1$ probability distribution is always possible, however it also means that the use of $W_q$ statistics is best suited for the cases where one has already a reasonable model of the distributions and can check potential bias due to the use of permutation test.   The multi-body $B$ and $D$ decays fall in this  category since one can use the fitted for amplitude models to estimate the potential bias in the permutation method for $W_q$ statistic. This was found to be small for the two $B$ and $D$ decays considered in the main text. 
 
A toy example that is closer to the case of  three body $B$ and $D$ decay Dalitz plots is the example of two displaced 2D Gaussian distributions, $g(x,y)\sim {\cal N}(x|\mu_x-\Delta\mu_x/2, \sigma){\cal N}(y|\mu_y-\Delta\mu_y/2, \sigma)$, and $\bar g(x,y)\sim {\cal N}(x|\mu_x+\Delta\mu_x/2, \sigma){\cal N}(y|\mu_y+\Delta\mu_y/2, \sigma)$. For simplicity we take the widths of all the Gaussian distributions to be the same, so that there is no CP violation (the two distributions are the same) if and only if $\Delta \mu_x=\Delta \mu_y=0$. The statistical analysis of this case is a trivial extension of the case of a 1D Gaussian toy model. Using the true ``CP-even'' $W_1$ distribution for $\Delta \mu_x=\Delta \mu_y=0$ leads to the correct coverage, while the permutation method gives some bias, as in the 1D case. 

For two-dimensional distributions there are additional observables and visualization tools that prove to be useful. First of all, for arbitrary two-sample 2D distributions one can define a Wasserstein distance asymmetry distribution  ${\mathcal W}^q_{\rm CP}$ in the same way as for the Dalitz plot, Eq.~\eqref{eq:Wq:asymm},
\begin{equation}
\label{eq:Wq:asymm:2D}
    \mathcal{W}^q_{\text{CP}} (x, y) = \frac{\sum_{\bar i} \delta \bar W_q(\bar i) - \sum_{i} \delta {W}_q(i)}{\sum_{\bar i} \delta \bar W_q(\bar i) + \sum_i \delta {W}_q(i)},
\end{equation}
where the summation over $i$ ($\bar i$) is only over the data-points from $g(x,y)$ sample contained in the bin centered at $(x, y)$ (from $\bar g(x,y)$ data in the  bin centered at $(x, y)$). In addition, we can also define a Wasserstein distance asymmetry heatmap
\begin{equation}
\label{eq:Wq:asymm:2D:heat}
    \omega^q_{\text{CP}} (x, y) = \frac{1}{s_i} \Big(\sum_{\bar i} \delta \bar W_q(\bar i) - \sum_{i} \delta {W}_q(i)\Big),
\end{equation}
where $s_{i}=\Delta x_i \Delta y_i $ is the area of the bin center at $(x, y)$. 

\begin{figure}[t]
\begin{center}
\includegraphics[width=\textwidth]{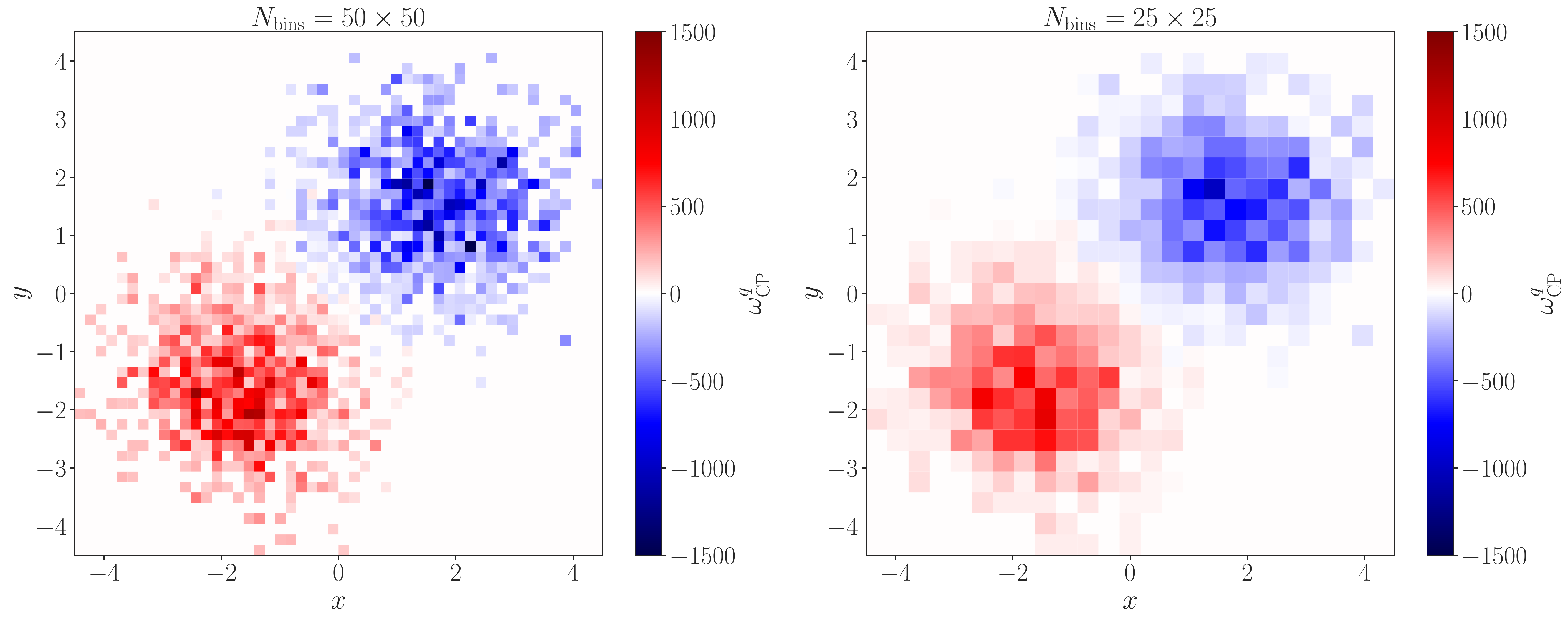}
\caption{The Wasserstein distance asymmetry heatmap $\omega_{\rm CP}^q$, $q=1$, for two 2D Gaussian distributions with equal widths, $\sigma=1$, but displaced by $\Delta \mu_x= \Delta\mu_y=3$, where the sample sizes are $N=\bar N=10^3$. Changing the binning size, from $50\times 50$ bins (left) in the range shown  to $25\times 25$ bins (right) does not change the overall size of the asymmetry heatmeap, just coarse-grains it, since the asymmetry is an intensive quantity normalized to the area. 
} 
\label{fig:heatmap_2D}
\end{center}
\end{figure}

Both ${\cal W}^q_{\rm CP} (x, y)$ and $\omega^q_{\rm CP} (x, y)$ are intensive quantities. That is, in the large $N_{\rm stat}$ limit and small bin sizes the  ${\cal W}^q_{\rm CP} (x, y)$ and $\omega^q_{\rm CP} (x, y)$ become independent of the sizes of the bins (that is as long as bins are small enough such that the variation of  ${\cal W}^q_{\rm CP} (x, y)$ and $\omega^q_{\rm CP} (x, y)$  from bin to bin is negligible). A numerical example for  $\omega^q_{\rm CP} (x, y)$ is shown in Fig.~\ref{fig:heatmap_2D}, where we see that changing the size of the bins simply corresponds to averaging the Wasserstein distance heatmap over a larger area. 

\begin{figure}[t]
\begin{center}
\includegraphics[width=0.50\textwidth]{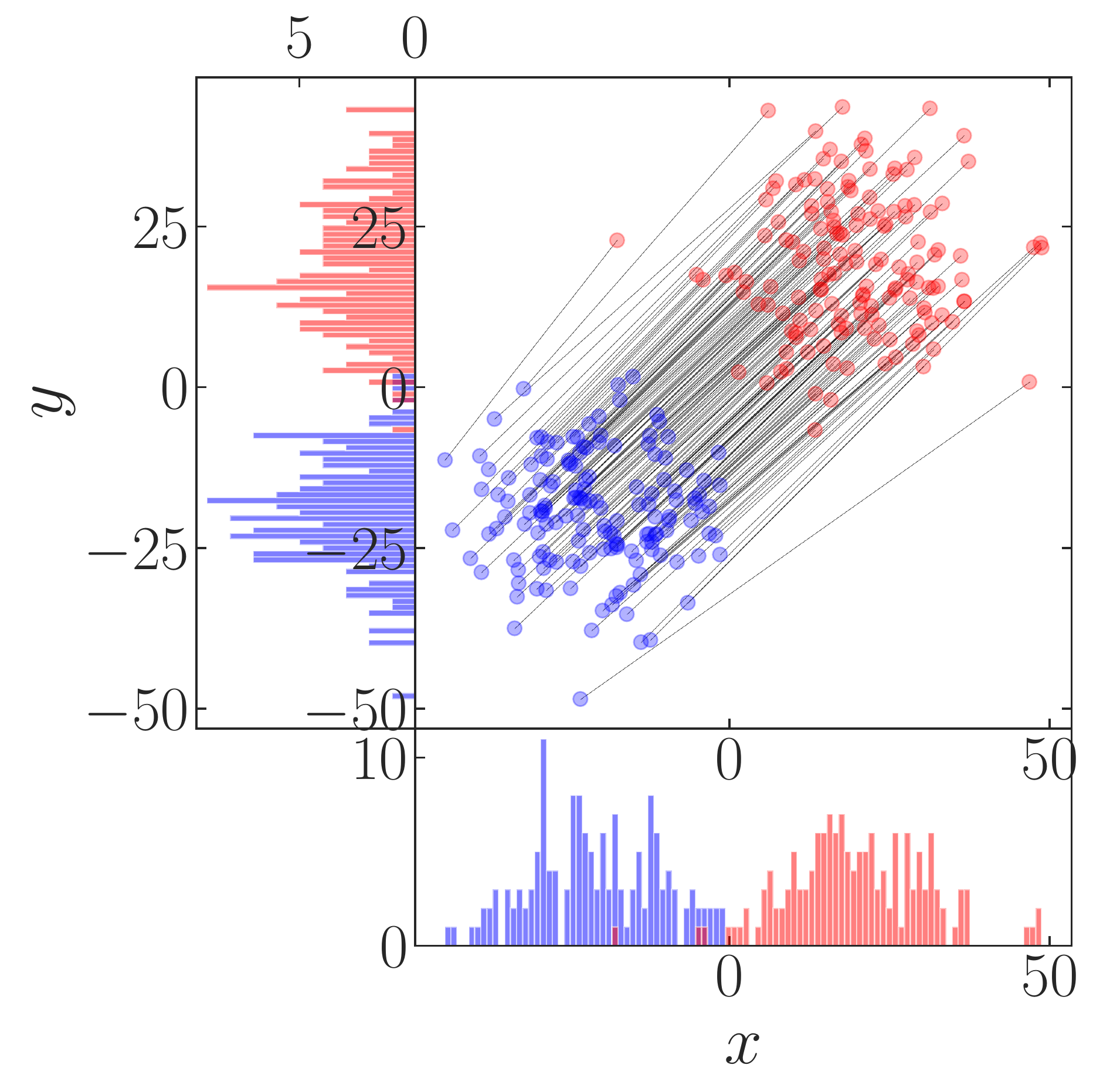}
\caption{The visualization of optimal transport of dataset sampled from $G$ (blue) to a dataset sampled from $\bar G$ (red), where $G$ and $\bar G$ are two 2D Gaussians displaced by $\Delta\mu_x=\Delta\mu_y=40$, and the sample size is $N=\bar N=150$.
}
\label{fig:2DGauss}
\end{center}
\end{figure}

For relatively small samples it is also possible to visualize the optimal transport for each individual point, an example of which is shown in Fig. \ref{fig:2DGauss} for two-sample 2D Gaussian distributions displaced by $\Delta \mu_x=40, \Delta\mu_y=40$, and a sample size of $N=\bar N=150$. The optimal transport moves the datapoints sampled from distribution $G$ (blue) to data sampled from $\bar G$ (red) shown with lines connecting pairwise the two samples. Since the two samples are of the same size, the optimal transport is a bijective map between $G$ and $\bar G$ datasets. We see that the typical shift is of order $\Delta\mu=(\Delta\mu_x^2+\Delta\mu_y^2)^{1/2}$, with datapoints on the far (near) side of $G$ distribution transported to near (far) side of $\bar G$ distribution, where near/far is defined with respect to the barycenter of $G$ and $\bar G$. 

\section{Details on EMD--test for three body decays}
\label{sec:app:further:EMD}

In this appendix we give further details on the implementation of Wasserstein distance as a measure of CPV in three body $B$ and $D$ decays.
  In Sect.~\ref{subsec:p_val_error_analysis} we discuss the details of the error analysis on the quoted $p-$values, in Sect.~\ref{sec:app:opt:q} the optimization of the $q$ parameter, while in Sect.~\ref{app:sec:q01:1:10} we collect the additional results for $q=\{0.1,1,10\}$, supplementing the results shown in the main text.

\subsection{The $p-$value error analysis}\label{subsec:p_val_error_analysis}

\begin{figure}[t]
\begin{center}
\includegraphics[width=0.49\textwidth]{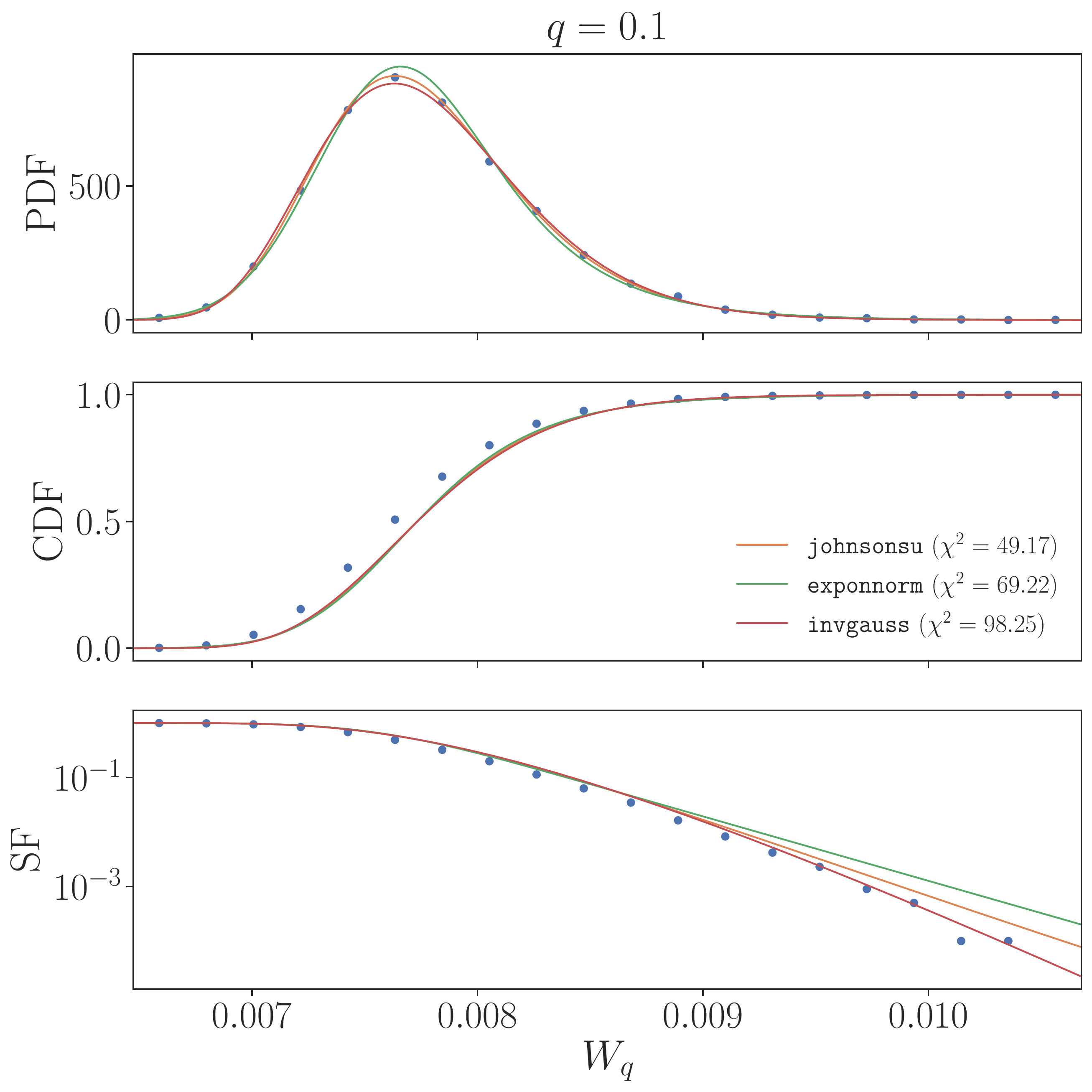}
\includegraphics[width=0.49\textwidth]{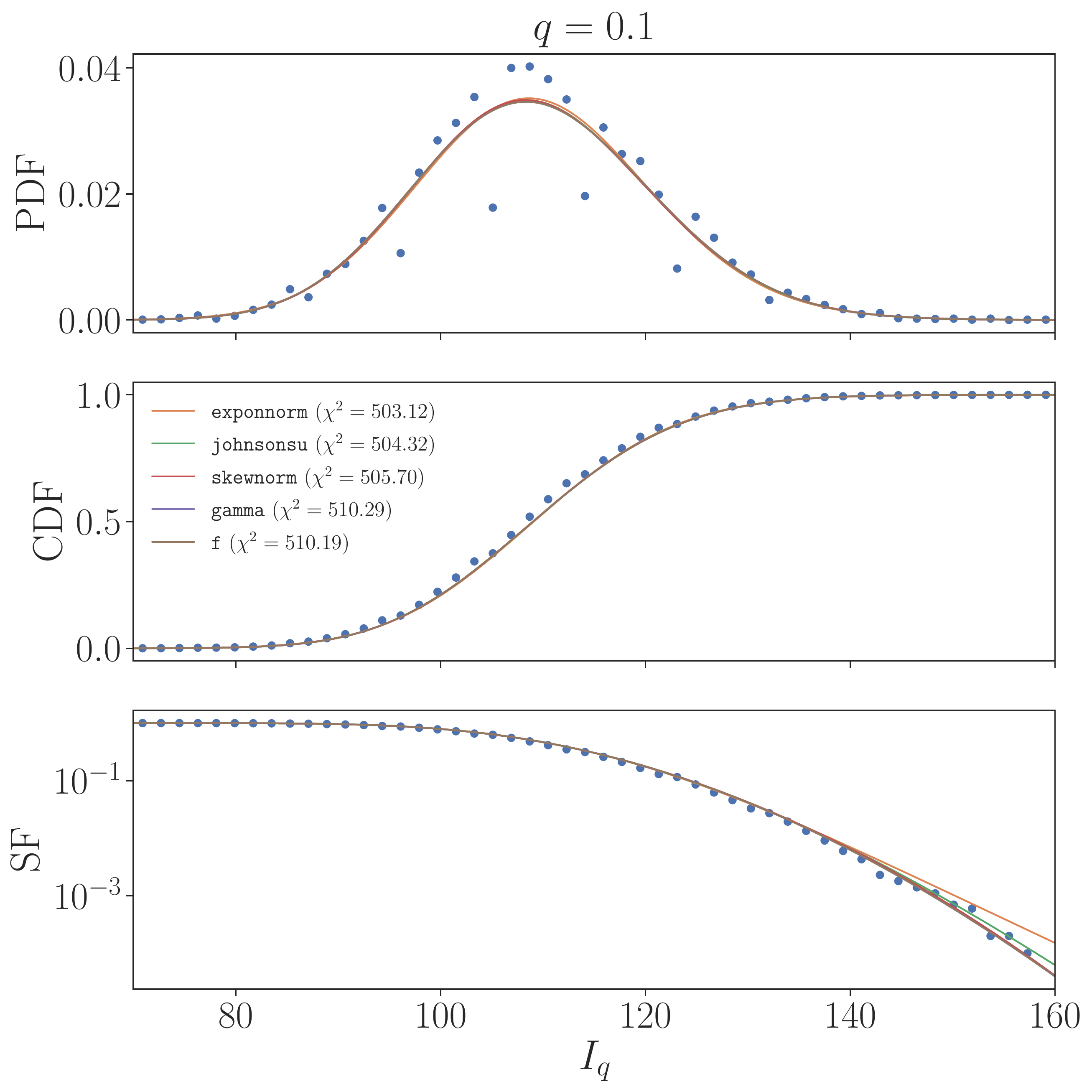}
\caption{The CP conserving PDF, CDF and SF distributions for $W_q$ (left) and $I_q$  (right) obtained using the master method for $B^0\to K^-\pi^+\pi^0$ decays with $q=0.1$ and $N=\bar N=10^3$. The different fit functions are denoted in the legend, along with respective $\chi^2$ values.}
\label{fig:pdf_cdf_sf_fits_q_0.1}
\end{center}
\end{figure}

\begin{figure}[t]
\begin{center}
\includegraphics[width=0.49\textwidth]{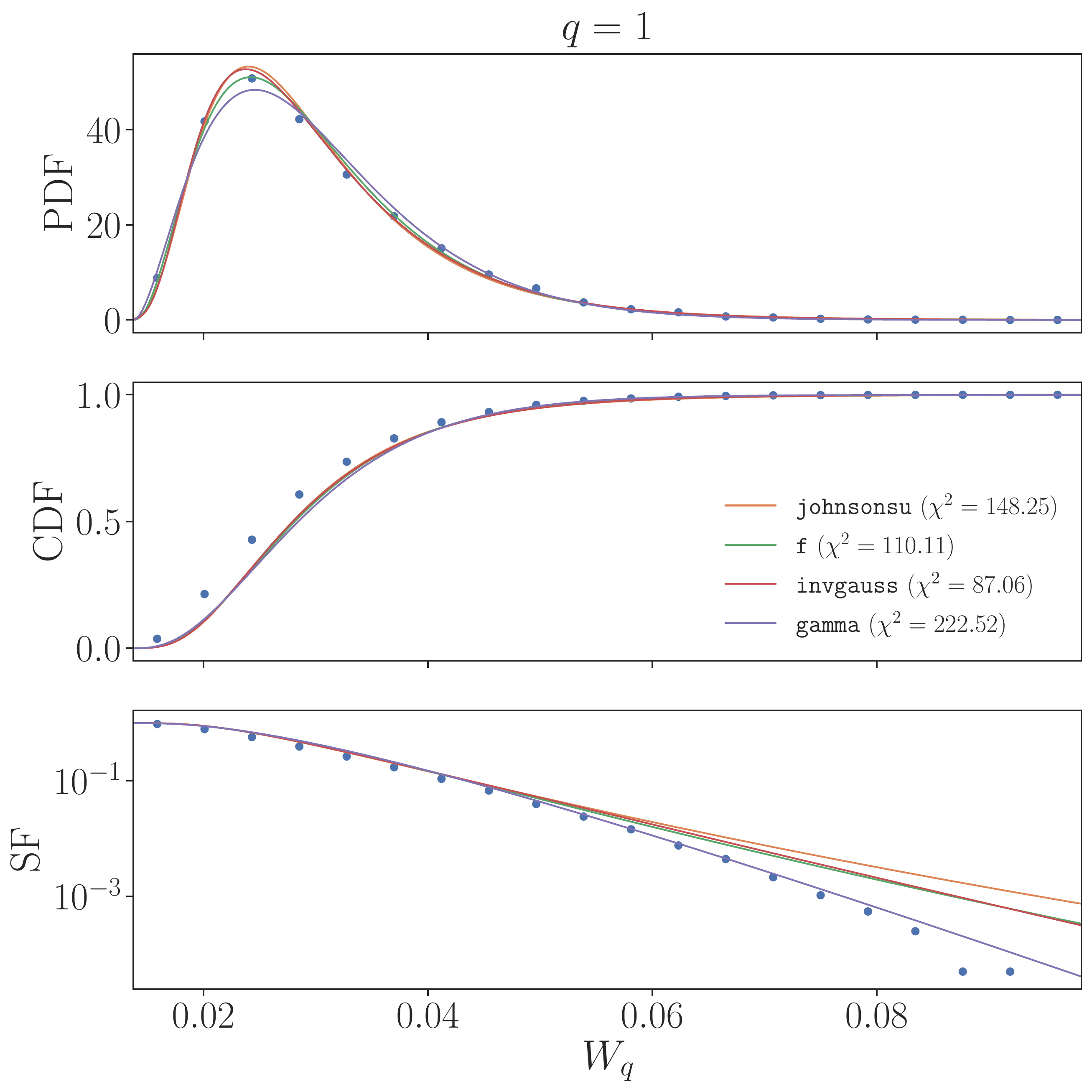}
\includegraphics[width=0.49\textwidth]{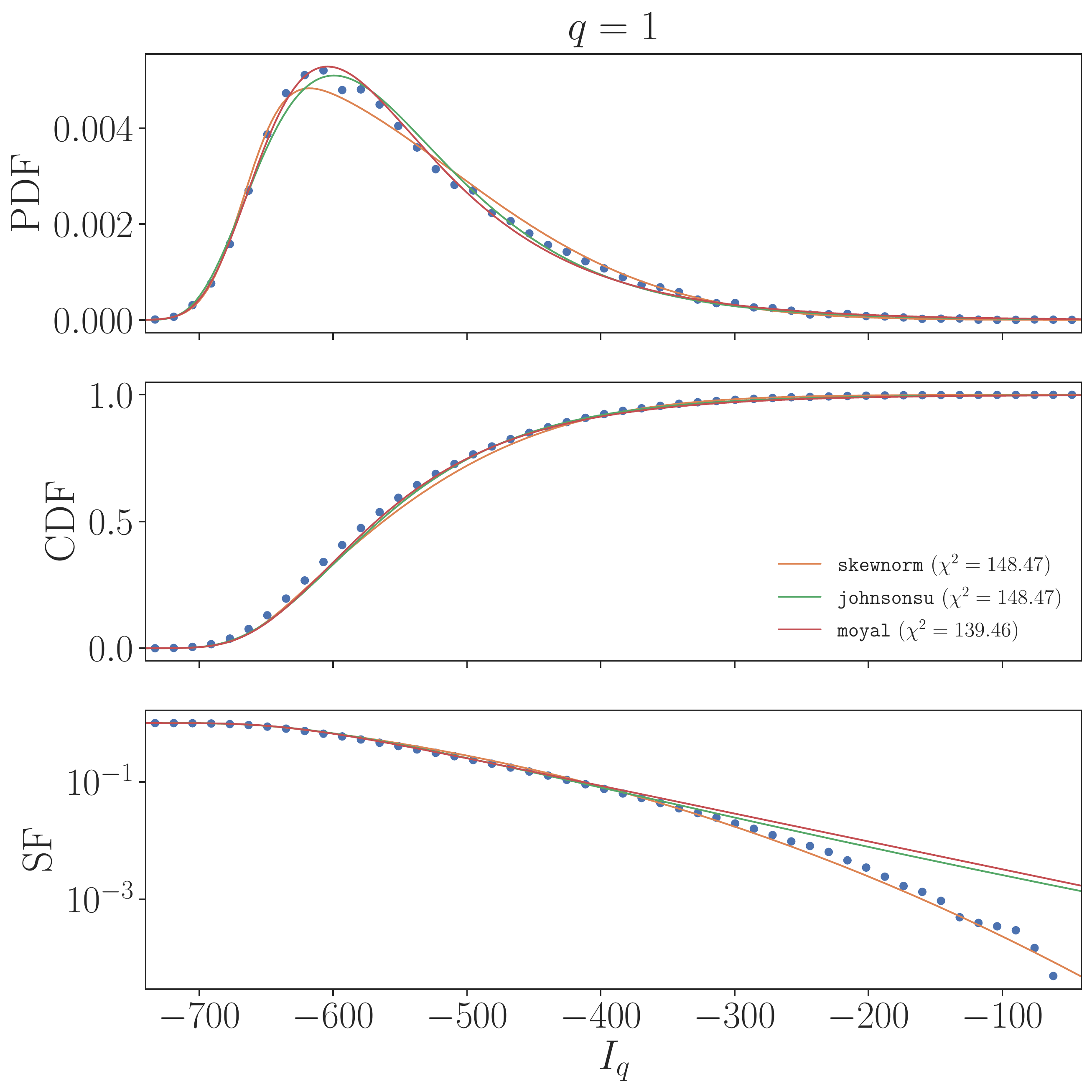}
\\[3mm]
\includegraphics[width=0.49\textwidth]{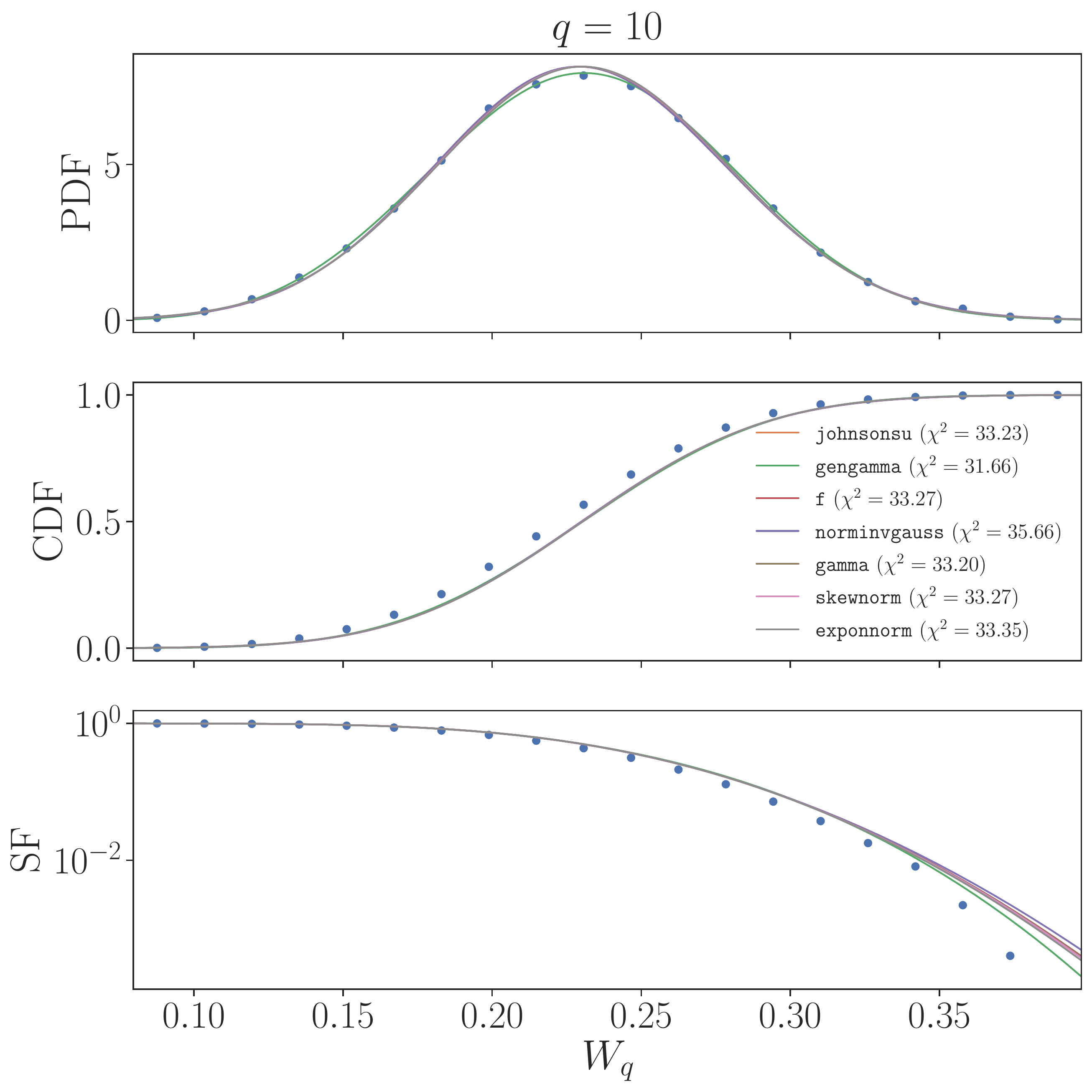}
\includegraphics[width=0.49\textwidth]{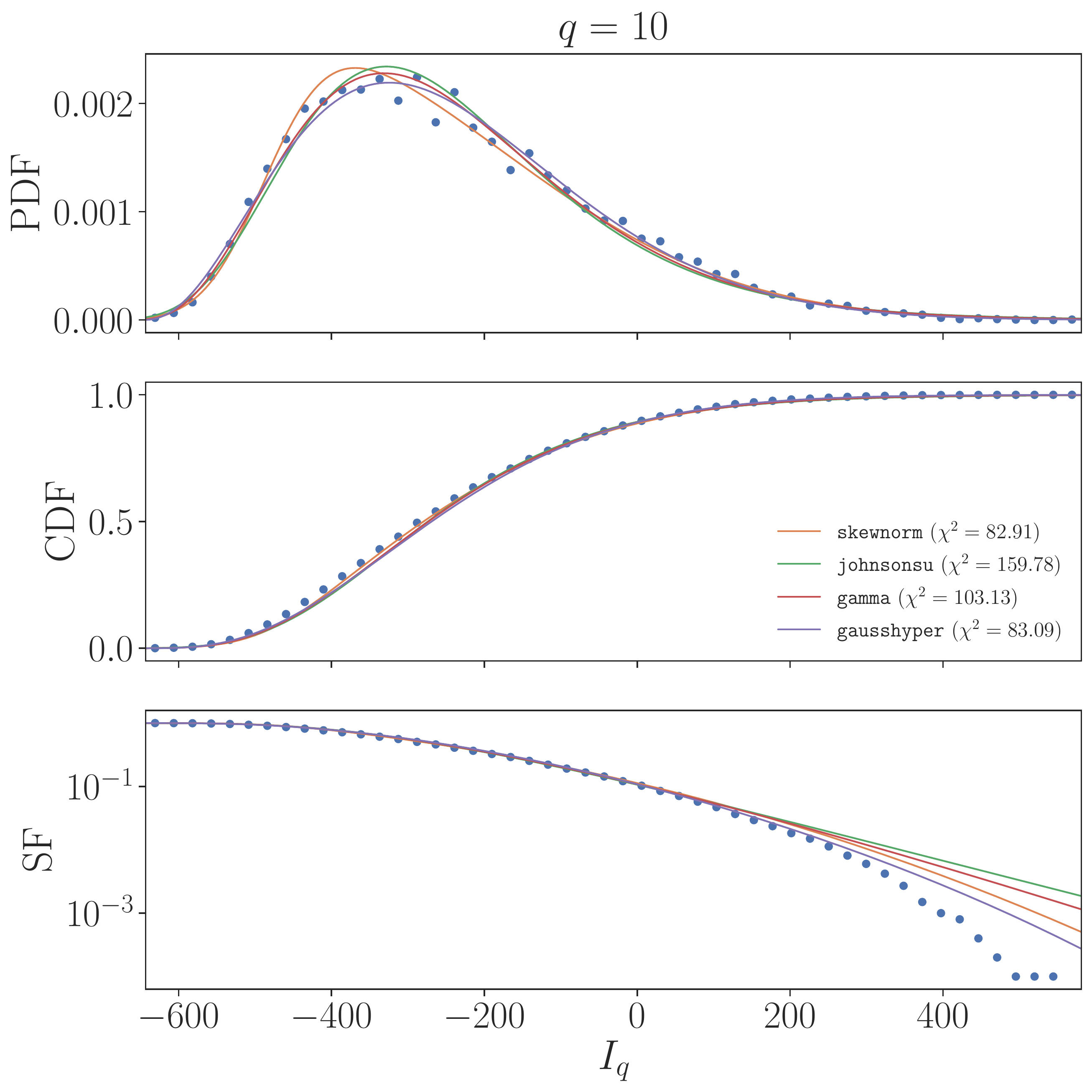}
\caption{Same as Fig. \ref{fig:pdf_cdf_sf_fits_q_0.1} but for
$q=1$ (top panels) and $q=10$ (bottom panels).}
\label{fig:pdf_cdf_sf_fits_q_10}
\end{center}
\end{figure}

In the numerical results in the main text we determine the $p-$value at which the hypothesis of CP conservation is excluded from the master $W_q$ distribution. This is numerically advantageous since it does not need to be recalculated for each dataset of $B$ and $\bar B$ or $D$ and $\bar D$ events, while given the results in Fig.~\ref{fig:scramble_vs_CPC_PDFs} we do not expect to introduce a significant difference to the estimates using the permutation method. 

For $B$ decays we calculate the master $W_q$ distribution from $10^4$ unique samplings of  $B$ and $\bar{B}$ datasets, each with $10^3$ events, while for $D$ decays we use $10^3$ unique samplings of  $D$ and $\bar{D}$ datasets, each containing $10^5$ events. The master $W_q$ PDF is fit with a smooth curve. First, the data is binned such that each bin is populated with at least one event. An initial fit is then performed using the $\texttt{SciPy}$'s non-linear least squares fitter, from which we obtain the initial values of the fit parameters. These parameters are then passed to the $\texttt{SciPy}$'s \texttt{curve\_fit} function along with statistical errors on the $i$th bin according to $\delta N_i = \sqrt{N_i (1-N_i / N)}$. This returns an updated list of fit parameters along with the parameter covariance matrix. The $1 \sigma$ parameter fit values are given as the square root of the diagonal elements of the covariance matrix. Errors on $p$--values are then estimated by considering the one sigma confidence bands on the SF distribution as shown in Fig. \ref{fig:count_difference}.  From the fit of the PDF we compute the survival factor distribution, SF=$1-$CDF, from which one can directly read off the $p-$value with which the no CPV hypothesis is excluded, for each value of the measured $W_{q}^{\text{exp}}$, as seen in Fig.~\ref{fig:PDF:CDF:SF}. 
For many of the CPV datasets we consider the value of the statistic $W_{q}^{\text{exp}}$ falls well outside the range for which the master distribution was computed.
For these cases we use the fit to extrapolate to smaller $p-$values, where the error on the extrapolation is estimated from errors on the fit parameters as described above. 

The results of the above procedure for $B^0\to K^+\pi^-\pi^0$ decay samples of size $N=\bar N=10^3$ are shown in Figs.~\ref{fig:pdf_cdf_sf_fits_q_0.1}, \ref{fig:pdf_cdf_sf_fits_q_10} for the Wasserstein statistic $W_q$ and windowed Wasserstein statistic ${I}_q$ for $q = 0.1, 1,$ and $10$. The CPC PDFs are iteratively fit 
to built-in continuous distributions contained in the \texttt{SciPy} statistics library, as listed in the legend of the corresponding panels. In most cases, the best fit is chosen according to the minimum of a $\chi^2$ statistic, however, in cases where multiple distributions achieve similar $\chi^2$ values, the distribution that best matches the tail of the distribution is chosen. In particular, for $I_q$ we use the \texttt{skewnorm} fit for the extrapolation to small $p-$values.

\subsection{Optimizing the $q$ value}
\label{sec:app:opt:q}
\begin{figure}[t]
\begin{center}
\includegraphics[width=0.5\textwidth]{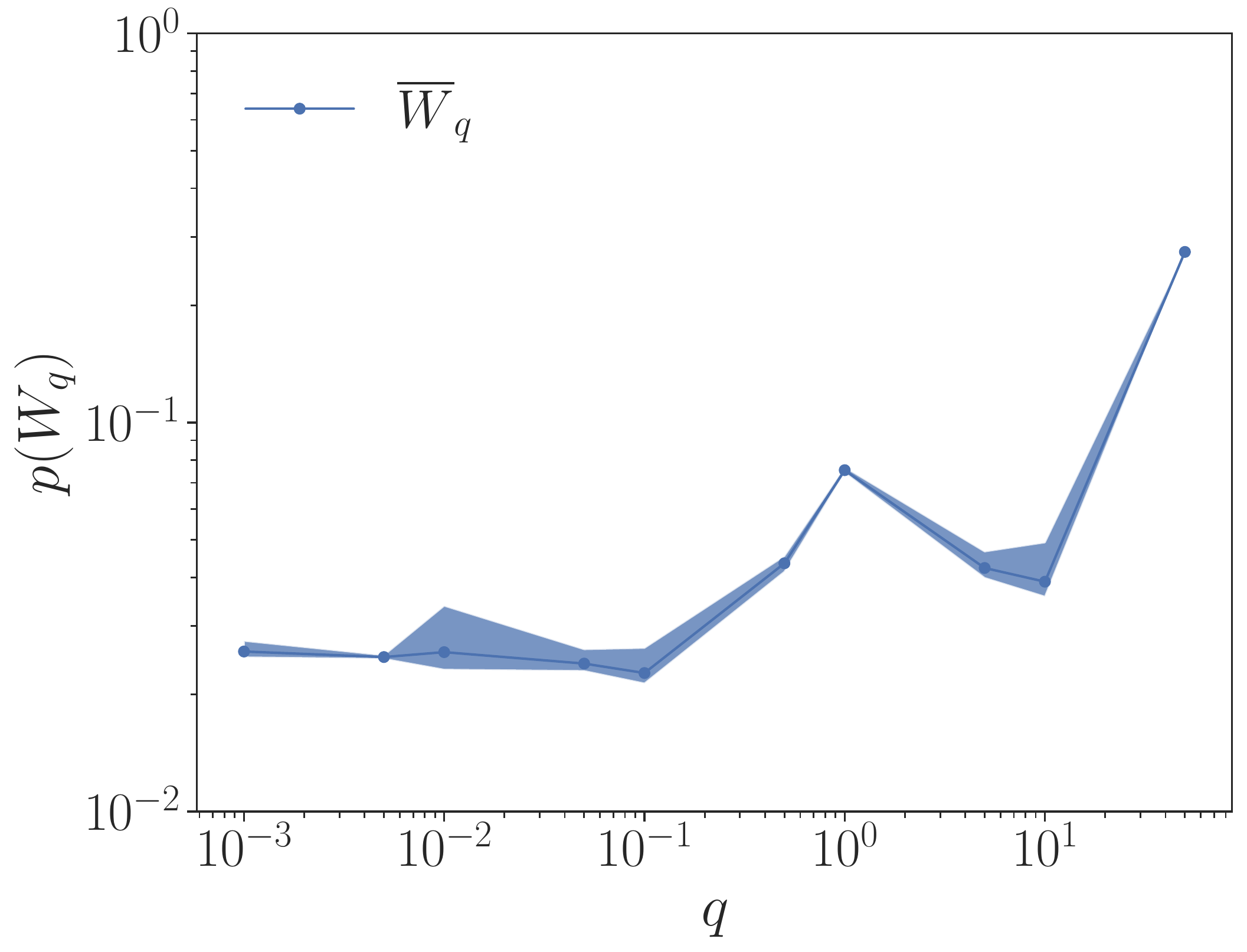}
\caption{The $q$ dependence of expected exclusion C.L., $p(W_q)$ (blue dots joined by a solid line), at which the CP conserving hypothesis is excluded for 
$N=\bar N=10^3$ event sample sizes, obtained by performing ensemble averages over $N_e=500$ CPV datasets, with the blue band indicate the $1\sigma$ spread of $p-$values over the ensemble. 
}
\label{fig:q_tuning}
\end{center}
\end{figure}

The Wasserstein distance 
weighting exponent parameter $q$, Eq.~\eqref{eq:def:Wp}, may be tuned to maximize the expected sensitivity to CPV in a particular distribution, such as the $B^0\to K^+\pi^-\pi^0$ Dalitz plot. Such an optimization of course depends on the assumed model for $B^0\to K^+\pi^-\pi^0$ decay amplitudes and in particular on the assumed values of the strong and weak phases that are hard to calculate but can in principle be fit from data.  

In the example shown in Fig.~\ref{fig:q_tuning} we used the nominal toy model for CPV in $B^0\to K^+\pi^-\pi^0$ Dalitz plot that we used throughout Sec. \ref{sec:three-bodyB},  with the amplitudes and phases for $B^0$ and $\bar B^0$ isobar models set to the central values of the measurement in Ref \cite{BABAR:2011ae}. Similarly, for the CPC datasets we use the central values of amplitudes and phases in the $B^0$ BaBar isobar model \cite{BABAR:2011ae} for both $B^0$ and $\bar B^0$ decays. Fig.~\ref{fig:q_tuning} shows the variation with $q$ of the expected C.L. $p(W_q)$ for exclusion of the CPC hypothesis, given our CPV model, for $N=\bar N=10^3$ event sample sizes. The blue bands give a $1\sigma$ range of expected C.L. exclusions as obtained form an ensemble of $N_e=500$ CPV samples. We see that for $q\lesssim 0.1$ the expected $p(W_q)$ remains unchanged when lowering $q$ within the range considered, while for higher $q$ there is in general diminished sensitivity to CPV, with the exception of the region around $q\sim {\mathcal O}(10)$. We suspect that these ranges of $q$ correspond to typical scales in the problem, i.e., the typical widths of the resonances (relative to the mass of $B$ quark), but did not explore this hypothesis further.  

In the numerics in the main text we chose $q = 0.1$, which roughly optimizes the sensitivity to CPV, but show in Sect.~\ref{app:sec:q01:1:10} below also the results for $q=1,10$.

\subsection{Further results for $q = 0.1,1,10$}
\label{app:sec:q01:1:10}

In this appendix we list further results for Wasserstein distances with $q=0.1,1,10$ both for $B$ and $D$ decays, supplementing the results discussed in Sects. \ref{sec:three-bodyB}, \ref{sec:D:dec}.

\begin{figure}[t]
\begin{center}
\includegraphics[width=0.8\textwidth]{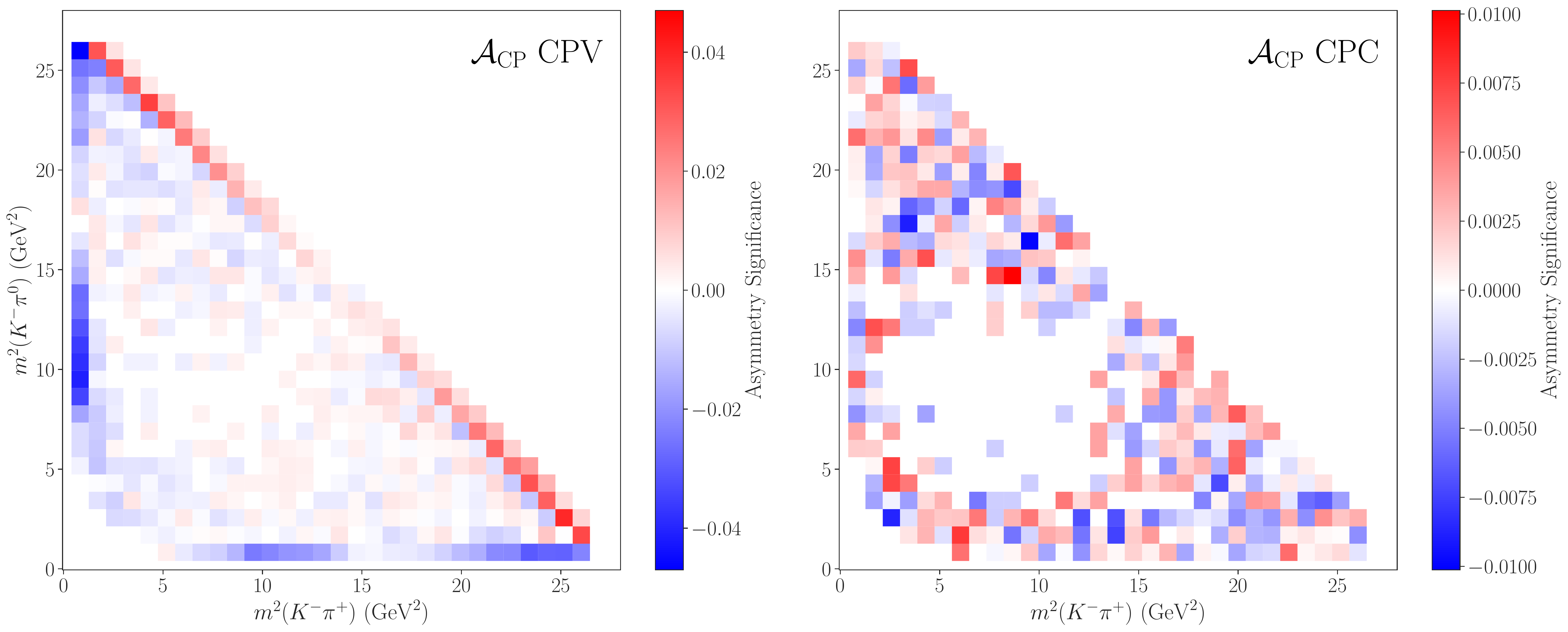}\\[1.0mm]
\includegraphics[width=0.8\textwidth]{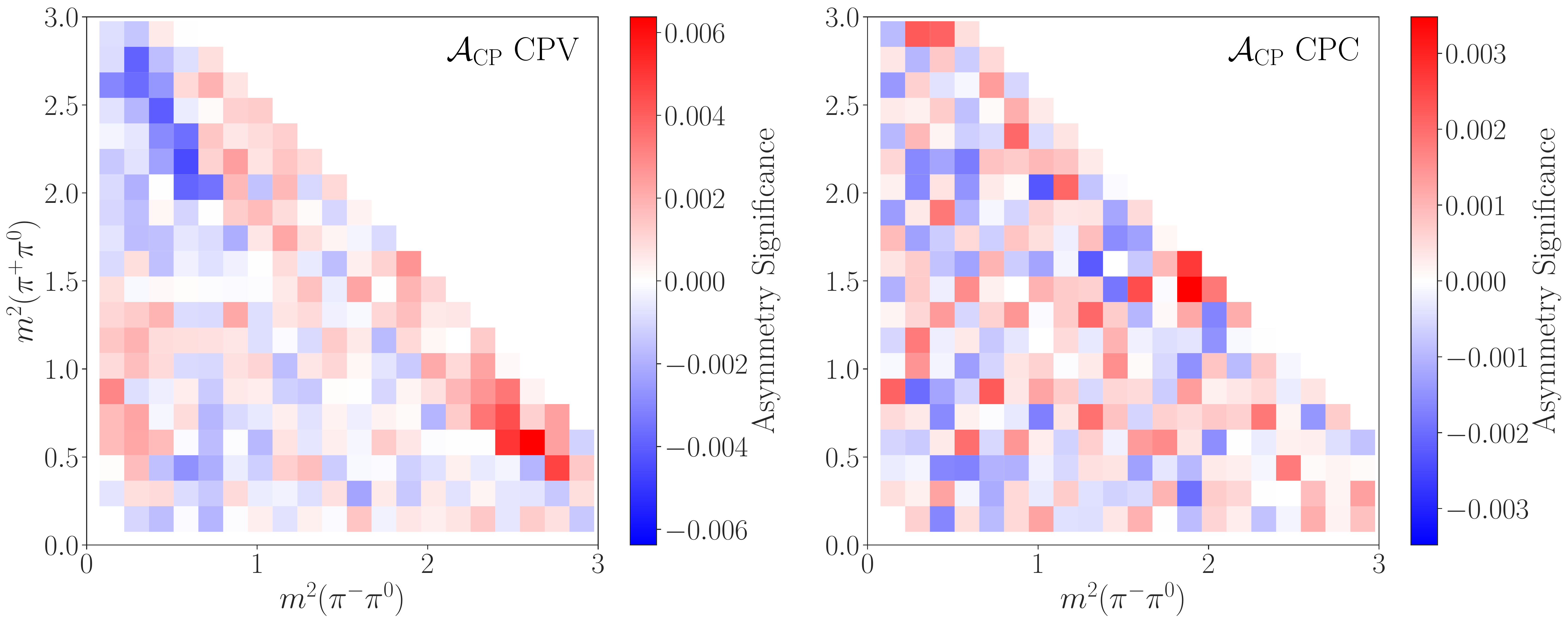}
\caption{The binned CP asymmetry significance $\mathcal{A}_{\text{CP}} / \delta \mathcal{A}_{\text{CP}}$, with $ \delta \mathcal{A}_{\text{CP}}$ given in Eq.~\eqref{eq:ACP_uncertainty}, for $B^0\to K^+\pi^-\pi^0$ ($D^0\to\pi^+\pi^-\pi^0$) Dalitz plot are shown in the top (bottom) panels for 
$N=\bar N=10^3 (10^6)$ event samples averaged over an ensemble of $N_e = 100$ ($N_e = 1$) datasets with CPV toy example show on the left and CP conserving datasets on the right. 
}
\label{fig:asymm_sig_B}
\end{center}
\end{figure}

Fig.~\ref{fig:asymm_sig_B} shows the CP asymmetry significance, $\mathcal{A}_{\text{CP}} / \delta \mathcal{A}_{\text{CP}}$, where the error 
on the CP asymmetry is given by
\begin{equation}\label{eq:ACP_uncertainty}
    \delta \mathcal{A}_{\text{CP}} = \sqrt{\frac{1 - \mathcal{A}^2_{\text{CP}}}{N+\bar{N}}}.
\end{equation}
The upper panels in Fig.~\ref{fig:asymm_sig_B} show the CP asymmetry for the case of $B^0\to K^+\pi^-\pi^0$ decay, with our toy example CPV amplitude model (left panel) leading to clearly identifiable regions in the Dalitz plot with CPV, and only noise in the Dalitz plot for the CPC case (right). The difference between CPV and CPC decays is less pronounced in the $D^0\to \pi^+\pi^-\pi^0$. Even so the Wasserstein distance based test statistics can still lead to exclusions of CPC hypothesis (cf. Fig.~\ref{fig:PDF:CDF:SF_D}, where the analysis was done for a sample size of $N=\bar N=10^5$).

\begin{figure}[t]
\begin{center}
\includegraphics[width=0.45\textwidth]{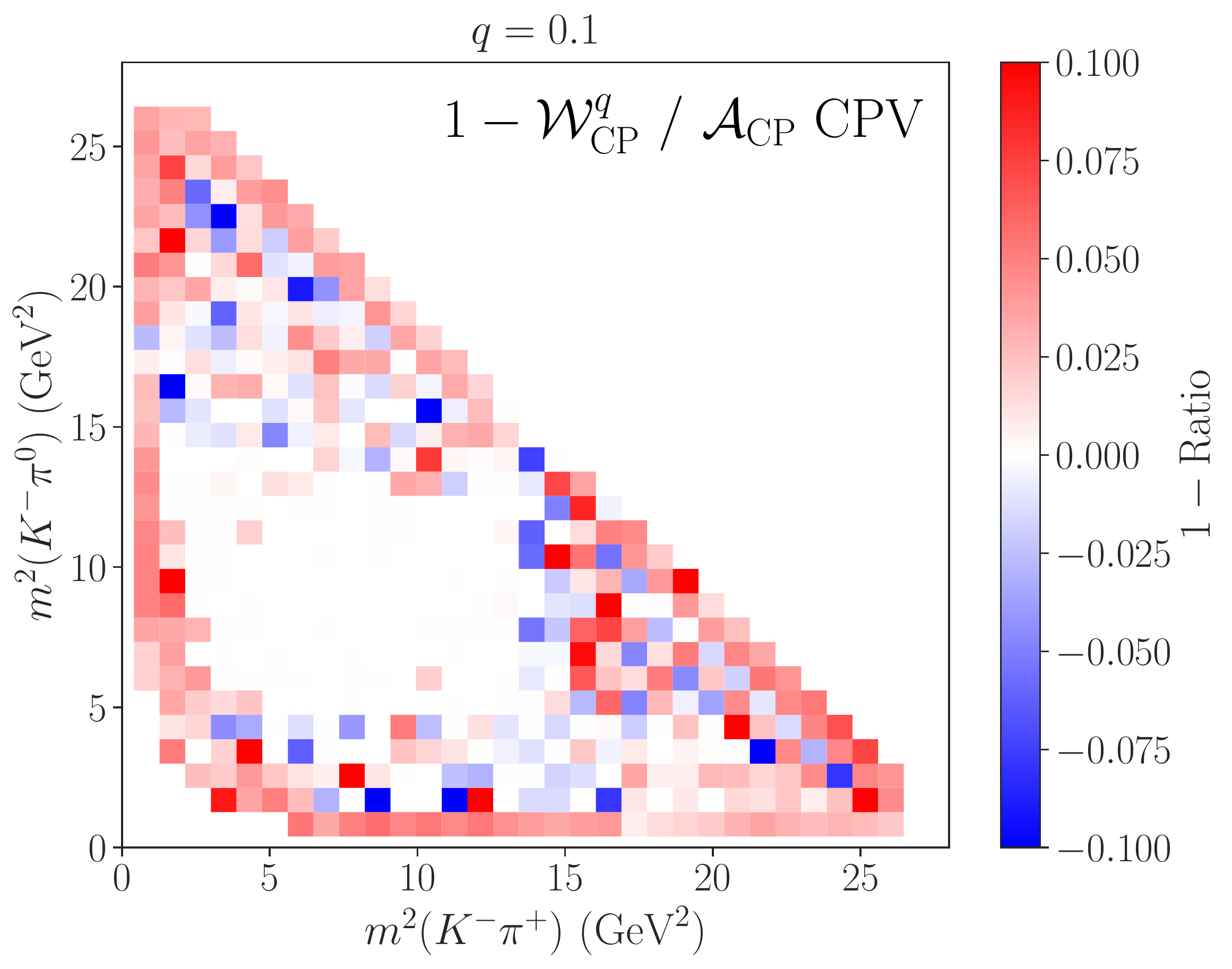}
\includegraphics[width=0.45\textwidth]{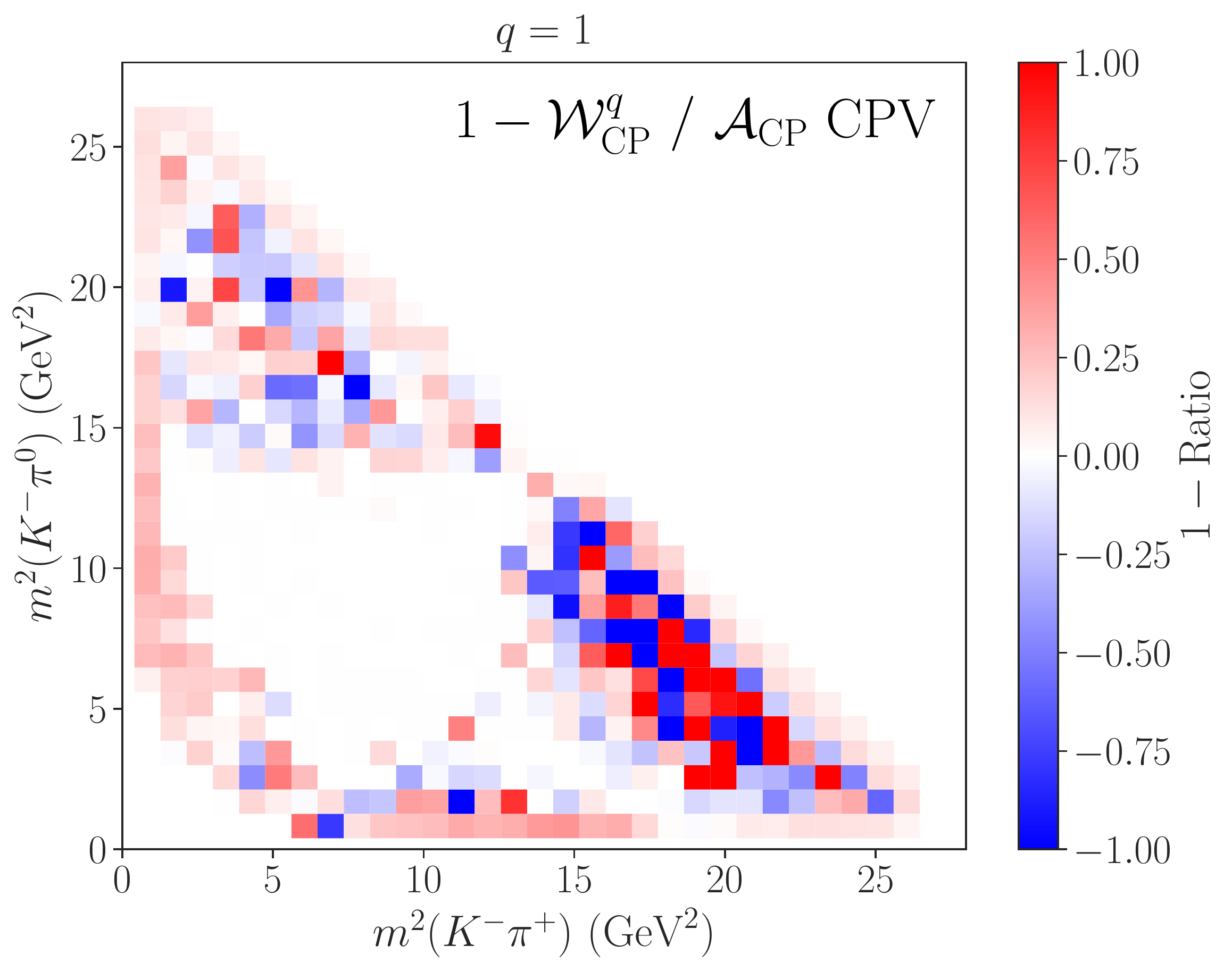}
\includegraphics[width=0.45\textwidth]{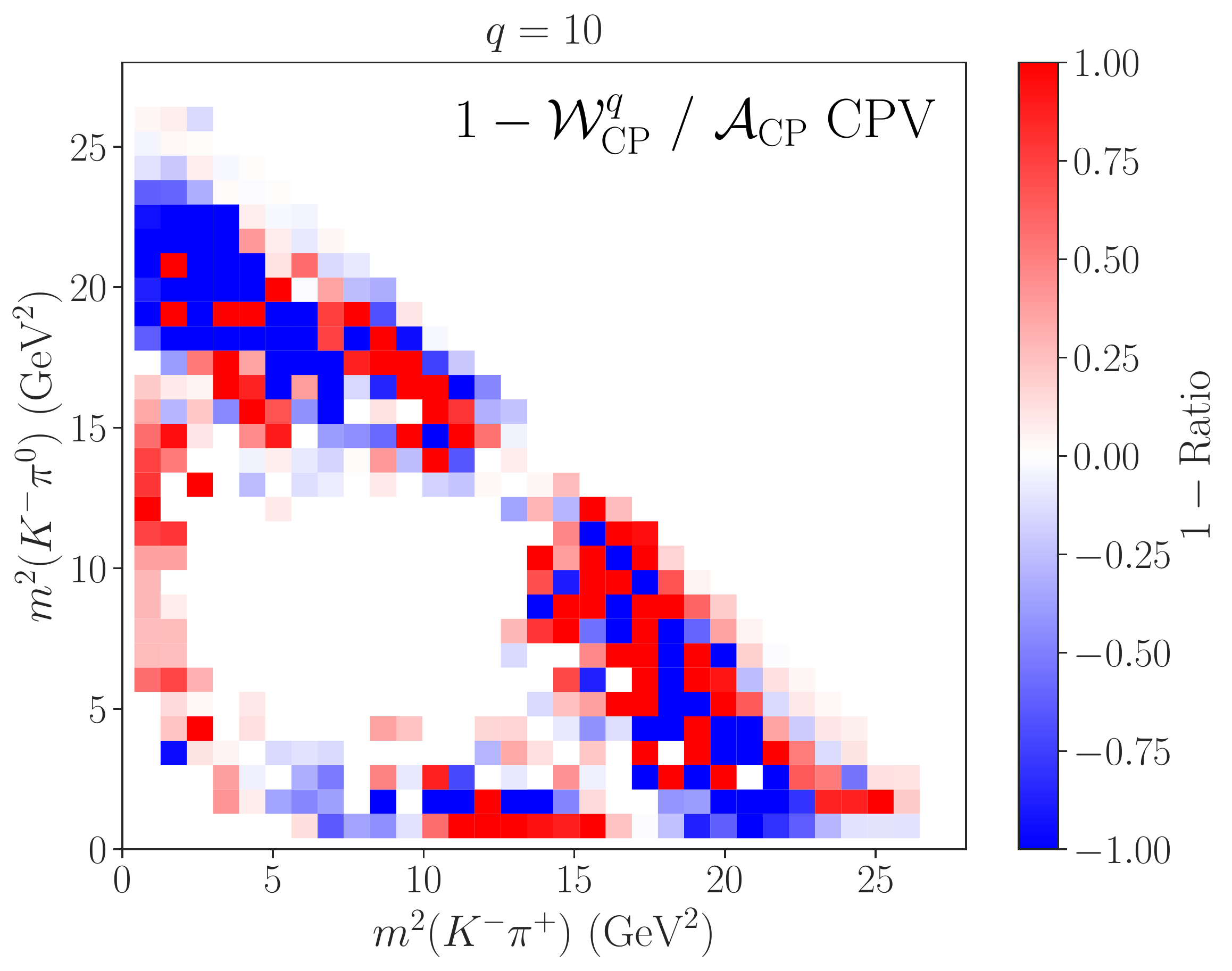}
\caption{We show $1-\mathcal{W}^q_{\text{CP}} / \mathcal{A}_{\text{CP}}$ 
for each bin within the two--dimensional Dalitz plot. Smaller values indicate better agreement. This is shown for CP violating $B^0 \rightarrow K^+ \pi^- \pi^0$ decays with $n_{\text{bins}} = 50$, $q = \{0.1, 1, 10\}$, and averaged over an $N_e=100$ dataset ensemble where each dataset contains 
$N=\bar{N} = 10^3$ $B^0$ and $\bar{B^0}$ events. 
}
\label{fig:dalitz_1-ratio_q_0.1}
\end{center}
\end{figure}

\begin{figure}[t]
\begin{center}
\includegraphics[width=0.85\textwidth]{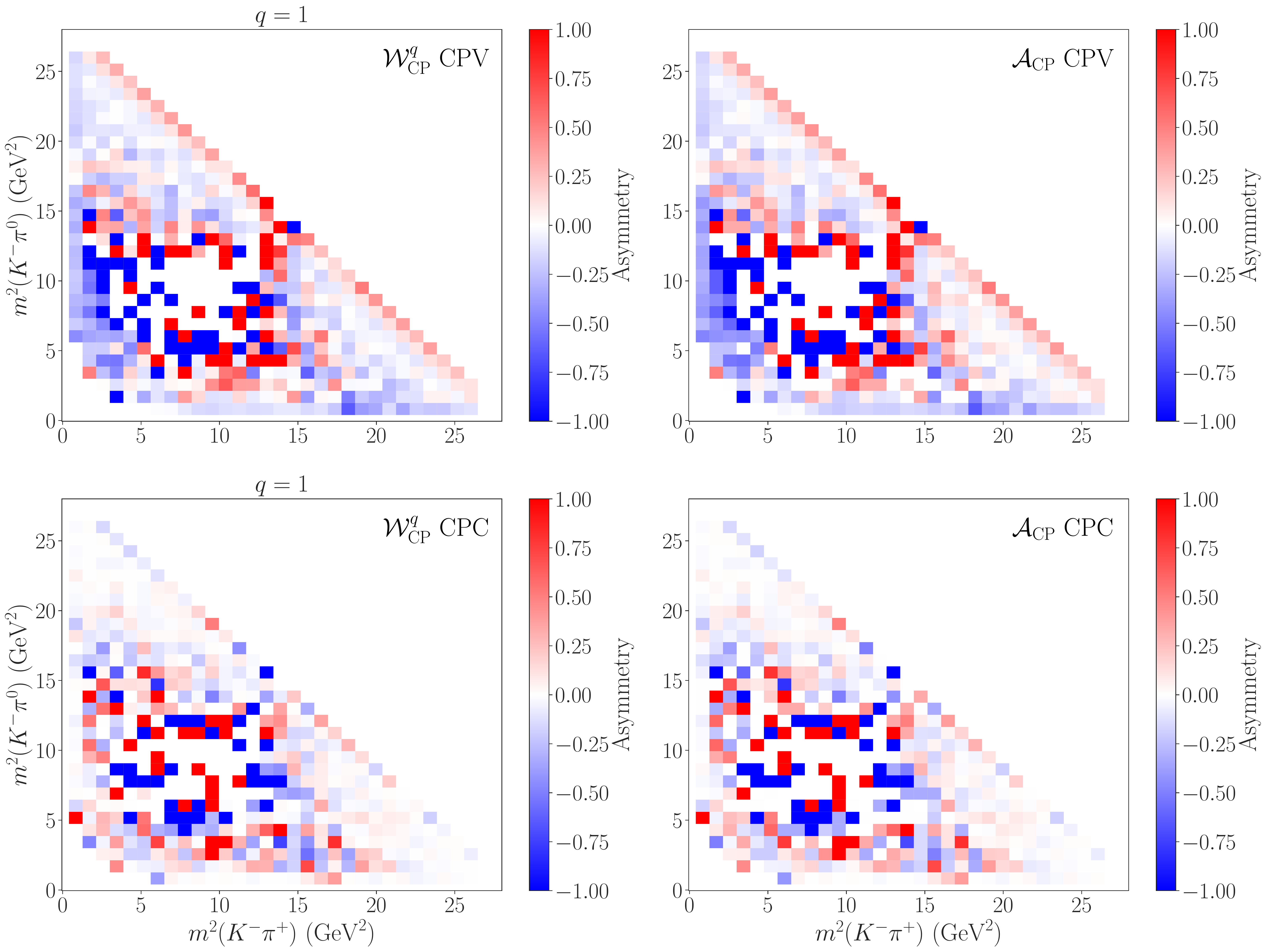}\\[0.1mm]
\includegraphics[width=0.85\textwidth]{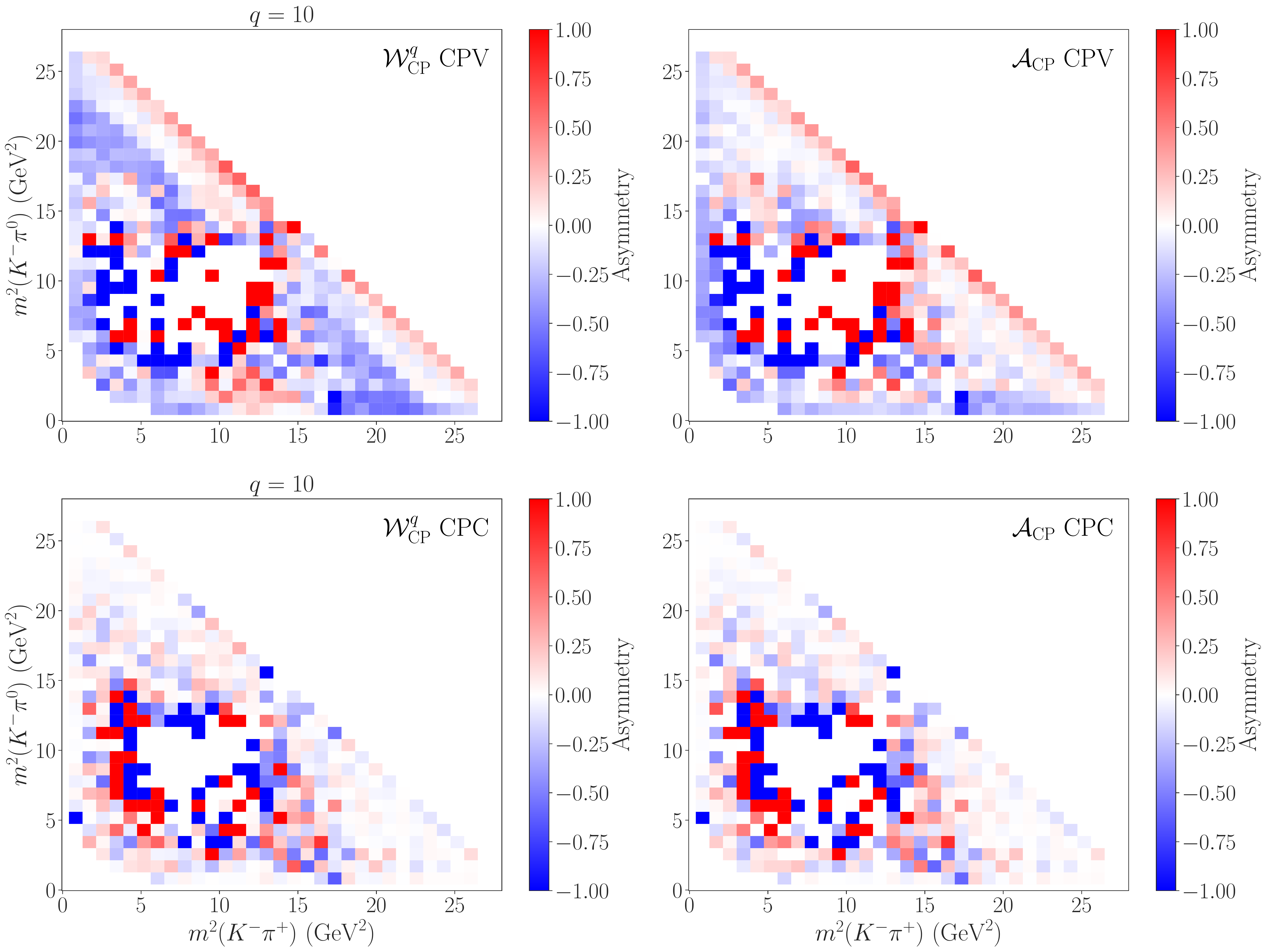}
\caption{Binned Dalitz plot comparison between the Wasserstein asymmetry $\mathcal{W}_{\text{CP}}^q$ (left) for $q = \{1 ,10\}$ (shown in top two and bottom two rows) and direct CP asymmetry $\mathcal{A}_{\text{CP}}$ (right), shown for CP violating $B^0\to K^+\pi^-\pi^0$ decays (1st and 3rd row) and CP conserving decays, i.e., the decays in which the asymmetries in the amplitude model were set to zero  (2nd and 4th row). The results shown are normalized and averaged over an ensemble of $N_e=100$ datasets, each containing $N=\bar N= 10^3$
events. 
}
\label{fig:WqCP:ACP:q:1:10}
\end{center}
\end{figure}

Fig.~\ref{fig:dalitz_1-ratio_q_0.1} shows the relative difference between the binned Wasserstein distance asymmetry, $\mathcal{W}^q_{\text{CP}} $, defined in Eq.~\eqref{eq:Wq:asymm}, and the CP asymmetry, $\mathcal{A}_{\text{CP}}$, Eq.~\eqref{eq:A_CP}. This complements Fig.~\ref{fig:asymmetry_plots:q0.1} and Fig. \ref{fig:WqCP:ACP:q:1:10}, which show the actual values of the  binned Wasserstein distance asymmetry, $\mathcal{W}^q_{\text{CP}} $,  and the CP asymmetry, $\mathcal{A}_{\text{CP}}$ for $q=0.1$ and $q=1,10$, respectively. We see that for the optimal value of $q=0.1$ the binned Wasserstein distance asymmetry, $\mathcal{W}^q_{\text{CP}}$ almost completely matches $\mathcal{A}_{\text{CP}}$ up to $\sim 10\%$ relative differences, where the differences are even closer to just a few percent in the regions of the Dalitz plot where the CP asymmetry significance is large, cf. Fig.~\ref{fig:asymm_sig_B}. The $\mathcal{W}^q_{\text{CP}}$ still tracks well the CP asymmetry $\mathcal{A}_{\text{CP}}$, however with exaggerated differences in the regions of the Dalitz plot with lower CP asymmetry significance.

\begin{figure}[t]
\begin{center}
\includegraphics[width=0.49\textwidth]{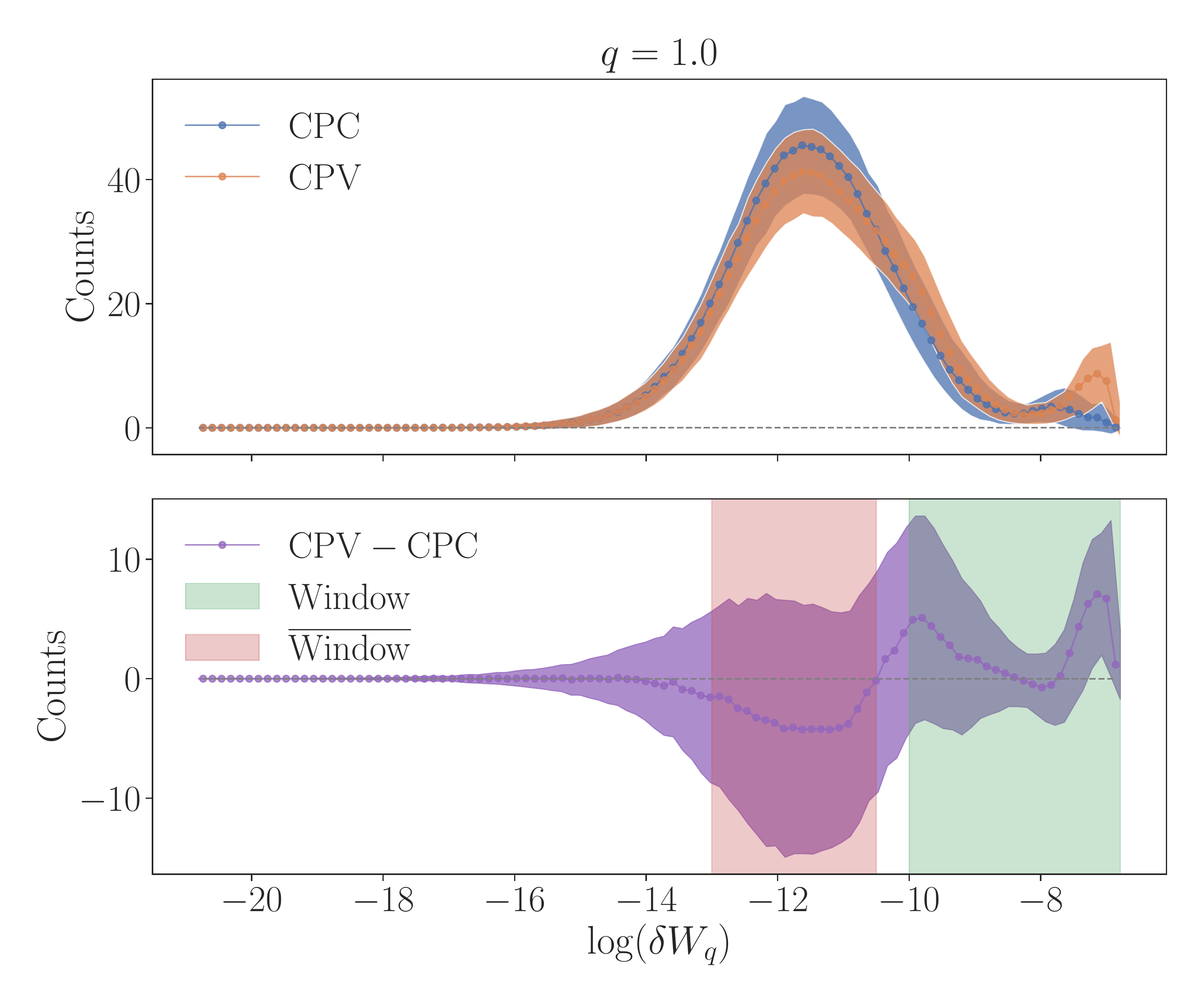}
\includegraphics[width=0.49\textwidth]{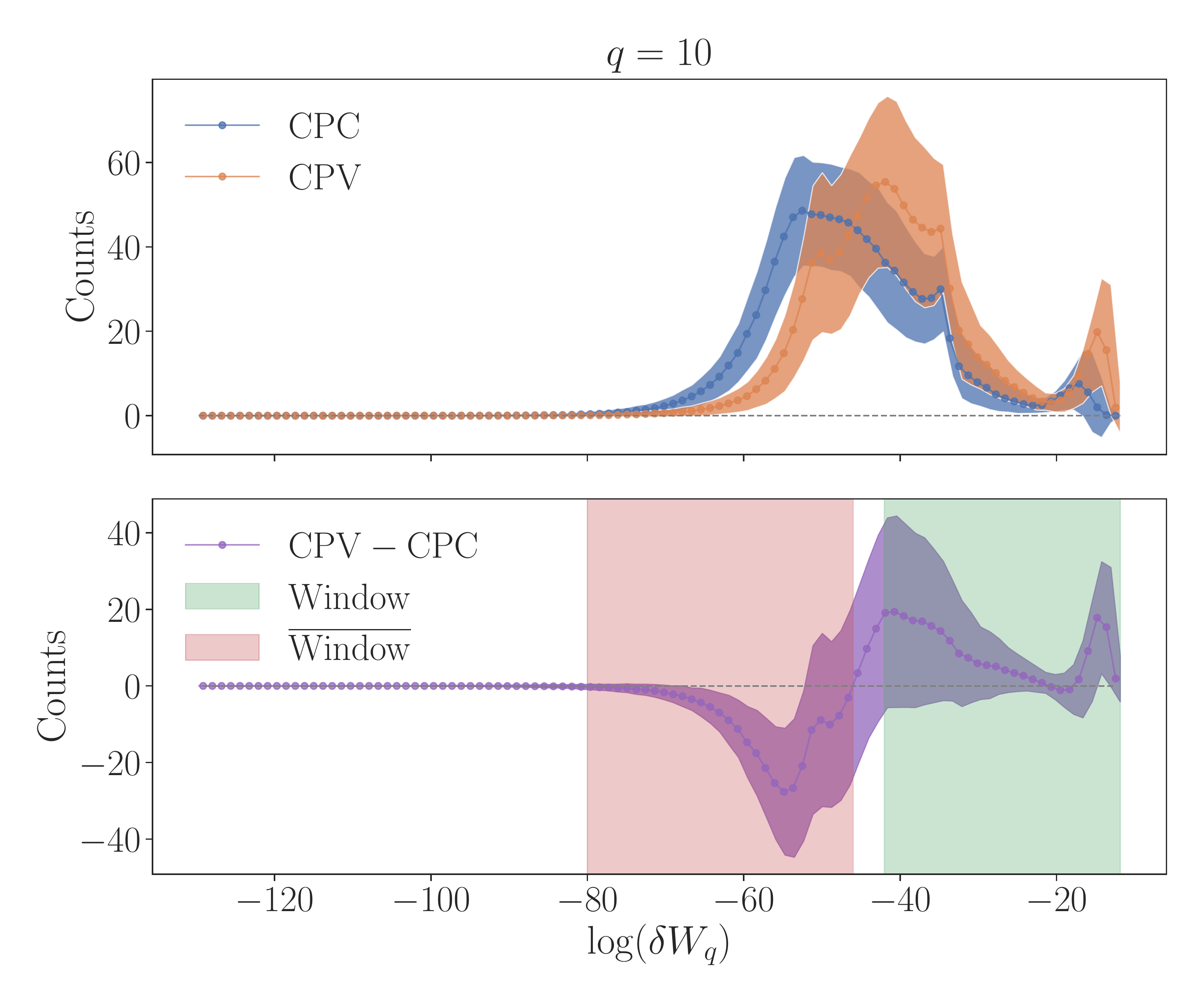}
\caption{The same as Fig.~\ref{fig:count_difference} but for $q=\{1,10\}$. On the lower panels the green bands denote the $[\delta W^{\text{win}}_{\text{min}}, \delta W^{\text{win}}_{\text{max}}]$ ranges and the red bands the $[\delta W^{\overline{\text{win}}}_{\text{min}}, \delta W^{\overline{\text{win}}}_{\text{max}}]$ range, each for the corresponding $q$ values.
}
\label{fig:count_difference_q_1_10}
\end{center}
\end{figure}

Fig.~\ref{fig:count_difference_q_1_10} shows the binned Wasserstein asymmetry $\mathcal{W}_{\text{CP}}^q$ and direct CP asymmetry $\mathcal{A}_{\text{CP}}$ for $q = \{1 ,10\}$  both for CPV and CPC $B^0\to K^+\pi^-\pi^0$ decays, where the same inputs for the $B^0$ decays were used as in the main text. The results shown were averaged over an ensemble of $N_e=100$ datasets, each containing $N=\bar N= 10^3$ events. 
 This figure supplements Fig.~\ref{fig:asymmetry_plots:q0.1} for $q=0.1$ in the main text. We see that in all cases, $q=\{0.1, 1, 10\}$, the  $\mathcal{W}_{\text{CP}}^q$ faithfully traces $\mathcal{A}_{\text{CP}}$ for $q = \{1 ,10\}$ throughout the Dalitz plot, especially where the CP asymmetries are statistically most significant. 

Fig.~\ref{fig:count_difference} shows $\log(\delta W_q)$ distributions and the difference between CPV and CPC $\log(\delta W_q)$ distributions for $q=1,10$. Compared to the $q=0.1$ case, shown in Fig.~\ref{fig:count_difference}, there is a more pronounced deficit of $\delta W_q$ counts in CPV distribution relative to the CPC one for the intermediate $\log(\delta W_q)$. The window function $w(x)$, Eq.~\eqref{eq:window}, is therefore chosen to have support both for the $+1$ (green bands in Fig.~\ref{fig:count_difference}) and $-1$ (red bands) weights. 

\begin{figure}[t]
\begin{center}
\includegraphics[width=0.65\textwidth]{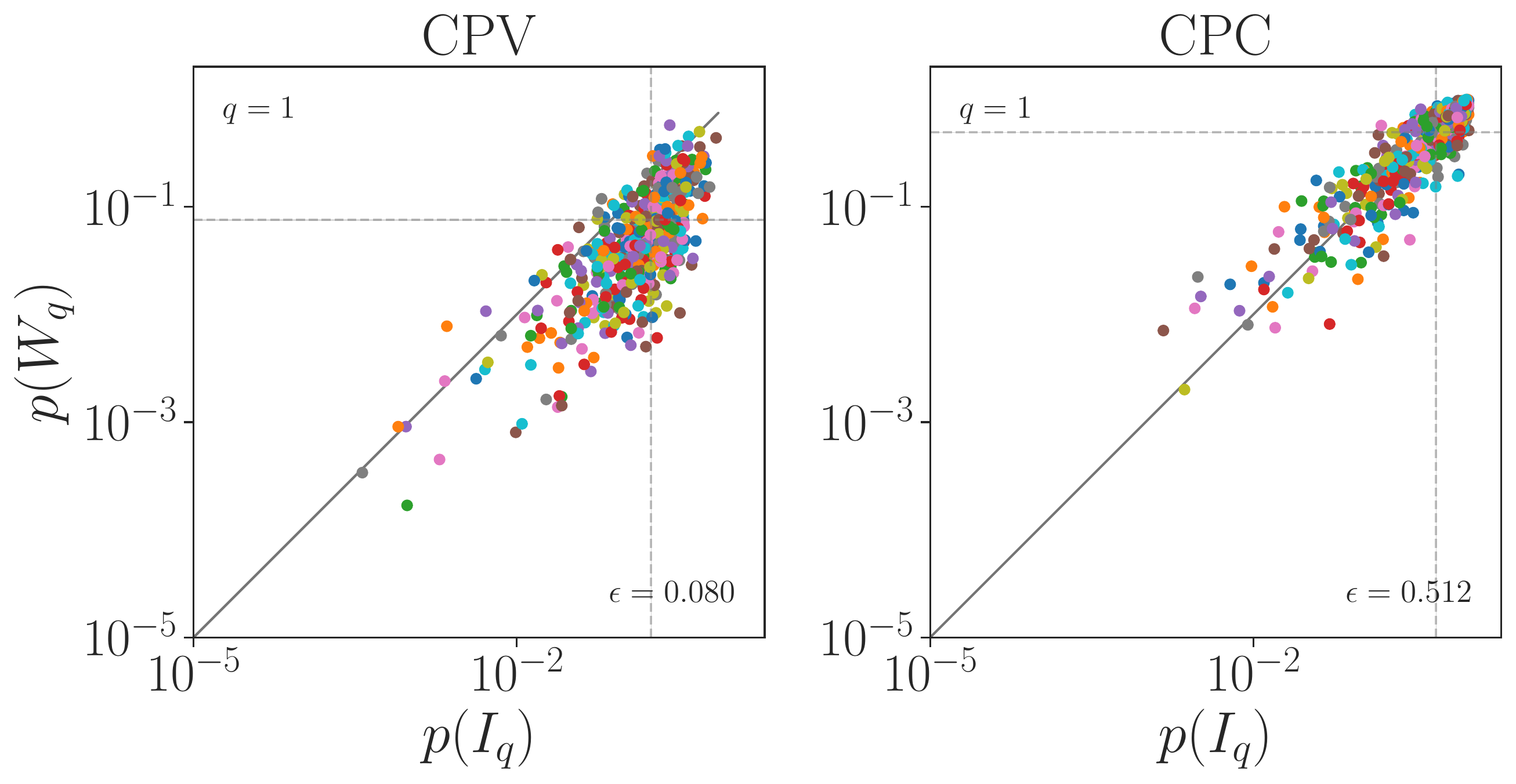}\\[0.1mm]
\includegraphics[width=0.65\textwidth]{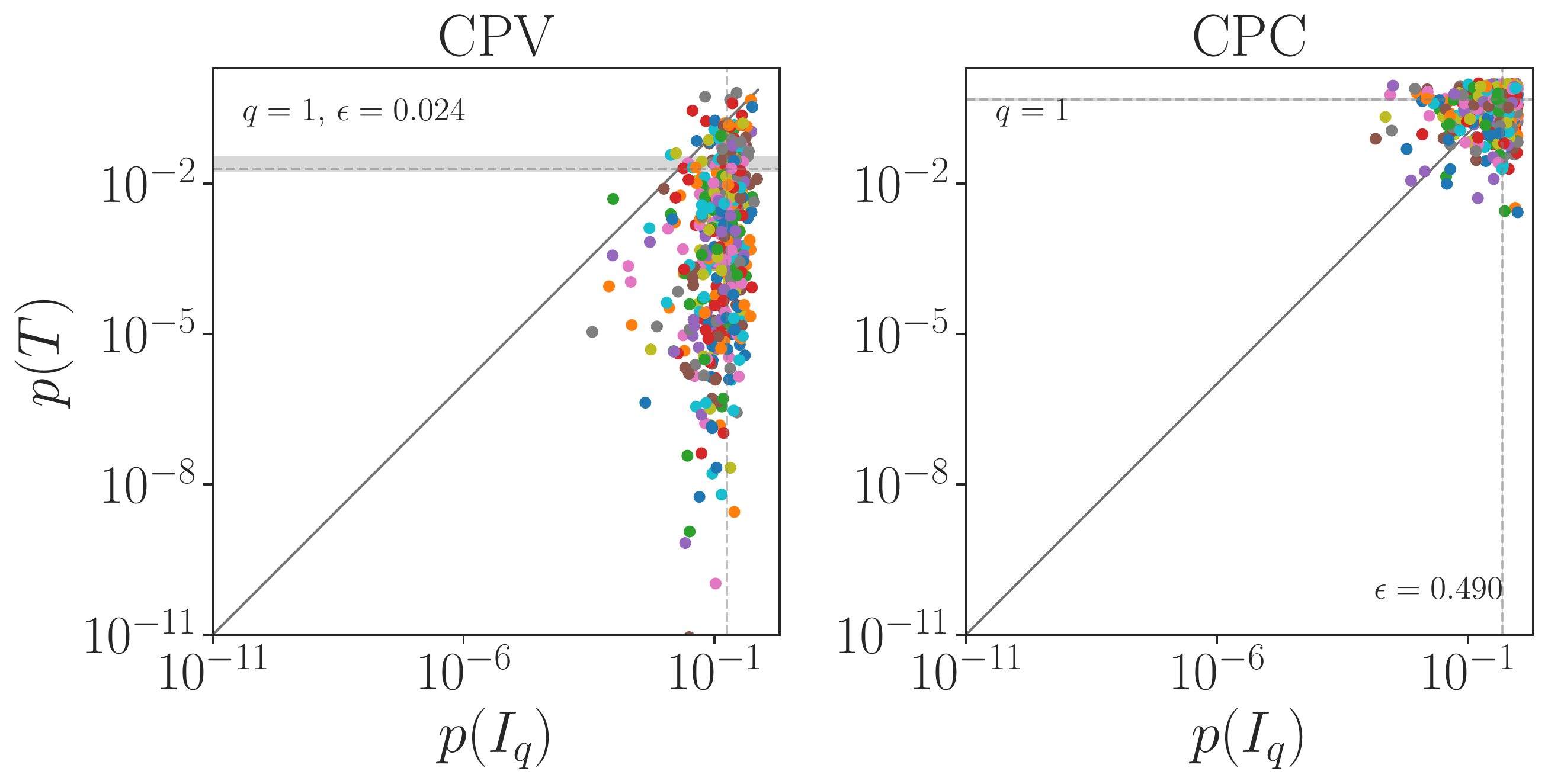}\\[1.0mm]
\includegraphics[width=0.65\textwidth]{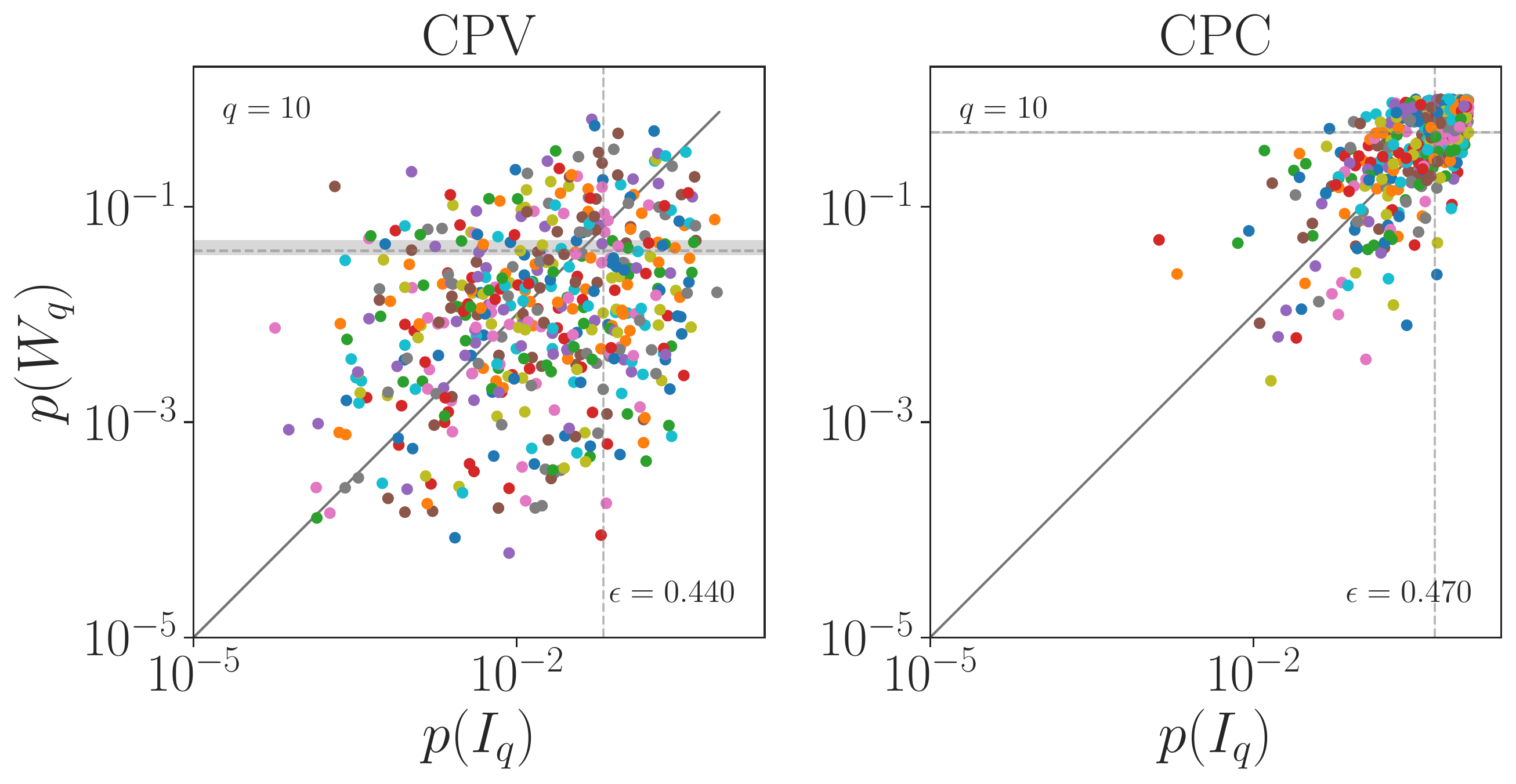}\\[1.0mm]
\includegraphics[width=0.65\textwidth]{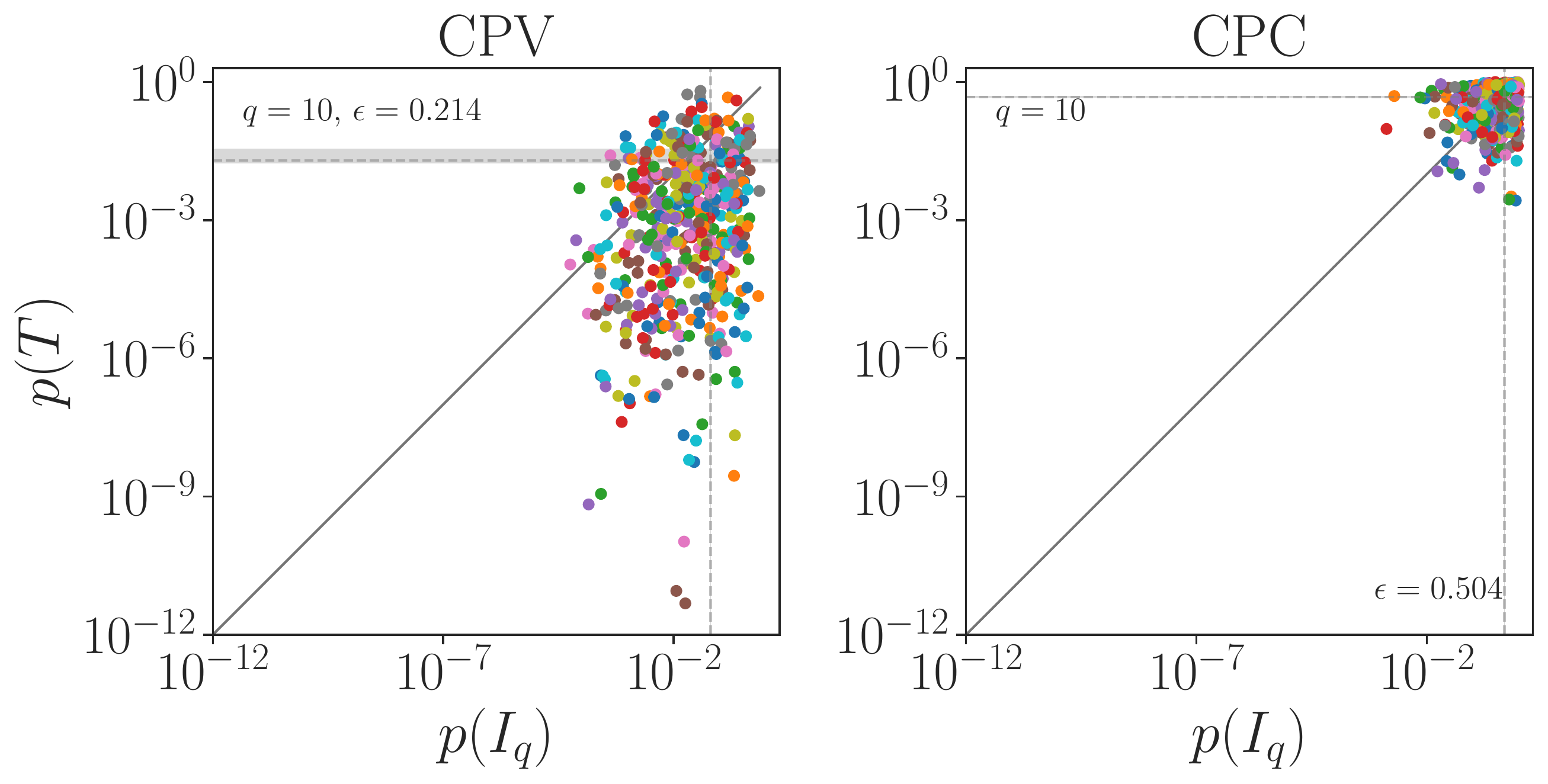}
\caption{The comparison of CL exclusion of the CP conserving hypothesis  either using energy test ($p(T)$), the Wasserstein distance ($p(W_q)$), or the windowed Wasserstein distance ($p(I_q)$), with the window function $w(x)$ as denoted in Fig.~\ref{fig:count_difference_q_1_10}. The top (bottom) two rows are for $q=1(10)$, with $500$ distinct CP violating (conserving) $B^0\to K^+\pi^-\pi^0$ decay datasets shown on the right (left). 
}
\label{fig:IWq_vs_Wq_vs_T_q_1}
\end{center}
\end{figure}

Fig. \ref{fig:IWq_vs_Wq_vs_T_q_1} is the $q=1,10$ complement of Fig.~\ref{fig:IWq_vs_Wq} in the main text. It shows, $500$ distinct datasets each with $N=\bar N=10^3$ events, the confidence levels with which the CP conserving hypothesis is excluded when applying different tests, either the energy test, giving CLs denoted with $p(T)$, the Wasserstein distance statistic test, giving $p(W_q)$, or the windowed Wasserstein distance statistic, giving $p(I_q)$. The windows and anti-windows for $q=1,10$ are shown in Fig.~\ref{fig:count_difference_q_1_10}. For $q=10$ the performance of the windowed Wasserstein distance is comparable yet slightly less sensitive than the energy test, while for $q=1$ the sensitivity of windowed Wasserstein distance statistic is significantly reduced. For $q=10$ the use of windowed Wasserstein distance is comparable to the simple Wasserstein distance statistic, i.e., it does not lead to any real gain, while for 
 $q=1$ case, the selected windows and anti-window reduce sensitivity of $I_q$ compared to $W_q$. However, additional tuning of the window and anti-window regions could be conducted to maximize significance. 

\section{Energy test}\label{app:energy_test}
The energy test, introduced in \cite{Aslan:2004}, is an unbinned two-sample test utilizing a test statistic, $T$, to analyze average distances between data points in phase space. The first proposal to utilize the energy test in searches for CP violation was described in \cite{Williams:2011cd}
and subsequent analyses performed in 
\cite{LHCb:2014nnj,Parkes:2016yie}. 

The statistic utilizes a weighting (distance) function $\psi_{ij}$ dependent on the distance $d_{ij}$ between the $i$th and $j$th event in the first and second sample, respectively. For searches of CP violation the two samples are distinguished by flavor ($B^0$ and $\bar{B}^0$). The test statistic is defined as \cite{Aslan:2004,Williams:2011cd}
\begin{equation}
\label{eq:T:def}
    T = \sum_{i,j > i}^N \frac{\psi_{ij}}{N(N-1)} + \sum_{i,j > i}^{\bar{N}} \frac{\psi_{ij}}{\bar{N}(\bar{N}-1)} - \sum_{i,j}^{N, \bar{N}}\frac{\psi_{ij}}{N\bar{N}},
\end{equation}  
where $N$, $\bar{N}$ denote the total number of events in the $B^0$ and $\bar B^0$ samples, respectively . The weighting function $\psi_{ij}$ is chosen such that the weight decreases with increasing distance, $d_{ij}$, between points. The summations in Eq.~\eqref{eq:T:def} are, from left to right, over $B^0$ sample, $\bar B^0$ sample and over both $B^0$ (index $i$) and $\bar B^0$ (index $j$) samples, respectively. The form of  the test statistic $T$ is motivated by the form of the electrostatic energy for overlapping distributions of positive and negative charges, in which case $\psi_{ij}\propto 1/d_{ij}$. If the two charge distributions are exactly the same, the net charge distribution is zero, and $T$ vanishes.   

The functional form on the weighting function $\psi_{ij}$ can be freely chosen, for instance in order to increase the sensitivity to local asymmetries at some typical length-scales, minimizing dilutions due to averaging over large Dalitz plot areas. We follow Refs.~\cite{LHCb:2014nnj,Parkes:2016yie} and choose a Gaussian weighting function
\begin{equation}
    \psi_{ij} \equiv \psi(d_{ij};\sigma) = e^{-d^2_{ij}/2\sigma^2},
\end{equation}
where $\sigma$ is a tunable parameter describing the effective radius between data points where asymmetry is measured, while $d_{ij}$ is the Euclidean distance in the Dalitz plot.

For events sampled from two identical distributions $T$ is expected to fluctuate close to zero, while for samples drawn from dissimilar distributions $T$ will tend to a nonzero value. To obtain the null hypothesis PDF for $T$ we use the master method described in Sect.~\ref{subsec:gen_CPC_dist_methods}. The labels for $B^0$ and $\bar B^0$ samples are ignored, and the $N+\bar N$ events randomly assigned to ${\cal E}$ and $\bar {\cal E}$ samples, each with $N=\bar N$ events, thus simulating the CP even datasets.
Repeating this 
$n$ times give a null hypothesis PDF for $T$, which is then fitted to a gamma 
distribution, used finally to obtain the $p-$values corresponding to the ``measured'' value of $T$. 

The computation of CP conserving $T$ distributions was done with the \texttt{Manet} software package \cite{Parkes:2016yie} (while for single computations our own implementation was used, see App.~\ref{sec:code}). The analysis was performed on $N = \bar{N} = 10^3$ $B^0$ and $\bar{B}^0$ events generated by \texttt{AmpGen} \cite{ampgen}. The null hypothesis $T$-distributions were computed with $N=10^3$ permutations, while the tunable parameter in the weighting function, $\sigma \approx 0.2 \text{ GeV}^2$, was chosen to maximize the significance (minimize $p-$value) in the case of a CP violating sample. 

\clearpage

\bibliographystyle{JHEP}
\bibliography{CPV_ML_biblio}
  
\end{document}